\documentstyle[11pt,epsfig,amsfonts]{article}
\renewcommand{\theequation}{\arabic{section}.\arabic{equation}}

\title{Radiation reaction\protect\\ in 2+1 electrodynamics}
\author{\bf Yurij Yaremko\footnote{Electronic mail: yar@ph.icmp.lviv.ua}}
\date{\it Institute for Condensed Matter Physics, \\
1 Svientsitskii St., 79011 Lviv, Ukraine}
\begin{document}

\maketitle
\begin{abstract}
A self-action problem for a pointlike charged particle arbitrarily moving 
in flat spacetime of three dimensions is considered. Outgoing waves carry 
energy-momentum and angular momentum; the radiation removes energy and 
angular momentum from the source which then undergoes a radiation reaction.
We decompose Noether quantities carried by electromagnetic field into bound 
and radiative components. The bound terms are absorbed by individual 
particle's characteristics within the renormalization procedure. Radiative 
terms together with already renormalized 3-momentum and angular momentum 
of pointlike charge constitute the total conserved quantities of our 
particle plus field system. Their differential consequences yield the 
effective equation of motion of radiating charge in an external 
electromagnetic field. In this integrodifferential equation the radiation 
reaction is determined by Lorentz force of pointlike charge acting upon 
itself plus nonlocal term which provides finiteness of the self-action.
\end{abstract}


\section{Introduction}\label{intro}
\setcounter{equation}{0}

Recently \cite{Gl,KLS}, there has been considerable interest in 
renormalization procedure in classical electrodynamics of a point particle 
moving in flat space-time of arbitrary dimensions. The main task is to 
derive the analog of the well-known Lorentz-Dirac equation \cite{Dir}. 
The Lorentz-Dirac equation is an equation of motion for a charged particle 
under the influence of an external force as well as its own electromagnetic 
field. (For a modern review see Refs. \cite{PsPr} and \cite{TVW}.)

A special attention in Refs. \cite{Gl} and \cite{KLS} is devoted to the 
mass renormalization in $2+1$ theory. (Note that electrodynamics in 
Minkowski space ${\mathbb M}_{\,3}$ is quite different from the 
conventional $3+1$ electrodynamics where one space dimension is reduced 
because of symmetry of a specific problem. For example, small charged balls 
on a plane are interacted inversely with the square of the distance between 
them, while in ${\mathbb M}_{\,3}$ the Coulomb field of a small static 
charged disk scales as $|{\bf r}|^{-1}$.) An essential feature of $2+1$ 
electrodynamics is that Huygens principle does not hold and radiation 
develops a tail, as it is in curved space-time of four dimensions \cite{WB} 
where electromagnetic waves propagate not just at the speed of light but 
all speeds smaller than or equal to it.

In Refs. \cite{Gl} and \cite{KLS} the self-force on a pointlike particle is 
calculated from the local fields in the immediate vicinity of its 
trajectory. The schemes involve some prescriptions for subtracting away the 
infinite contributions to the force due to the singular nature of the field 
on the particle's world line. In Ref. \cite{KLS} the procedure of 
regularization is based on the methods of functional analysis which are 
applied to the Tailor expansion of the retarded Green's function. The 
authors derive the covariant analog of the Lorentz-Dirac equation which 
is something other than that obtained in Ref. \cite{Gl}. Both the divergent 
self-energy absorbed by the ``bare'' mass of pointlike charge and the 
radiative term which leads an independent existence are nonlocal. (They 
depend not only on the current state of motion of the particle but also on 
its past history.)

In this paper we develop a consistent regularization procedure which 
exploits the symmetry properties of $2+1$ electrodynamics. It can be 
summarized as a simple rule which obeys the spirit of the Dirac scheme of 
decomposition of the vector potential of a pointlike charge.

According to the scheme proposed by Dirac in his classical paper 
\cite{Dir}, one can decompose retarded Green's function associated with 
the four-dimensional Maxwell field equation $G^{ret}(x,z)=G^{sym}(x,z)+
G^{rad}(x,z)$. The first term, $G^{sym}(x,z)$, is one-half sum of the 
retarded and the advanced Green's functions; it is just singular as 
$G^{ret}(x,z)$. The second one, $G^{rad}(x,z)$, is one-half of the retarded 
minus one-half of the advanced Green's functions; it satisfies the 
homogeneous wave equation. Convolving the source with the Green's functions
$G^{sym}(x,z)$ and $G^{rad}(x,z)$ yields the singular and the radiative 
parts of the vector potential of a pointlike charge, respectively.

The analogous decomposition of Green's function in curved space-time is 
much more delicate because of richer causal structure. Detweiler and 
Whiting \cite{DW} modified the singular Green's function by means of
two-point function $v(x,z)$ which is symmetric in its arguments. It is 
constructed from the solutions of the homogeneous wave equation in such a 
way that a new symmetric Green's function 
$G^S(x,z)=G^{sym}(x,z)+1/(8\pi)v(x,z)$ has no support within the null cone.

It is obvious that the physically relevant solution to the wave equation is 
the retarded solution. In Ref. \cite{Teit} the Lorentz-Dirac equation is 
derived within the framework of retarded causality. Teitelboim substituted
the retarded Li\'enard-Wiechert fields in the electromagnetic field's 
stress-energy tensor and computed the flow of energy-momentum which flows 
across a tilted hyperplane which is orthogonal to particle's four-velocity 
at the instant of observation. The effective equation of motion is obtained 
in Ref. \cite{Teit} via consideration of energy-momentum conservation. 
Similarly, L\'opez and Villarroel \cite{LV} found the total angular 
momentum carried by the electromagnetic field of a pointlike charge. It was 
shown \cite{Yar03} that the Lorentz-Dirac equation can be derived from the 
energy-momentum and angular momentum balance equations.  In Ref. 
\cite{Yar04} the analog of the Lorentz-Dirac equation in six dimensions 
is obtained via analysis of 21 conserved quantities which correspond to the 
symmetry of an isolated point particle coupled with an electromagnetic 
field. (First this equation was obtained by Kosyakov in Ref. \cite{Kos} via 
the consideration of energy-momentum conservation. An alternative 
derivation was produced by Kazinski {\it et al} in Ref. \cite{KLS}.)

In nonlocal theories, the computation of Noether quantities is highly 
nontrivial. Quinn and Wald \cite{QW} studied the energy-momentum 
conservation for a point charge moving in curved space-time. The Stokes
theorem is applied to the integral of flux of electromagnetic energy over 
the compact region $V(t^+,t^-)$. It is expanded to the limits $t^-\to 
-\infty$ and $t^+\to +\infty$, so that finally the boundary of the 
integration domain involves smooth spacelike hypersurfaces at the remote 
past and in the distant future. The space-time is asymptotically flat here. 
The authors proved that the net energy radiated to infinity is equal to the 
total work done on the particle by the electromagnetic self-force. 
(DeWitt-Brehme \cite{WB,Hb} radiation-reaction force is meant.) It was 
shown also \cite{QW} that the total work done by the gravitational 
self-force is equal to the energy radiated (in gravitational waves) by the 
particle. (The effective equation of motion of a point mass undergoing 
radiation reaction is obtained in Ref. \cite{MST}; see also review 
\cite{Pois} where the motion of a point electric charge, a point scalar 
charge, and a point mass in curved space-time is considered in details.)

In Ref. \cite{Q} Quinn derived the effective equation of motion of a point 
particle coupled with a scalar field moving in curved space-time. The 
author establishes that the total work done by the scalar self-force 
matches the amount of energy radiated away by the particle.

In the present paper we calculate the total flows of energy-momentum and 
angular momentum of the retarded field which flow across a hyperplane 
$\Sigma_t=\{y\in{\mathbb M}_{\,3}:y^0=t\}$ associated with an unmoving 
observer. The field is generated by a pointlike charge arbitrarily moving 
in flat Minkowski space ${\mathbb M}_{\,3}$ of three dimensions. This paper 
is organized as follows. In Section \ref{field} we recall the retarded and 
the advanced Green's functions associated with the three-dimensional 
D'Alembert operator. Convolving them with the point source, we derive the 
retarded and the advanced vector potential and field strengths. In Section 
\ref{trace} we trace a series of stages in the calculation of a surface 
integral which gives the energy-momentum carried by the retarded 
electromagnetic field. (Details are given in the appendixes.) We introduce 
appropriate coordinate system centered on an accelerated world line and we 
express the components of the Maxwell energy-momentum tensor density in 
terms of these curvilinear coordinates. We integrate it over the variables 
which parametrize the surface of integration $\Sigma_t$. The resulting 
expression becomes a combination of two-point functions depending on the 
state of the particle's motion at instants $t_1$ and $t_2$ before the 
observation instant $t$. They are integrated over the particle's world line 
twice. We arrange them in Section \ref{meq}. We split the momentum 
three-vector carried by electromagnetic field into singular and radiative 
parts by means of the Dirac scheme which deals with fields defined on the 
world line only. All diverging quantities have disappeared into the 
procedure of mass renormalization while radiative terms lead independent 
existence. In an analogous way we analyze the angular momentum of the
electromagnetic field. Total energy-momentum and total angular momentum of 
our particle plus field system depend on already renormalized particle's 
individual characteristics and radiative parts of ``electromagnetic''
Noether quantities. Having differentiated the conserved quantities we 
derive the effective equation of motion of a radiating charge. In 
Section \ref{concl}, we discuss the result and its implications.

\section{Electromagnetic potential and electromagnetic field\\ 
in 2+1 theory}\label{field}
\setcounter{equation}{0}

We consider an electromagnetic field produced by a particle with 
$\delta$-shaped distribution of the electric charge $e$ moving on a world 
line $\zeta\subset {\mathbb M}_{\,3}$ described by functions $z^\alpha(\tau)$ 
of proper time $\tau$. The Maxwell equation
\begin{equation}\label{Me}
F^{\alpha\beta}{}_{,\beta}=2\pi j^\alpha
\end{equation}
where current density $j^\alpha$ is given by
\begin{equation}\label{je}
j^\alpha=e\int_{-\infty}^{+\infty}d\tau 
u^\alpha (\tau)\delta^{(3)}(y-z(\tau))
\end{equation} 
governs the propagation of the electromagnetic field. $u^\alpha (\tau)$ 
denotes the (normalized) three-velocity vector $dz^\alpha(\tau)/d\tau$ and 
$\delta^{(3)}(y-z)=\delta(y^0-z^0)\delta(y^1-z^1)\delta(y^2-z^2)$ is a 
three-dimensional Dirac distribution supported on the particle's world 
line $\zeta$. Both the strength tensor $F^{\alpha\beta}$ and the 
current density $j^\alpha$ are evaluated at a field point $y\in{\mathbb 
M}_{\,3}$. (We choose Minkowski metric tensor 
$\eta_{\alpha\beta}={\rm diag}(-1,1,1)$ which we shall use to raise and 
lower indices. Greek indices run from 0 to 3, and Latin indices from 1 to 
2; summation over repeated indices understood throughout the paper.)

We express the electromagnetic field in terms of a vector potential, 
$\hat F=d\hat A$. We impose the Lorentz gauge $A^\alpha{}_{,\alpha}=0$; then 
the Maxwell field equation (\ref{Me}) becomes
\begin{equation} 
\square A^\alpha =-2\pi j^\alpha .
\end{equation} 
In {\it 2+1 theory} the retarded Green's function associated with the 
D'Alembert operator 
$\square:=\eta^{\alpha\beta}\partial_\alpha\partial_\beta$ 
has the form \cite{Gl,KLS}
\begin{equation} \label{G3}
G_{2+1}^{ret}(y,x)=\frac{\theta(y^0-x^0-|{\mathbf y}-{\mathbf 
x}|)}{\sqrt{-2\sigma(y,x)}}.
\end{equation}
$\theta(y^0-x^0-|{\mathbf y}-{\mathbf x}|)$ is step function defined to 
be one if $y^0-x^0\ge |{\mathbf y}-{\mathbf x}|$, and defined to be zero 
otherwise. Synge's world function $\sigma(y,x)$ is numerically equal to half the squared 
distance between $y$ and $x$:
\begin{equation}\label{sigma}
\sigma(y,x)=\frac12\eta_{\alpha\beta}(y^\alpha-x^\alpha)(y^\beta-x^\beta).
\end{equation}
The first is $y$, to which we refer the ``field point'', while the second 
is $x$, to which we refer the ``emission point''. 

While in four-dimensional Minkowski space-time the retarded Green's 
function has support on the future light cone of the emission point $x$, 
in $2+1$ electrodynamics its support extends inside the light cone as well.

Using the retarded Green function (\ref{G3}) and the charge-current density 
(\ref{je}) we construct the retarded Li\'enard-Wiechert potential in three 
dimensions. Denoting $K^\mu=y^\mu - z^\mu(\tau)$ as the unique timelike 
(or null) geodesic connecting a field point $y$ to the emission point
$z(\tau)\in\zeta$, we arrive at
\begin{equation} \label{A}
A_\mu^{ret}(y)=e\int_{-\infty}^{+\infty}d\tau 
\theta(K^0-|{\mathbf K}|)\frac{u_\mu(\tau)}{\sqrt{-(K\cdot K)}}
\end{equation}
where the dot denotes the scalar product of three-vector $K$ on itself 
(it is equal to double Synge's function of field point $y$ and 
emission point $z(\tau)$).

We now turn to the calculation of electromagnetic field 
$F_{\mu\nu}^{ret}=\partial_\mu A^{ret}_\nu -\partial_\nu A^{ret}_\mu$ 
generated by an arbitrarily moving pointlike charge. It consists of two 
quite different terms. The first term is due to differentiation of 
$\theta$-function involved in the vector potential (\ref{A}):
\begin{equation}\label{Fd}
F_{\mu\nu}^{(\delta)}=\lim_{\tau\to\tau^{ret}}\frac{e}{\sqrt{-(K\cdot 
K)}}\frac{u_\mu K_\nu - u_\nu K_\mu}{-(K\cdot u)}.
\end{equation}
$\tau^{ret}(y)$ denotes the proper-time parameter at the point on the 
world line which links with $y$ by the unique future-directed null 
geodesic. Since $\tau^{ret}(y)$ is the root of algebraic equation 
$K^0-|{\mathbf K}|=0$ the $\delta$-term (\ref{Fd}) diverges.

\begin{figure}[t]
\begin{center}
\epsfclipon
\epsfig{file=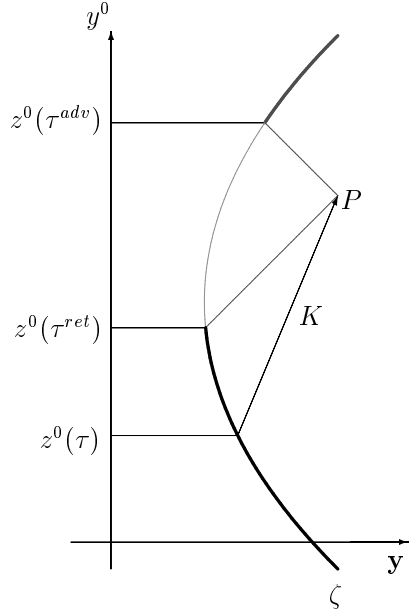,width=5.5cm}
\end{center}
\caption{\label{fld}
\small In four dimensions the retarded (advanced) field at observation 
point $P(y)$ is generated by a single event in space-time: the 
intersection of the world line and $P$'s past (future) light cone. In 
three dimensions the retarded field depends also on the particle's history 
before $\tau^{ret}(y)$. The advanced field depends on the particle's 
history after $\tau^{adv}(y)$. The vector $K$ is a vector pointing from 
the emission point $z(\tau)\in\zeta$ to field point $P$.
}
\end{figure} 

The second term is
\begin{eqnarray}\label{Fth}
F_{\mu\nu}^{(\theta)}&=&-e\int_{-\infty}^{+\infty}d\tau 
\theta(K^0-|{\mathbf K}|)
\frac{u_\mu K_\nu - u_\nu K_\mu}{[-(K\cdot K)]^{3/2}}
\nonumber\\
&=&-e\int_{-\infty}^{\tau^{ret}(y)}d\tau 
\frac{u_\mu K_\nu - u_\nu K_\mu}{[-(K\cdot K)]^{3/2}}.
\end{eqnarray}
We see that the strength tensor $F_{\mu\nu}^{ret}$ of the adjunct 
electromagnetic field consists of terms proportional to $\delta-$ and 
$\theta-$functions: ${\hat F}^{ret}={\hat F}^{(\delta)}+{\hat 
F}^{(\theta)}$. The terms separately are singular. The 
singularity, however, can be removed from the sum of ${\hat 
F}^{(\delta)}$ and ${\hat F}^{(\theta)}$. Using the identity
\begin{equation}
\frac{1}{[-(K\cdot K)]^{3/2}}=\frac{1}{-(K\cdot 
u)}\frac{d}{d\tau}\frac{1}{\sqrt{-(K\cdot K)}}
\end{equation}
in eq.(\ref{Fth}) yields
\begin{eqnarray}\label{Ft}
F_{\mu\nu}^{(\theta)}&=&-\frac{e}{\sqrt{-(K\cdot K)}}\left.
\frac{u_\mu K_\nu - u_\nu K_\mu}{-(K\cdot u)}
\right|_{\tau\to -\infty}^{\tau\to\tau^{ret}(y)}
\\
&+&e\int_{-\infty}^{\tau^{ret}(y)}
\frac{d\tau}{\sqrt{-(K\cdot K)}}\left\{
\frac{a_\mu K_\nu - a_\nu K_\mu}{-(K\cdot u)}
+\frac{u_\mu K_\nu - u_\nu K_\mu}{[-(K\cdot u)]^2}\left[1+(K\cdot a)\right]
\right\}\nonumber
\end{eqnarray}
after integration by parts. Summing up (\ref{Fd}) and (\ref{Ft}) and 
taking into account that $1/\sqrt{-(K\cdot K)}$ vanishes whenever 
$\tau\to -\infty$\footnote{We assume that average velocities are not 
large enough to initiate particle creation and annihilation, so that 
``space contribution'' $|{\mathbf K}|$ can not match with an extremely 
large zeroth component $K^0$.}, we finally obtain the expression
\begin{equation}\label{Fret}
{\hat F}^{ret}(y)= e\int_{-\infty}^{\tau^{ret}(y)}
\frac{d\tau }{\sqrt{-(K\cdot K)}}\left\{
\frac{a\wedge K}{r}
+\frac{u\wedge K}{r^2}\left[1+(K\cdot a)\right]
\right\}
\end{equation}
which is regular on the light cone. (It diverges on the particle's 
trajectory only.) The symbol $\wedge$ denotes the wedge product. The 
invariant quantity
\begin{eqnarray}\label{rint}
r&=&-(K\cdot u)\\
&=&-\eta_{\alpha\beta}(y^\alpha-z^\alpha(\tau))u^\beta(\tau)\nonumber
\end{eqnarray}
is an affine parameter on the time-like (null) geodesic that links $y$ to 
$z(\tau)$; it can be loosely interpreted as the time delay between $y$ and
$z(\tau)$ as measured by an observer moving with the particle.
When $\tau=\tau^{ret}(y)$, the parameter $r$ is also the spatial distance 
between $z(\tau^{ret})$ and $y$ as measured in this momentarily comoving 
Lorentz frame.

In three dimensions the advanced Green's function 
\begin{equation} \label{G3a}
G_{2+1}^{adv}(y,x)=\frac{\theta(-y^0+x^0-|{\mathbf y}-{\mathbf 
x}|)}{\sqrt{-2\sigma(y,x)}}
\end{equation}
is nonzero in the past of $x$. The advanced strength tensor
\begin{equation}\label{Fadv}
{\hat F}^{adv}(y)= e\int^{+\infty}_{\tau^{adv}(y)}
\frac{d\tau }{\sqrt{-(K\cdot K)}}\left\{
\frac{a\wedge K}{r}
+\frac{u\wedge K}{r^2}\left[1+(K\cdot a)\right]
\right\}
\end{equation}
is generated by the point charge during its entire future history following 
the advanced time associated with $y$ (see figure \ref{fld}).

\section{Equation of motion of radiating charge}\label{meq}
\setcounter{equation}{0}

In this section we derive the ``three-dimensional'' analog of the 
Lorentz-Dirac equation via analysis of the energy-momentum and angular 
momentum balance equations. The momentum three-vector carried by the
electromagnetic field is calculated in the next Section and in 
Appendixes D, E, and F; the total angular momentum is obtained in 
\ref{angul}. We split the Noether quantities into bound (singular) 
and radiative (regular) parts. The energy-momentum and angular momentum of 
bare sources absorb the bound terms within regularization procedure. 
Already renormalized characteristics of charged particles are proclaimed 
to be finite. Together with radiative terms, they constitute the total
energy-momentum and angular momentum of our particle plus field system
which are properly conserved.

To find out electromagnetic field's energy-momentum, we integrate the 
Maxwell energy-momentum tensor density over the plane
$\Sigma_t=\{y\in{\mathbb M}_{\,3}:y^0=t\}$. The resulting expressions 
(\ref{p0em}) and (\ref{piem}), presented in the next Section, can be 
rewritten in manifestly covariant fashion:
\begin{eqnarray}\label{prom}
p^\mu_{\rm em}(\tau)&=&\frac{e^2}{2}\int_{-\infty}^\tau 
d\tau_2\frac{u^\mu(\tau_2)}{\sqrt{2\sigma(\tau,\tau_2)}}\\ 
&+&\left.
\begin{array}{c}
\displaystyle 
e^2\int_{-\infty}^\tau d\tau_1\int_{-\infty}^{\tau_1}d\tau_2\\
\\[-1em]
\displaystyle
e^2\int_{-\infty}^\tau d\tau_2\int_{\tau_2}^\tau d\tau_1
\end{array}
\right\}
\left[
-\frac{\partial^2\sigma}{\partial 
\tau_1\partial\tau_2}\frac{q^\mu}{(2\sigma)^{3/2}}+\frac12
\frac{\partial}{\partial \tau_1}\left(\frac{u_2^\mu}{\sqrt{2\sigma}}\right)
-\frac12
\frac{\partial}{\partial \tau_2}\left(\frac{u_1^\mu}{\sqrt{2\sigma}}\right)
\right].\nonumber
\end{eqnarray}
(We omit structureless terms which arise due to choice of non-covariant 
surface of integration.) Index $1$ indicates that particle's 
velocity or position is referred to the instant $\tau_1\in ]-\infty,\tau]$ 
while index $2$ says that the particle's characteristics are evaluated at 
instant $\tau_2\le\tau_1$.
Here $q^\mu=z_1^\mu-z_2^\mu$ defines the unique timelike geodesic 
connecting a field point $z(\tau_1)\in\zeta$ to an emission point
$z(\tau_2)\in\zeta$; by $\sigma$ we mean the Synge's world function 
(\ref{sigma}) of $z_1$ and $z_2$, taken with opposite sign:
\begin{equation}\label{sgm}
\sigma(\tau_1,\tau_2)=-\frac12(q\cdot q).
\end{equation}
Two double integrals over (proper) time variables (one about the 
other) describe integration over the domain 
$D_\tau=\{(\tau_1,\tau_2)\in{\mathbb R}^{\,2}:
\tau_1\in ]-\infty,\tau],\tau_2\le\tau_1\}$.

We have to decompose the expression eq.(\ref{prom}) into singular and 
regular parts. Following Ref.\cite{KLS}, we postulate that splitting 
should satisfy the conditions:
\begin{itemize}
\item
proper non-accelerating limit of singular and regular parts;
\item
proper short-distance behavior of regular part;
\item
Poincar\'e invariance and reparametrization invariance.
\end{itemize}
By ``proper short-distance behavior'' we mean the finiteness of integrand 
at the edge $\tau_2=\tau_1$ of the domain $D_\tau$.

So, we take one-half of the first term in between the square brackets 
under the double integral signs in eq.(\ref{prom}):
\begin{equation}
-\frac12\frac{\partial^2\sigma}{\partial 
\tau_1\partial\tau_2}\frac{q^\mu}{(2\sigma)^{3/2}}=
-\frac12\frac{(u_1\cdot u_2)q^\mu}{[-(q\cdot q)]^{3/2}}
\end{equation}
and add the second term
\begin{equation}\label{d1}
\frac12\frac{\partial}{\partial 
\tau_1}\left(\frac{u_2^\mu}{\sqrt{2\sigma}}\right)
=\frac12\frac{(u_1\cdot q)u_2^\mu}{[-(q\cdot q)]^{3/2}}.
\end{equation}
It is convenient to rewrite the resulting expression as follows:
\begin{equation}\label{p12mu}
p_{12}^\mu=\frac12u_{1,\alpha}\frac{-u_2^\alpha q^\mu +u_2^\mu 
q^\alpha}{[-(q\cdot q)]^{3/2}}.
\end{equation} 
We introduce the function
\begin{equation}\label{Gret}
G_{\rm ret}^\mu (\tau_1)=e^2u_{1,\alpha}
\int_{-\infty}^{\tau_1}d\tau_2
\frac{-u_2^\alpha q^\mu +u_2^\mu 
q^\alpha}{[-(q\cdot q)]^{3/2}}
\end{equation}
which is the convolution $u_\mu(\tau_1) F^{\mu\nu}_{(\theta)}$ of 
three-velocity and non-local part (\ref{Fth}) of the retarded strength 
tensor evaluated at point $z(\tau_1)\in\zeta$. It is intimately 
connected with the retarded field (\ref{Fret}) generated at point 
$z(\tau_1)\in\zeta$ by the portion of the world line that corresponds to 
the interval $-\infty<\tau_2\le\tau_1$:
\begin{equation}\label{GFret}
G_{\rm ret}^\mu(\tau_1)=\frac{e^2}{2}a_1^\mu - eu_{1,\alpha}F_{\rm 
ret}^{\mu\alpha}(\tau_1).
\end{equation}
(It may be checked via integration by parts.)

Next we take the remaining one-half of the first term and add the third 
term:
\begin{equation}
-\frac12\frac{(u_1\cdot u_2)q^\mu}{[-(q\cdot q)]^{3/2}}
+\frac12\frac{(u_2\cdot q)u_1^\mu}{[-(q\cdot q)]^{3/2}}
=\frac12u_{2,\alpha}\frac{-u_1^\alpha q^\mu +u_1^\mu 
q^\alpha}{[-(q\cdot q)]^{3/2}}.
\end{equation}
We change the order of integration of this term over $D_\tau$: 
\begin{equation}
\frac{e^2}{2}\int_{-\infty}^\tau d\tau_1\int_{-\infty}^{\tau_1}d\tau_2
u_{2,\alpha}\frac{-u_1^\alpha q^\mu +u_1^\mu 
q^\alpha}{[-(q\cdot q)]^{3/2}}=
\frac{e^2}{2}\int_{-\infty}^\tau d\tau_2u_{2,\alpha}
\int_{\tau_2}^{\tau}d\tau_1
\frac{-u_1^\alpha q^\mu +u_1^\mu 
q^\alpha}{[-(q\cdot q)]^{3/2}}
\end{equation}
and interchange indices ``first'' and ``second''. Taking into account that
$q(\tau_2,\tau_1)=-q(\tau_1,\tau_2)$, we finally obtain:
\begin{equation}
-\frac{e^2}{2}\int_{-\infty}^\tau d\tau_1u_{1,\alpha}
\int_{\tau_1}^{\tau}d\tau_2
\frac{-u_2^\alpha q^\mu +u_2^\mu q^\alpha}{[-(q\cdot q)]^{3/2}}.
\end{equation}
The integrand coincides with that under integral sign in 
the right-hand side of eq.(\ref{Gret}) while the domain of inner 
integration corresponds to the interval $\tau_1\le\tau_2\le\tau$. We 
introduce the function
\begin{equation}\label{Gadv}
G_{\rm adv}^\mu (\tau_1,\tau)=e^2u_{1,\alpha}
\int_{\tau_1}^{\tau}d\tau_2
\frac{-u_2^\alpha q^\mu +u_2^\mu q^\alpha}{[-(q\cdot q)]^{3/2}}.
\end{equation}
which is the convolution $u_\mu(\tau_1) F^{\mu\nu}_{(\vartheta)}$ of 
three-velocity and non-local part of the advanced strength 
tensor evaluated at point $z(\tau_1)\in\zeta$. If the instant of 
observation $\tau$ tends to $+\infty$, this function can be rewritten as
\begin{equation}\label{GFadv}
G_{\rm adv}^\mu(\tau_1)=
\frac{e^2}{2}a_1^\mu 
- eu_{1,\alpha}F_{\rm adv}^{\mu\alpha}(\tau_1).
\end{equation}
The second term is convolution of velocity with the advanced field 
(\ref{Fadv}) generated at point $z(\tau_1)\in\zeta$ by the portion of the 
world line that corresponds to the interval $[\tau_1,+\infty[$.
The relations (\ref{GFret}) and (\ref{GFadv}) are symmetric upon future 
and past.

\begin{figure}[t]
\begin{center}
\epsfclipon
\epsfig{file=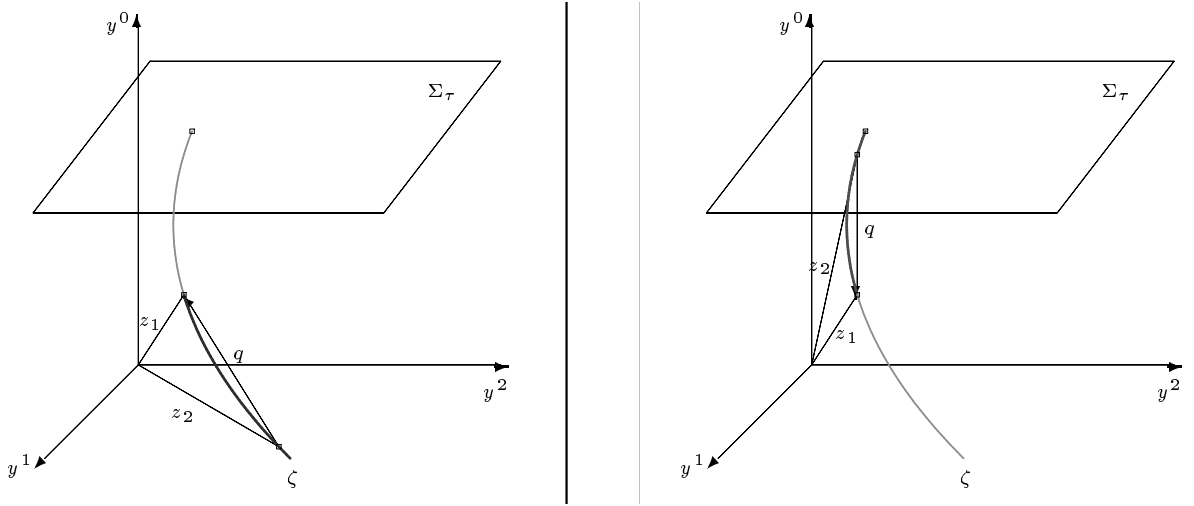,width=16cm}
\end{center}
\caption{\label{Dscheme}
\small We call ``retarded'' the term (\ref{Gret}) with integration over the 
portion of the world line {\it before} $\tau_1$. We call ``advanced'' the 
term (\ref{Gadv}) with integration over the portion of the world line {\it 
after} $\tau_1$. For an observer placed at point $z(\tau_1)\in\zeta$ the 
regular part (\ref{pR}) of electromagnetic field momentum looks as the 
combination of incoming and outgoing radiation, and yet the retarded 
causality is not violated. We still consider the interference of outgoing 
waves presented at the observation instant $\tau$. The electromagnetic 
field carries information about the charge's past.
}
\end{figure}

We see that  the double integral in eq.(\ref{prom}) can be expressed as 
one-half of $G_{\rm ret}$ minus one-half of $G_{\rm adv}$ integrated over 
the world line $\zeta$. For this reason we proclaim the expression
\begin{eqnarray}\label{pR}
p^\mu_{\rm R}(\tau)&=&\frac12\int_{-\infty}^\tau d\tau_1\left[
G_{\rm ret}^\mu(\tau_1) - G_{\rm adv}^\mu(\tau_1,\tau)
\right]\\
&=&
e^2\int_{-\infty}^\tau d\tau_1\int_{-\infty}^{\tau_1}d\tau_2
\left[
-\frac{(u_1\cdot u_2)q^\mu}{[-(q\cdot q)]^{3/2}}+
\frac12\frac{(u_1\cdot q)u_2^\mu}{[-(q\cdot q)]^{3/2}}+
\frac12\frac{(u_2\cdot q)u_1^\mu}{[-(q\cdot q)]^{3/2}}
\right]\nonumber
\end{eqnarray}
the radiative part of energy-momentum carried by the electromagnetic field. 
The situation is pictured in figure \ref{Dscheme}. 

We evaluate the short-distance behavior of the expression under the double 
integral in eq.(\ref{pR}). Let $\tau_1$ be fixed and 
$\tau_1-\tau_2:=\Delta$ be a small parameter. With a degree of accuracy 
sufficient for our purposes
\begin{eqnarray}
\sqrt{-(q\cdot q)}&=&\Delta\\
q^\mu&=&\Delta\left[u_1^\mu-a_1^\mu\frac{\Delta}{2}+
{\dot a}_1^\mu\frac{\Delta^2}{6}\right]\nonumber\\
u_2^\mu&=&u_1^\mu-a_1^\mu\Delta+{\dot a}_1^\mu\frac{\Delta^2}{2}.\nonumber
\end{eqnarray}
Substituting these into integrand of the double integral of eq.(\ref{pR}) 
and passing to the limit $\Delta\to 0$ yields regular expression
\begin{equation}\label{lim}
\lim_{\tau_2\to\tau_1}\left[
\frac12u_{1,\alpha}\frac{-u_2^\alpha q^\mu +u_2^\mu 
q^\alpha}{[-(q\cdot q)]^{3/2}}+
\frac12u_{2,\alpha}\frac{-u_1^\alpha q^\mu +u_1^\mu 
q^\alpha}{[-(q\cdot q)]^{3/2}}
\right]
=\frac13(a_1)^2u_1^\mu-\frac{1}{12}{\dot a}_1^\mu.
\end{equation}
Therefore the subscript ``R'' stands for ``regular'' as well as for 
``radiative''.

Alternatively, choosing the linear superposition
\begin{equation}\label{pS}
p^\mu_{\rm S}=
\frac12\int_{-\infty}^\tau d\tau_1\left[
G_{\rm ret}^\mu(\tau_1) + G_{\rm adv}^\mu(\tau_1,\tau)
\right]
\end{equation}
we restore the first term in the right-hand side of eq.(\ref{prom}). 
Indeed, having integrated the half-sum of the functions (\ref{Gret}) and 
(\ref{Gadv}) over $\zeta$, we obtain
\begin{eqnarray}
p^\mu_{\rm S}(\tau)&=&\frac{e^2}{2}\int_{-\infty}^\tau d\tau_2 
\left.
\frac{u_2^\mu}{\sqrt{2\sigma(\tau_1,\tau_2)}}
\right|_{\tau_1=\tau_2}^{\tau_1=\tau}
+\frac{e^2}{2}\int_{-\infty}^\tau d\tau_1
\left.
\frac{u_1^\mu}{\sqrt{2\sigma(\tau_1,\tau_2)}}
\right|_{\tau_2\to -\infty}^{\tau_2=\tau_1}\nonumber\\
&=&\frac{e^2}{2}
\int_{-\infty}^\tau 
d\tau_2\frac{u^\mu(\tau_2)}{\sqrt{2\sigma(\tau,\tau_2)}}.
\end{eqnarray}
Since this non-local term diverges, in eq.(\ref{pS}) the subscript ``S''
stands for ``singular'' as well as ``symmetric''. 

In the specific case of uniformly moving charge 
$\sqrt{2\sigma(\tau,\tau_2)}=\tau-\tau_2$. Hence $p^\mu_{\rm S}(\tau)$
coincides with that obtained in \ref{unif} where rectilinear uniform 
motion is considered (see eq.(\ref{Apem})). Since the bracketed integrand 
in (\ref{pR}) vanishes if $u^\mu=const$, nonaccelerating charge does 
not radiate.

We therefore introduce the radiative part $p_{\rm R}$ of energy-momentum 
and postulate that it, and it alone exerts a force on the particle. 
The singular part $p_{\rm S}$ should be coupled with the particle's 
three-momentum, so that ``dressed'' charged particle would not undergo any 
additional radiation reaction. The already renormalized particle's 
individual three-momentum, say $p_{\rm part}$, together with $p_{\rm R}$ 
constitute the total energy-momentum of our particle plus field system:  
$P=p_{\rm part}+p_{\rm R}$. Since $P$ does not change with time, its time 
derivative yields 
\begin{eqnarray}\label{pdot}
{\dot p}^\mu_{\rm part}(\tau)&=&-{\dot p}^\mu_{\rm R}\\
&=&-\frac{e^2}{2}\int_{-\infty}^\tau ds
\left[
u_{\tau,\alpha}\frac{-u_s^\alpha q^\mu +u_s^\mu
q^\alpha}{[2\sigma(\tau,s)]^{3/2}}+
u_{s,\alpha}\frac{-u_\tau^\alpha q^\mu +u_\tau^\mu 
q^\alpha}{[2\sigma(\tau,s)]^{3/2}}
\right].\nonumber
\end{eqnarray} 
(The overdot means the derivation with respect to proper time $\tau$.)
Here index $\tau$ indicates that the particle's velocity or position is 
referred to the observation instant $\tau$ while index 
$s$ indicates that the particle's characteristics are evaluated at instant 
$s\le\tau$.

Our next task is to derive an expression which explain how three-momentum 
of a ``dressed'' charged particle depends on its individual characteristics 
(velocity, position, mass etc.). We do not make any assumptions about the 
particle structure, its charge distribution and its size. We only assume 
that the particle 3-momentum $p_{\rm part}$ is finite. To find out 
the desired expression we analyze conserved quantities corresponding to the 
invariance of the theory under proper homogeneous Lorentz transformations. 
The total angular momentum, say $M$, consists of particle's 
angular momentum $z\wedge p_{\rm part}$ and radiative part of angular 
momentum carried by electromagnetic field:
\begin{equation}\label{Mtot}
M^{\mu\nu}=z_\tau^\mu p_{\rm part}^\nu(\tau) 
- z_\tau^\nu p_{\rm part}^\mu(\tau) + M^{\mu\nu}_{\rm R}(\tau).
\end{equation}
(Singular part is absorbed by $p_{\rm part}$.) The last term is calculated 
in \ref{angul}:
\begin{equation}\label{M_R}
M_R^{\mu\nu}=\frac{e^2}{2}\int_{-\infty}^\tau 
d\tau_1\int_{-\infty}^{\tau_1}d\tau_2\left(
z_1^\mu p_{12}^\nu - z_1^\nu p_{12}^\mu 
+z_2^\mu p_{21}^\nu - z_2^\nu p_{21}^\mu 
\right)
\end{equation}
where two-point function $p_{12}^\alpha$ is given by eq.(\ref{p12mu}).

Having differentiated eq.(\ref{Mtot}) and inserting eq.(\ref{pdot}) 
we arrive at the equation
\begin{equation}\label{Mdot}
u_\tau\wedge p_{\rm part}=
\frac{e^2}{2}\int_{-\infty}^\tau ds\frac{u_\tau\wedge 
u_s}{\sqrt{2\sigma(\tau,s)}}
\end{equation}
where the symbol $\wedge$ denotes the wedge product. We obtain the system 
of tree linear equations in three components of the particle's momentum. 
Its rank is equal to 2. Therefore, an arbitrary scalar function $m(\tau)$ 
arises:
\begin{equation}\label{part}
p_{\rm part}^\mu(\tau)=mu^\mu(\tau)+\frac{e^2}{2}\int_{-\infty}^\tau 
ds\frac{u^\mu(s)-u^\mu(\tau)}{\sqrt{2\sigma(\tau,s)}}.
\end{equation}
(We choose the simplest expression that is finite near the point of 
observation.) We see that, apart from the usual velocity term, the 
particle's 3-momentum contains also nonlocal contribution from the 
particle's electromagnetic field.

The scalar product of the particle three-velocity on the first-order 
time-derivative of the particle three-momentum (\ref{pdot}) is as follows:
\begin{equation}\label{udp}
({\dot p}_{\rm part}\cdot u_\tau)=\frac{e^2}{2}\int_{-\infty}^\tau ds
\left[
(u_\tau\cdot u_s)\frac{(u_\tau\cdot q)}{[2\sigma]^{3/2}} +
\frac{(u_s\cdot q)}{[2\sigma]^{3/2}}
\right].
\end{equation}
Since $(u\cdot a)=0$, the scalar product of particle acceleration on the 
particle three-momentum (\ref{part}) is given by
\begin{equation}\label{ap}
(p_{\rm part}\cdot a_\tau)=\frac{e^2}{2}\int_{-\infty}^\tau ds
\frac{(a_\tau\cdot u_s)}{\sqrt{2\sigma}}.
\end{equation}
Summing up eqs.(\ref{udp}) and (\ref{ap}) we obtain
\begin{equation}\label{dpdp}
\frac{d}{d\tau}(p_{\rm part}\cdot u_\tau)=
\frac{e^2}{2}\int_{-\infty}^\tau ds\left\{
\frac{\partial}{\partial\tau}\left[\frac{(u_\tau\cdot u_s)}{\sqrt{2\sigma}}
\right] +
\frac{(u_s\cdot q)}{[2\sigma]^{3/2}}
\right\}.
\end{equation}

Alternatively, the scalar product of 3-momentum (\ref{part}) and 3-velocity 
is as follows:
\begin{equation}\label{pu}
(p_{\rm part}\cdot u_\tau)=-m+\frac{e^2}{2}\int_{-\infty}^\tau ds
\frac{(u_\tau\cdot u_s)+1}{\sqrt{2\sigma}}.
\end{equation}
Further we compare its differential consequence with eq.(\ref{dpdp}). A 
surprising feature of the already renormalized {\it dynamical} mass $m$ 
is that it depends on $\tau$: 
\begin{equation}\label{dotm}
\dot m=\frac{e^2}{2}\int_{-\infty}^\tau ds
\frac{(q\cdot u_\tau)-(q\cdot u_s)}{[2\sigma]^{3/2}}.
\end{equation}
It is interesting that a similar phenomenon occurs in the theory which 
describes a pointlike charge coupled with massless scalar field in flat 
space-time of three dimensions \cite{Br}. The charge loses its mass through 
the emission of monopole radiation.

Having integrated derivative (\ref{dotm}) over the world line $\zeta$, we 
obtain
\begin{eqnarray}
m&=&m_0+\frac{e^2}{2}\int_{-\infty}^\tau 
d\tau_1\int_{-\infty}^{\tau_1} d\tau_2 
\left[
\frac{\partial}{\partial\tau_1}\left(\frac{1}{\sqrt{2\sigma}}\right)+
\frac{\partial}{\partial\tau_2}\left(\frac{1}{\sqrt{2\sigma}}\right)
\right]\\
&=&m_0+\frac{e^2}{2}\int_{-\infty}^\tau \frac{ds}{\sqrt{2\sigma(\tau,s)}}
\nonumber
\end{eqnarray}
where $m_0$ is an infinite bare mass of the particle. Inserting this 
into eq.(\ref{part}), we arrive at the equality $p_{\rm 
part}^\mu(\tau)=m_0u^\mu_\tau+p_S^\mu$ which shows that the particle's 
momentum renormalization agrees with the renormalization of mass.

The main goal of the present paper is to compute the effective equation of 
motion of radiating charge in $2+1$ dimensions. To do it we replace ${\dot 
p}_{\rm part}^\mu$ in the left-hand side of eq.(\ref{pdot}) by differential 
consequence of eq.(\ref{part}) where the right-hand side of eq.(\ref{dotm}) 
substitutes for $\dot m$. At the end of a straightforward calculations, we 
obtain 
\begin{eqnarray}\label{me}
ma^\mu_\tau&=&\frac{e^2}{2}a^\mu_\tau-e^2\int_{-\infty}^\tau ds\left[
u_{\tau,\alpha}\frac{-u^\alpha_sq^\mu + u^\mu_sq^\alpha}{[2\sigma]^{3/2}}
-\frac12\frac{a^\mu_\tau}{\sqrt{2\sigma}}
\right]\\
&=&\frac{e^2}{2}a^\mu_\tau-G^\mu_{\rm ret}(\tau)
+\frac{e^2}{2}a^\mu_\tau\int_{-\infty}^\tau 
\frac{ds}{\sqrt{2\sigma(\tau,s)}}.\nonumber
\end{eqnarray}
The first term in the right-hand side of this equation looks horribly 
irrelevant. Relation (\ref{GFret}) prompts that the retarded Lorentz 
self-force should be substituted for the combination of the first (local) 
and of the second (non-local) terms in the right-hand side of 
eq.(\ref{me}). If an external electromagnetic field $\hat F_{\rm ext}$ is 
applied, the equation of motion of radiating charge in $2+1$ theory becomes
\begin{equation}\label{mext}
ma^\mu_\tau=eu_{\tau,\alpha} F^{\mu\alpha}_{\rm ret}(\tau)+ 
\frac{e^2}{2}a^\mu_\tau\int_{-\infty}^\tau 
\frac{ds}{\sqrt{2\sigma(\tau,s)}}
+eu_{\tau,\alpha} F^{\mu\alpha}_{\rm ext}
\end{equation}
where
\begin{eqnarray}\label{retF}
F^{\mu\alpha}_{\rm ret}(\tau)&=&
e\int_{-\infty}^\tau\frac{ds}{\sqrt{2\sigma(\tau,s)}}
\left\{\frac{u_s^\mu q^\alpha -u_s^\alpha 
q^\mu}{r^2}\left[1+(a_s\cdot q)\right]
+\frac{a_s^\mu q^\alpha -a_s^\alpha q^\mu}{r}
\right\}\\
&=&\int_{-\infty}^\tau dsf^{\mu\alpha}(\tau,s)\nonumber
\end{eqnarray}
is the field strengths at point $z(\tau)\in\zeta$ generated by a portion of 
the world line {\it before} the observation instant $\tau$. The non-local 
term in eq.(\ref{mext}) which is proportional to the particle's 
acceleration $a(\tau)$ arises also in Ref.\cite{KLS}. It provides proper 
short-distance behavior of the radiation backreaction. If $s\to\tau$, the 
integrand tends to three-dimensional analog of the Abraham radiation 
reaction vector:
\begin{equation}
\lim_{s\to\tau}\left[
eu_{\tau,\alpha}f^{\mu\alpha}(\tau,s)+
\frac{e^2}{2}\frac{a^\mu_\tau}{\sqrt{2\sigma(\tau,s)}}
\right]=\frac23e^2\left({\dot a}^\mu-a^2u^\mu\right).
\end{equation}
(All quantities on the right-hand side refer to the instant of observation 
$\tau$.)

If one moves the second term to the left-hand side of eq.(\ref{mext}), 
they restore unphysical motion equation which follows from variational 
principle: it involves an infinite bare mass and divergent Lorentz 
self-force.

\section{Energy-momentum of electromagnetic field in 2+1 
dimensions}\label{trace}
\setcounter{equation}{0}

In this section we trace a series of stages in calculation of the surface 
integral 
\begin{equation}\label{pem}
p^\nu_{\rm em}(\tau)=\int_\Sigma d\sigma_\mu T^{\mu\nu}
\end{equation}
which gives the energy-momentum carried by electromagnetic field of a 
point-like source arbitrarily moving in ${\mathbb M}_{\,3}$. In 
\ref{pink}, \ref{blue} and \ref{yellow} we perform the computation in 
detail.

In eq.(\ref{pem}) $d\sigma_\mu$ is the vectorial surface element on an 
arbitrary space-like surface $\Sigma$. The electromagnetic field's 
stress-energy tensor $\hat T$ has the components
\begin{equation}\label{T}
2\pi T^{\mu\nu}=F^{\mu\lambda}F^\nu{}_{\lambda}- 
1/4\eta^{\mu\nu}F^{\kappa\lambda}F_{\kappa\lambda}
\end{equation}
where $\hat F$ is the nonlocal strength tensor (\ref{Fret}).

\subsection{Coordinate system}
\renewcommand{\theequation}{\arabic{section}.\arabic{subsection}.\arabic{equation}}
\setcounter{equation}{0}

In general, the rate of radiation does not depend on the shape of $\Sigma$. 
We choose the simplest plane $\Sigma_t=\{y\in{\mathbb M}_{\,3}:y^0=t\}$ 
associated with unmoving inertial observer. 
If parametrization of the world line is provided by a disjoint union of 
planes $\Sigma_t$, particle's velocity takes the form $u^\mu=\gamma 
v^\mu, v^\mu=(1,{\dot z}^i)$, and acceleration $a^\mu=\gamma^4({\bf 
v}{\bf\dot v})v^\mu+\gamma^2{\dot v}^\mu;$ factor $\gamma=1/\sqrt{1-{\bf 
v}^2}$. (The overdot indicates differentiation with respect to $t$.) 
Electromagnetic field (\ref{Fret}) takes the form
\begin{equation}\label{Fi}
{\hat F}^{ret}(y)= e\int_{-\infty}^{t^{ret}(y)}
\frac{dt}{\sqrt{-(K\cdot K)}}\left\{
\frac{{\dot v}\wedge K}{r}
+\frac{v\wedge K}{r^2}\left[\gamma^{-2}+(K\cdot {\dot v})\right]
\right\}
\end{equation}
where ${\dot v}^\mu=(0,{\dot v}^i)$ and $r=-(K\cdot v)$. (Although we use 
the same notation, $r$ should not be confused with manifestly covariant 
parameter (\ref{rint}).)

\begin{figure}[t]
\begin{center}
\epsfclipon
\epsfclipon
\epsfig{file=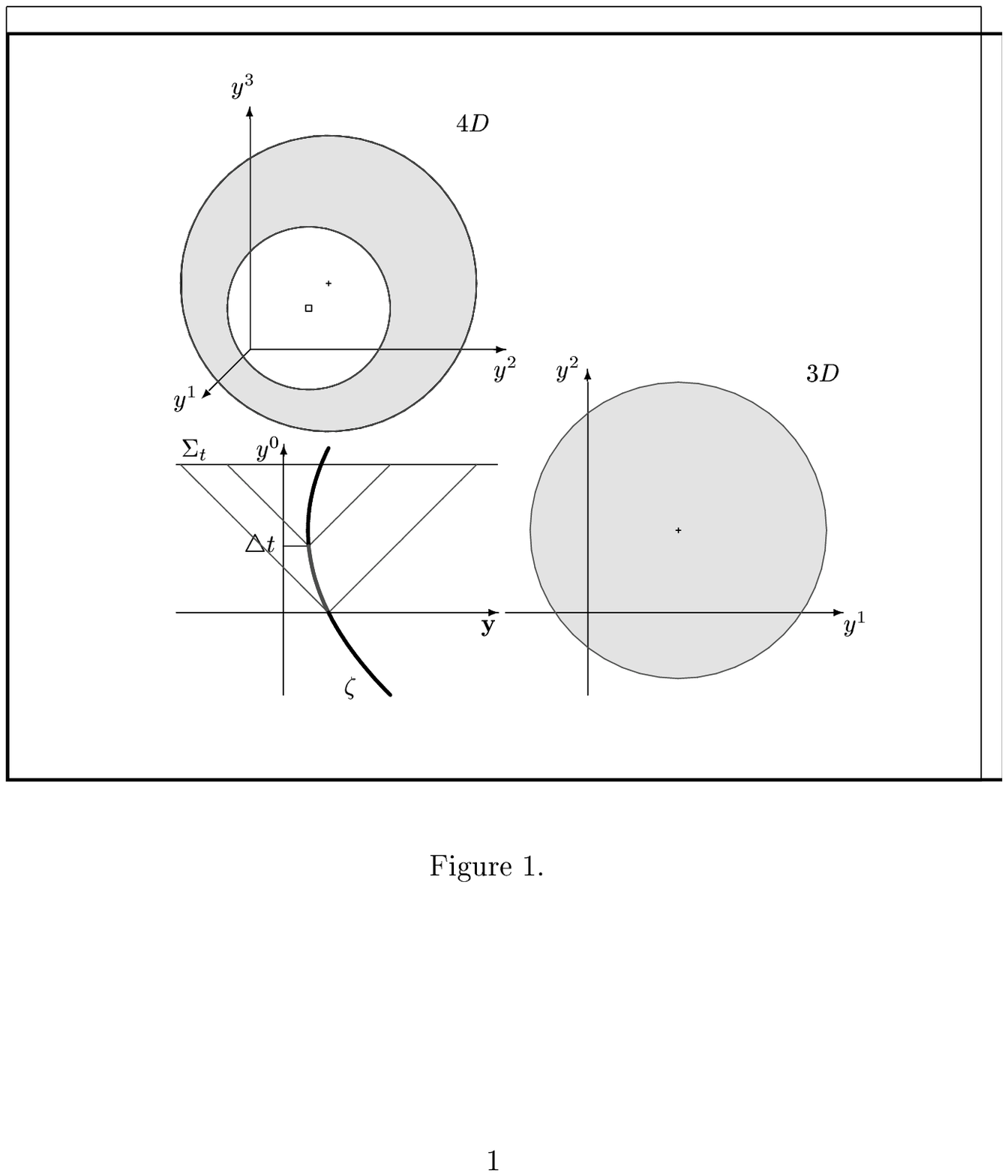,width=9cm}
\end{center}
\caption{\label{Hg}
\small Let the point source radiates within the interval $[0,\triangle 
t]$. In four dimensions the support of the Maxwell energy-momentum tensor 
density in hyperplane $y^0=t$ is in between two spheres centered at 
points $z^i(0)$ (cross symbol) and $z^i(\triangle t)$ (box symbol) with 
radii $t$ and $t-\triangle t$, respectively. In three dimensions the 
radiation fills the disk with radius $t$ centered at point $z^i(0)$ 
(cross symbol) even if the interval shrinks to zero.
}
\end{figure}

Huygens principle does not hold in three dimensions and radiation 
develops a tail (see figure \ref{Hg}). In $3D$ the circle $C(z(0),t)=
\{y\in{\mathbb M}_3: (y^0)^2=\sum_{i=1}^2(y^i-z^i(0))^2, y^0=t\}$ 
is filled up by electromagnetic radiation even if interval $\triangle t\to 
0$. (The period of time during which the point source emanates 
is meant.) So, a point $z(t_1)\in\zeta$ produces the disk of radius 
$t-t_1$ in the observation plane $\Sigma_t=\{y\in{\mathbb M}_3: y^0=t\}$. 
This property reflects the fact that in ${\mathbb M}_3$ the 
electromagnetic field at $y$ is generated by the portion of the world 
line that corresponds to the interval $-\infty <\tau <\tau^{ret}(y)$; 
this represents the past history of the particle.

\begin{figure}[t]
\begin{center}
\epsfclipon
\epsfclipon
\epsfig{file=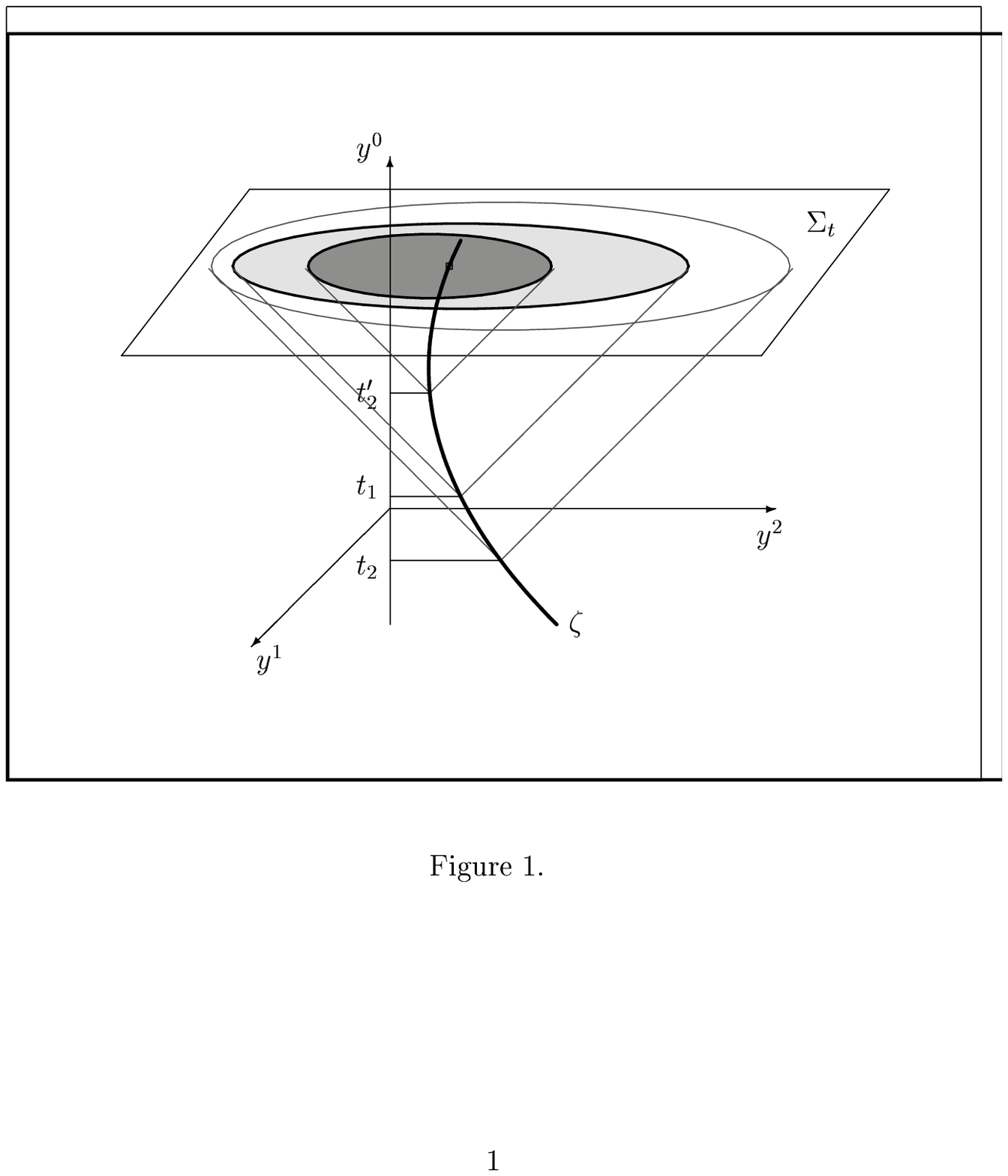,width=9cm}
\end{center}
\caption{\label{f_ab}
\small  Outgoing electromagnetic waves generated by the portion of the 
world line that corresponds to the interval $-\infty <t_2<t_1$ combine 
within the gray disk with radius $k_1^0=t-t_1$. Their contribution is 
given by the first fourfold integral in eq.(\ref{pint}). If the domain of 
integration $t_1<t_2\le t$ the waves joint together inside the dark disk 
with radius $k_2^0=t-t_2'$. The second fourfold integral in 
eq.(\ref{pint}) describes them.}
\end{figure} 

We introduce coordinate system associated with two points on a 
particle's world line labelled by instants $t_1$ and $t_2$ (see figure 
\ref{f_ab}). Flat space-time ${\mathbb M}_3$ becomes a disjoint union of 
planes $\Sigma_t=\{y\in{\mathbb M}_3:y^0=t\}$. A plane $\Sigma_t$ is a 
union of (retarded) disks centered on a world line of the particle. The 
disk
\begin{equation}\label{C}
C(z(t_a),t-t_a)=\{y\in{\mathbb 
M}_3:y^0-t_a\ge\sqrt{\sum_i(y^i-z^i(t_a))^2},y^0=t\}
\end{equation}
is bounded by the intersection of the future light cone generated by null 
rays emanating from $z(t_a)\in\zeta$ in all possible directions, and plane 
$\Sigma_t$. The circular spot (\ref{C}) is filled up by circles of radii 
$R\in[0,t-t_a]$ centered on points on a line connecting points $z^i(t_1)$ 
and $z^i(t_2)$. Points in an $R-$circle are distinguished by polar angle 
$\varphi$. We define the coordinate transformation locally written as
\begin{eqnarray}\label{ct}
y^0&=&t\\
y^i&=&\alpha z^i(t_1)+\beta z^i(t_2) + R\omega^i{}_j n^j\nonumber
\end{eqnarray} 
where $\alpha +\beta =1$ and $n^j=(\cos\varphi,\sin\varphi)$. Orthogonal 
matrix $\omega$ is given by eq.(\ref{om}) (see \ref{coord}). It rotates 
space axes till new $y^1-$axis be directed along two-vector ${\bf q}:={\bf 
z}(t_1)-{\bf z}(t_2)$.

The integration of energy and momentum densities over two-dimensional 
plane $y^0=const$ means the study of interference of outgoing 
electromagnetic waves emitted by different points on particle's world 
line (see figure \ref{f_ab}). Note that the retarded field is generated by 
portion of the world line $\zeta$ that corresponds to the particle's 
history {\it before} $t^{ret}(y)$. Since the stress-energy tensor is 
quadratic in field strengths, we should {\it twice} integrate it over 
$\zeta$. There are also two variables which parametrize $\Sigma_t$. In 
curvilinear coordinates $(t,t_1,t_2,s,\varphi)$ the surface integral 
(\ref{pem}) becomes 
\begin{eqnarray}\label{pint}
p_{\rm em}^\alpha&=&\int_{\Sigma_t}d\sigma_0 T^{0\alpha}\\
&=&\int\limits_{-\infty}^tdt_1\int\limits_{-\infty}^{t_1}dt_2
\int\limits_{0}^{k_1^0}dR 
\int\limits_0^{2\pi}d\varphi Jt^{0\alpha}_{12}
+
\int\limits_{-\infty}^tdt_1\int\limits_{t_1}^tdt_2\int\limits_{0}^{k_2^0}dR
\int\limits_0^{2\pi}d\varphi Jt^{0\alpha}_{12}\nonumber
\end{eqnarray}
with Jacobian 
\begin{eqnarray}\label{J}
J&=&\left(1-q\frac{\partial\beta}{\partial R}\cos\varphi\right)R\\
&=&\left(1+q\frac{\partial\alpha}{\partial R}\cos\varphi\right)R.
\nonumber
\end{eqnarray}
The integrand
\begin{equation}
2\pi t^{\alpha\beta}_{12}=
f_{(1)}^{\alpha\lambda}f_{(2)\lambda}^\beta -\frac14\eta^{\alpha\beta}
f_{(1)}^{\mu\nu}f^{(2)}_{\mu\nu}
\end{equation}
describes the combination of field strength densities at $y\in\Sigma_t$
\begin{equation}\label{F1F2}
{\hat f}_{(a)}(y)=\frac{e}{\sqrt{-(K_a\cdot K_a)}}\left(
\frac{{\dot v}_a\wedge K_a}{r_a}+\frac{v_a\wedge K_a}{(r_a)^2}c_a
\right)
\end{equation}
generated by emission points $z(t_1)\in\zeta$ and $z(t_2)\in\zeta$. Symbol 
$c_a$ denotes the factor $\gamma_a^{-2}+(K_a\cdot {\dot v}_a)$ 
involved in eq.(\ref{Fi}).

The first multiple integral calculates the interference of the disk 
emanated by fixed point $z(t_1)\in\zeta$ with radiation generated by 
all the points {\it before} the instant $t_1$. The second fourfold 
integral gives the contribution of points {\it after} $t_1$ (see figure 
\ref{f_ab}).  

It is worth noting that time variables $t_1$ and $t_2$ parametrize the 
same world line $\zeta$. Coordinate transformation (\ref{ct}) is 
invariant with respect to the following reciprocity:
\begin{equation}
\Upsilon: t_1\leftrightarrow t_2, \alpha\leftrightarrow\beta, 
\varphi\mapsto\varphi+\pi.
\end{equation}
This symmetry provides identity of domains of fourfold integrals in 
energy-momentum (\ref{pint}).

It is obvious that the support of double integral 
$\int_{-\infty}^tdt_1\int_{t_1}^{t}dt_2$ coincides with the support of the 
integral $\int_{-\infty}^tdt_2\int_{-\infty}^{t_2}dt_1$. Since instants 
$t_1$ and $t_2$ label different points at the same world line $\zeta$, one 
can interchanges the indices ``first'' and ``second'' in the second 
fourfold integral of eq.(\ref{pint}). Via interchanging of these indices 
we finally obtain $\int_{-\infty}^tdt_1\int_{-\infty}^{t_1}dt_2$
instead of initial $\int_{-\infty}^tdt_1\int_{t_1}^{t}dt_2$.
Taking into account these circumstances in the expression (\ref{pint}) for 
energy-momentum carried by electromagnetic field we finally obtain
\begin{equation}\label{pnt}
p_{\rm em}^\alpha=\int\limits_{-\infty}^tdt_1\int\limits_{-\infty}^{t_1}dt_2
\int\limits_{0}^{k_1^0}dR 
\int\limits_0^{2\pi}d\varphi Jt^{0\alpha}
\end{equation}
where new stress-energy tensor is symmetric in 
the pair of indices 1 and 2:
\begin{eqnarray}\label{tfin}
2\pi t^{0\alpha}&=& 
2\pi\left(t^{0\alpha}_{12}+t^{0\alpha}_{21}\right)\\
&=&f_{(1)}^{0\lambda}f_{(2)\lambda}^\alpha +
f_{(2)}^{0\lambda}f_{(1)\lambda}^\alpha
 -\frac14\eta^{0\alpha}\left[
f_{(1)}^{\mu\nu}f^{(2)}_{\mu\nu} +f_{(2)}^{\mu\nu}f^{(1)}_{\mu\nu}
\right].\nonumber
\end{eqnarray}

\subsection{Angular integration of energy-momentum tensor density}
\setcounter{equation}{0}

\begin{figure}[t]
\begin{center}
\epsfclipon
\epsfclipon
\epsfig{file=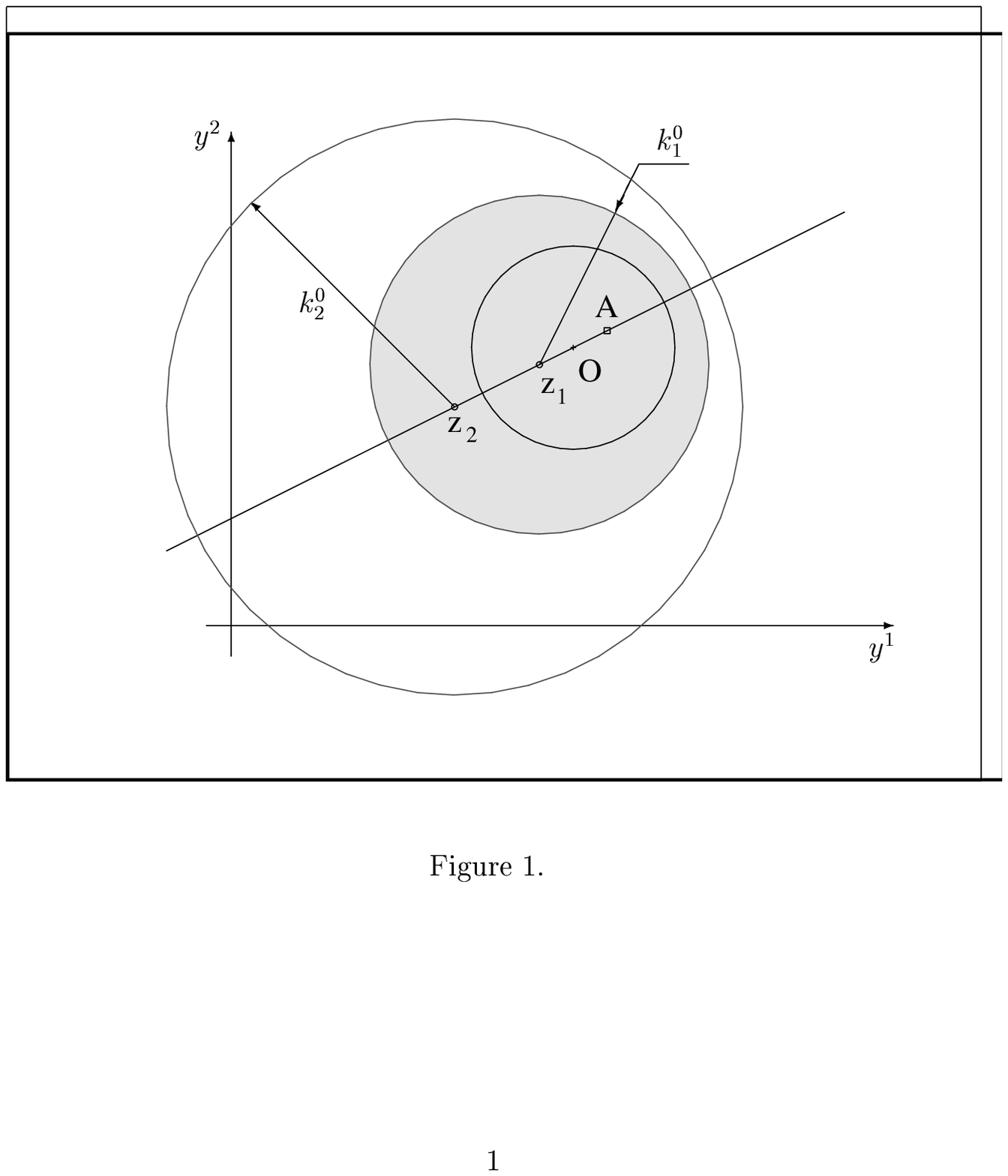,width=9cm}
\end{center}
\caption{\label{bt}
\small The interference picture in a plane $\Sigma_t$. The points 
$z(t_1)\in\zeta$ and $z(t_2)\in\zeta, t_2<t_1,$ emanate the radiation which 
filled up the disks centered at $z_1$ and $z_2$, respectively. The 
gray disk with radius $k_1^0=t-t_1$ is filled up by nonconcentric 
circles centered at the line crossing both the points $z_1$ and $z_2$. 
If parameter $\beta$ vanishes the circle is centered at $z_1$; its 
radius is equal to $k_1^0$. If $\beta=\beta_0<0$ the circle reduces to the 
point $A$ labeled by the box symbol. In case of intermediate value 
$\beta_0<\beta<0$ we have the circle of radius $R$ with center 
at point $O$ between $z_1$ and $A$.}
\end{figure}

We see that it is sufficient to consider the situation when $t_1\ge t_2$.
The smaller disk $C_1(z_1,t-t_1)\subset C_2(z_2,t-t_2)$ is filled up by 
nonconcentric circles with radii $R\in[0,k_1^0]$ (see figures 
\ref{f_ab} and \ref{bt}). To calculate the total flows (\ref{pnt}) of 
electromagnetic field energy and momentum  which flow across the plane 
$\Sigma_t$ we should integrate the Maxwell energy-momentum tensor density 
(\ref{tfin}) over angular variable $\varphi$, over radius $R$ and, 
finally, over time variables $t_1$ and $t_2$. Integration over 
$\varphi$ is not a trivial matter. The difficulty resides mostly 
with norms  $\|K_a\|^2=-\eta_{\alpha\beta}K_a^{\alpha}K_a^{\beta}$ of 
separation vectors $K_a=y-z_a$ which result in elliptic integrals. To avoid 
dealing with them we modify the coordinate transformation (\ref{ct}). We 
fix the parameter $\beta$ in such a way that the norm $\|K_1\|^2$ becomes 
proportional to the norm $\|K_2\|^2$:
\begin{equation} 
\|K_1\|^2=-\frac{\beta}{\alpha}\|K_2\|^2.
\end{equation} 
Keeping in mind identity $\alpha+\beta=1$, we arrive at the quadratic 
algebraic equation on $\beta$ which does not contain the angle variable:
\begin{equation}\label{r2}
R^2=\alpha (k_1^0)^2+\beta (k_2^0)^2 -\alpha\beta{\bf q}^2.
\end{equation}
We choose the root which vanishes when $R=k_1^0$:
\begin{equation}\label{beta}
\beta=\frac{1}{2{\bf q}^2}\left(
-(k_2^0)^2+(k_1^0)^2+{\bf q}^2+\sqrt{D}
\right)\qquad
D=\left[(k_2^0)^2-(k_1^0)^2-{\bf q}^2\right]^2 -
4{\bf q}^2\left[(k_1^0)^2-R^2\right].\nonumber
\end{equation}
If ${\bf q}^2$ tends to zero while $t_1\ne t_2$, it becomes the unique 
root of the linear equation on $\beta$ originated from eq.(\ref{r2}) with 
${\bf q}^2=0$.

If $R=0$ the $R-$circle reduces to point $A$ with coordinates 
$(z_1^i-\beta_0q^i)$ where $\beta_0=\left.\beta\right|_{R=0}$. If 
$R=k_1^0$ then $\beta=0$ and the circle is centered at $z^i(t_1)$
(see figure \ref{bt}). 

Changing the variable of integration from $R$ to $\beta$ transforms two 
inner integrals in the fourfold integral (\ref{pnt}) as follows:
\begin{equation}\label{rb}
\int\limits_0^{k_1^0}dR\int\limits_0^{2\pi}d\varphi \left(
1-q\frac{\partial\beta}{\partial R}\cos\varphi\right)R
=\int\limits_{\beta_0}^{0}d\beta\int\limits_0^{2\pi}d\varphi \left(
\frac12\frac{\partial R^2}{\partial\beta}-qR\cos\varphi\right).
\end{equation}
Having differentiated eq.(\ref{r2}) with respect to $\beta$ we obtain new 
Jacobian
\begin{equation}\label{Jc}
J=(1/2)\left[(k_2^0)^2-(k_1^0)^2-{\mathbf q}^2\right] +\beta {\mathbf 
q}^2-qR\cos\varphi.
\end{equation}

It is straightforward to substitute the field densities (\ref{F1F2}) 
evaluated at instants $t_1$ and $t_2$ into expression (\ref{tfin})
to calculate the integrand of the multiple integral (\ref{pnt}).
It is of great importance that the square of norm $\|K_a\|$ is 
proportional to $J$ (see eqs.(\ref{KKJ}) derived in \ref{coord}).
For the first term, $Jt_{12}^{\alpha\beta}$, we obtain the following 
cumbersome expression
\begin{eqnarray}\label{ff12}
Jt^{\alpha\beta}_{12}&=&
\frac{e^2}{2}I\left\{
{\hat T}^{\alpha\beta}_{12}\left(\frac{\partial^2\sigma}{\partial 
t_1\partial t_2}\right)
+{\hat T}^\alpha_1\left(v_2^\beta\frac{\partial\sigma}{\partial 
t_1}\right) 
+{\hat T}^\beta_2\left(v_1^\alpha\frac{\partial\sigma}{\partial 
t_2}\right)
+{\hat T}^0(v_1^\alpha v_2^\beta\sigma)
\right.\nonumber\\
&-&\left. C_1^\alpha v_2^\beta\frac{\partial^2\sigma}{\partial 
t_1\partial t_2}
- D_1^\alpha v_2^\beta\frac{\partial^3\sigma}{\partial t_1^2\partial t_2}
- B_2^\beta v_1^\alpha\frac{\partial^2\sigma}{\partial 
t_1\partial t_2}
- D_2^\beta v_1^\alpha\frac{\partial^3\sigma}{\partial t_1\partial t_2^2}
\right.\nonumber\\
&-&\left.
B^0v_1^\alpha v_2^\beta\frac{\partial\sigma}{\partial t_1}
-C^0v_1^\alpha v_2^\beta\frac{\partial\sigma}{\partial t_2}
-D^0\left(
{\dot v}_1^\alpha v_2^\beta\frac{\partial\sigma}{\partial t_2}
+v_1^\alpha {\dot v}_2^\beta\frac{\partial\sigma}{\partial t_1}
+v_1^\alpha v_2^\beta\frac{\partial^2\sigma}{\partial t_1\partial t_2}
\right)
\right\}\nonumber\\
&-&\frac{e^2}{2}I'{\hat T}^J(v_1^\alpha v_2^\beta)
-\frac{e^2}{4}\eta^{\alpha\beta}\left\{
I{\hat T}^0(\lambda)
-I'{\hat T}^J(\lambda_0)\right\}.
\end{eqnarray}
The functions
\begin{equation}\label{lbd}
\lambda=\sigma\frac{\partial^2\sigma}{\partial t_1\partial t_2} -
\frac{\partial\sigma}{\partial t_1}\frac{\partial\sigma}{\partial 
t_2},\qquad \lambda_0=\frac{\partial^2\sigma}{\partial 
t_1\partial t_2}
\end{equation}
depend on Synge's world function (\ref{sgm}) of two timelike related 
points, $z(t_1)\in\zeta$ and $z(t_2)\in\zeta$, taken with opposite sign. 
Symbols $I$ and $I'$ denote $\beta$-dependent factors 
\begin{equation}\label{Ibar}
I=\frac{1}{\sqrt{-\beta\alpha}},\qquad 
I'=\sqrt{\frac{-\beta}{\alpha}}+
\sqrt{\frac{\alpha}{-\beta}}.
\end{equation}

Each second order differential operator
\begin{equation}\label{Ta}
{\hat T}^a=
D^a\frac{\partial^2 }{\partial t_1\partial t_2}+
B^a\frac{\partial}{\partial t_1}
+C^a\frac{\partial}{\partial t_2}+A^a
\end{equation}
has been labeled according to its dependence on the combination of 
components of the separation vectors $K_1$ and $K_2$ pictured in figure
\ref{K1K2} (see \ref{coord}) or on Jacobian (\ref{Jc}). The components of 
these vectors are involved in $\varphi$-dependent coefficients
\begin{eqnarray}\label{ABC}
D^a&=&\frac{1}{2\pi}\frac{a}{r_1r_2},\qquad
B^a=\frac{1}{2\pi}\frac{ac_2}{r_1(r_2)^2}\\
C^a&=&\frac{1}{2\pi}\frac{ac_1}{(r_1)^2r_2},
\qquad
A^a=\frac{1}{2\pi}\frac{ac_1c_2}{(r_1)^2(r_2)^2}
\nonumber
\end{eqnarray}
where factor $a$ is replaced by $K_1^\alpha K_2^\beta, K_1^\alpha, 
K_2^\beta, J$ or $1$ for $\hat T^{\alpha\beta}_{12}$, $\hat T^\alpha_1$, 
$\hat T^\beta_2$, $\hat T^J$ or $\hat T^0$, respectively.

Integration of the electromagnetic field's stress-energy tensor over the 
angular variable is the key to the problem. All the $\varphi$-dependent 
constructions are concentrated in the coefficients of differential operators 
of the type in eq.(\ref{Ta}). We introduce a new operator 
\begin{equation}\label{calTa}
{\hat{\cal T}}^a=
{\cal D}^a\frac{\partial^2 }{\partial t_1\partial t_2}+
{\cal B}^a\frac{\partial}{\partial t_1}
+{\cal C}^a\frac{\partial}{\partial t_2}+{\cal A}^a
\end{equation}
where the script letters denote the coefficients (\ref{ABC}) integrated 
over $\varphi$. (The integration is performed in \ref{varphi}).

To distinguish the partial derivatives in time variables we rewrite the 
operator (\ref{calTa}) as the sum of the second-order differential operator
\begin{equation}\label{Pa}
{\hat\Pi}^a=\frac{\partial^2 }{\partial t_1\partial t_2}{\cal D}^a+
\frac{\partial}{\partial t_1}\left({\cal B}^a-
\frac{\partial{\cal D}^a}{\partial t_2}\right)
+\frac{\partial}{\partial t_2}\left({\cal C}^a-
\frac{\partial{\cal D}^a}{\partial t_1}\right)
\end{equation}
and the tail
\begin{equation}\label{pa}
\pi^a=\frac{\partial^2 {\cal D}^a}{\partial t_1\partial t_2}-
\frac{\partial{\cal B}^a}{\partial t_1}-
\frac{\partial{\cal C}^a}{\partial t_2}+{\cal A}^a.
\end{equation}
For a smooth function $f(t_1,t_2)$ we have
\begin{equation}
{\hat{\cal T}}^a(f)={\hat\Pi}^a(f)+f\pi^a.
\end{equation}

In \ref{varphi} we derive the relations
\begin{eqnarray}\label{cli}
\pi^0&=&0,\qquad \pi^J=0\\
\pi_1^\alpha&=&v_1^\alpha\left({\cal B}^0-\frac{\partial{\cal 
D}^0}{\partial t_2}\right),\qquad
\pi_2^\beta=v_2^\beta\left({\cal C}^0-\frac{\partial{\cal D}^0}{\partial 
t_1}\right)
\nonumber\\
\pi_1^{\alpha J}&=&v_1^\alpha\left({\cal B}^J-\frac{\partial{\cal 
D}^J}{\partial t_2}\right),\qquad
\pi_2^{\beta J}=v_2^\beta\left({\cal C}^J-\frac{\partial{\cal 
D}^J}{\partial t_1}\right)
\nonumber\\
\pi_{12}^{\alpha\beta}&=&
v_1^\alpha\left({\cal B}_2^\beta-\frac{\partial {\cal D}_2^\beta}{\partial 
t_2}\right)
+v_2^\beta\left({\cal C}_1^\alpha-\frac{\partial {\cal 
D}_1^\alpha}{\partial t_1}\right)-v_1^\alpha v_2^\beta {\cal D}^0
\nonumber
\end{eqnarray}
which allow us to rewrite the integral of eq.(\ref{ff12}) over $\varphi$ in 
terms of differential operators ${\hat\Pi}^a$:
\begin{eqnarray}\label{f12fi}
\int_0^{2\pi}d\varphi Jt^{\alpha\beta}_{12}&=&
\frac{e^2}{2}I\left\{
{\hat\Pi}^{\alpha\beta}_{12}\left(\frac{\partial^2\sigma}{\partial 
t_1\partial t_2}\right)
+{\hat\Pi}^\alpha_1\left(v_2^\beta\frac{\partial\sigma}{\partial 
t_1}\right) 
+{\hat\Pi}^\beta_2\left(v_1^\alpha\frac{\partial\sigma}{\partial 
t_2}\right)
+{\hat\Pi}^0(v_1^\alpha v_2^\beta\sigma)
\right.\nonumber\\
&-&\left. 
\frac{\partial}{\partial t_1}\left(
D_1^\alpha v_2^\beta\frac{\partial^2\sigma}{\partial t_1\partial t_2}
\right)
-\frac{\partial}{\partial t_2}\left( 
D_2^\beta v_1^\alpha\frac{\partial^2\sigma}{\partial t_1\partial t_2}
\right)
\right.\nonumber\\
&-&\left. 
\frac{\partial}{\partial t_1}\left(
D^0v_1^\alpha v_2^\beta\frac{\partial\sigma}{\partial t_2}
\right)
-\frac{\partial}{\partial t_2}\left( 
D^0 v_1^\alpha v_2^\beta\frac{\partial\sigma}{\partial t_1}
\right)
\right\}\nonumber\\
&-&\frac{e^2}{2}I'{\hat\Pi}^J(v_1^\alpha v_2^\beta)
-\frac{e^2}{4}\eta^{\alpha\beta}\left\{
I{\hat\Pi}^0(\lambda)
-I'{\hat\Pi}^J(\lambda_0)\right\}.
\end{eqnarray}
Since the operator in eq.(\ref{Pa}) is the combination of partial 
derivatives in time variables, the angular integration gives the key to the 
problem.

Setting $\alpha=0$ and $\beta=i$ in eq.(\ref{f12fi}) we obtain the first 
term of the mixed space-time components of the stress-energy tensor 
({\ref{tfin}). We add the term where indices $1$ and $2$ are 
interchanged. Since zeroth components of the separation three-vectors 
$K_1$ and $K_2$ do not depend on $\varphi$, the final expression get
simplified:
\begin{eqnarray} \label{t0if}
\int_0^{2\pi}d\varphi Jt^{0i}&=&\frac{e^2}{2}I\left[
{\hat\Pi }_1^i\left(\frac{\partial\lambda_2}{\partial t_1}\right) +
{\hat\Pi }_2^i\left(\frac{\partial\lambda_1}{\partial t_2}\right)+
{\hat\Pi }^0\left(v_2^i\lambda_1+v_1^i\lambda_2\right)\right.\\
&-&\left.\frac{\partial}{\partial t_1}\left(v_2^i
\frac{\partial\lambda_1}{\partial t_2}{\cal D}^0
\right)-
\frac{\partial}{\partial t_2}\left(v_1^i
\frac{\partial\lambda_2}{\partial t_1}{\cal D}^0
\right)
\right]\nonumber\\
&-&\frac{e^2}{2}I'
{\hat\Pi}^J(v_1^i+v_2^i)\nonumber
\end{eqnarray} 
where 
\begin{equation}\label{l1-2}
\lambda_1=k_1^0\frac{\partial\sigma}{\partial t_1}+\sigma,\qquad
\lambda_2=k_2^0\frac{\partial\sigma}{\partial t_2}+\sigma.
\end{equation}

Similarly we derive the zeroth component $t^{00}$. Setting $\alpha=0$ and 
$\beta=0$ in eq.(\ref{f12fi}) we obtain the first term of energy density. 
Since it is symmetric in the pair of indices 1 and 2, the second term, 
$t_{21}^{00}$, doubles it. The integral of energy density $t^{00}$ over 
the angular variable has the form
\begin{equation} \label{t00f}
\int_0^{2\pi}d\varphi Jt^{00}=e^2\left[I{\hat\Pi
}^0(\kappa) - I'{\hat\Pi}^J(\mu)\right]
\end{equation}
where
\begin{eqnarray}\label{km}
\kappa&=&k_1^0k_2^0\frac{\partial^2\sigma}{\partial t_1\partial t_2} +
k_1^0\frac{\partial\sigma}{\partial t_1}+
k_2^0\frac{\partial\sigma}{\partial t_2}-
\frac12\frac{\partial\sigma}{\partial 
t_1}\frac{\partial\sigma}{\partial t_2}
+\sigma\mu\\
\mu&=&\frac12\frac{\partial^2\sigma}{\partial t_1\partial t_2} +1 .
\nonumber
\end{eqnarray} 

We see that the integration of the electromagnetic field's stress-energy 
tensor over $\varphi$ yields integrals that are functions of the end points 
only. In the next subsection we classify them and consider the problem of 
integration over the remaining variables.

\subsection{Integration over time variables and $\beta$}
\setcounter{equation}{0}

Our purpose in this section is to develop the mathematical tools required in 
a surface integration of energy-momentum tensor density in 
2+1 electrodynamics. Integration over angle variable results the 
combination of partial derivatives in time variables:
\begin{equation}\label{triple}
p_{\rm em}^\alpha(t)=\left.
\begin{array}{c}
\displaystyle 
e^2\int_{-\infty}^t dt_1\int_{-\infty}^{t_1}dt_2\\
\\[-1em]
\displaystyle
e^2\int_{-\infty}^t dt_2\int_{t_2}^t dt_1
\end{array}
\right\}
\left(
\frac{\partial^2 G^\alpha_{12}}{\partial t_1\partial t_2}+
\frac{\partial G^\alpha_1}{\partial t_1}+
\frac{\partial G^\alpha_2}{\partial t_2}
\right).
\end{equation}
By virtue of the equality
\begin{equation}\label{dG}
\int_{\beta_0}^0d\beta\frac{\partial G(\beta,t_1,t_2)}{\partial t_a}=
\frac{\partial}{\partial t_a}\left[\int_{\beta_0}^0d\beta
G(\beta,t_1,t_2)\right]+G(\beta_0,t_1,t_2)
\frac{\partial \beta_0(t_1,t_2)}{\partial t_a}
\end{equation}
the triple integral (\ref{triple}) can be rewritten as follows:
\begin{eqnarray}\label{pG}
p_{\rm em}^\alpha(t)&=&e^2\left[\lim_{k_1^0\to 0}
\int_{\beta_0}^0d\beta G_{12}^\alpha\right]_{t_2\to -\infty}^{t_2=t}
+e^2\int_{-\infty}^tdt_2\lim_{k_1^0\to 0}\left[
\left.G_{12}^\alpha\right|_{\beta=\beta_0}
\frac{\partial \beta_0}{\partial t_2}\right]
\\
&-&e^2\int_{-\infty}^tdt_2\lim_{\triangle t\to 0}
\int_{\beta_0}^0d\beta\left[\frac{\partial G_{12}^\alpha}{\partial t_2} 
+G_1^\alpha\right]_{k_1^0=k_2^0-\triangle t}
+e^2\int_{-\infty}^tdt_2\lim_{k_1^0\to 0}\left[
\int_{\beta_0}^0d\beta G_1^\alpha\right]
\nonumber\\
&+&e^2\int_{-\infty}^tdt_1\lim_{\triangle t\to 0}\left[
\int_{\beta_0}^0d\beta G_2^\alpha\right]_{k_2^0=k_1^0+\triangle t 
}
-e^2\int_{-\infty}^tdt_1\lim_{t_2\to -\infty}\left[
\int_{\beta_0}^0d\beta G_2^\alpha\right]
\nonumber\\
&+&\left.
\begin{array}{c}
\displaystyle 
e^2\int_{-\infty}^t dt_1\int_{-\infty}^{t_1}dt_2\\
\\[-1em]
\displaystyle
e^2\int_{-\infty}^t dt_2\int_{t_2}^t dt_1
\end{array}
\right\}\left(
\left[\frac{\partial G_{12}^\alpha}{\partial t_2}+
G_1^\alpha\right]_{\beta=\beta_0}
\frac{\partial\beta_0}{\partial t_1} + 
\left.G_2^\alpha\right|_{\beta=\beta_0}\frac{\partial\beta_0}{\partial t_2}
\right).\nonumber
\end{eqnarray}
In  \ref{varphi} we calculate the functions
\begin{eqnarray}\label{G00}
G_{12}^0&=&I{\cal D}^0\kappa -
I'{\cal D}^J\mu \\
G_1^0&=&-\sqrt{\frac{-\beta}{\alpha}}\kappa\frac{{\bf v}_1^2}{\|r_1\|^3}-
I'\mu\frac{\partial}{\partial\beta}
\left(\frac{\beta}{\|r_1\|}
\right)\nonumber\\
G_2^0&=&\sqrt{\frac{\alpha}{-\beta}}\kappa\frac{{\bf v}_2^2}{\|r_2\|^3}-
I'\mu\frac{\partial}{\partial\beta}
\left(\frac{\alpha}{\|r_2\|}
\right)\nonumber\\\label{Gi}
G_{12}^i&=&\frac{I}{2}\left[
\frac{\partial\lambda_1}{\partial t_2}{\cal D}_2^i
+\frac{\partial\lambda_2}{\partial t_1}{\cal D}_1^i
+\left(v_1^i\lambda_2+v_2^i\lambda_1\right){\cal D}^0\right]
- \frac{I'}{2}\left(v_1^i+v_2^i\right)
{\cal D}^J\\
G_1^i&=&\frac{I}{2}\frac{\beta}{\|r_1\|^3}\left[
\frac{\partial\lambda_1}{\partial t_2}
\left(\alpha q^i{\bf v}_1^2+r_1^0v_1^i\right)+
\frac{\partial\lambda_2}{\partial t_1}
\left(-\beta q^i{\bf v}_1^2+r_1^0v_1^i\right)
+\left(v_1^i\lambda_2+v_2^i\lambda_1\right)
{\bf v}_1^2\right]\nonumber\\
&-&
\frac{I'}{2}\left(v_1^i+v_2^i\right)
\frac{\partial}{\partial\beta}
\left(\frac{\beta}{\|r_1\|}
\right)\nonumber\\
G_2^i&=&\frac{I}{2}\frac{\alpha}{\|r_2\|^3}\left[
\frac{\partial\lambda_1}{\partial t_2}
\left(\alpha q^i{\bf v}_2^2+r_2^0v_2^i\right)+
\frac{\partial\lambda_2}{\partial t_1}
\left(-\beta q^i{\bf v}_2^2+r_2^0v_2^i\right)
+\left(v_1^i\lambda_2+v_2^i\lambda_1\right)
{\bf v}_2^2\right]\nonumber\\
&-&
\frac{I'}{2}\left(v_1^i+v_2^i\right)
\frac{\partial}{\partial\beta}
\left(\frac{\alpha}{\|r_2\|}
\right)\nonumber
\end{eqnarray}
involved in these integrals.

{\bf $\mathbf 1^{\rm o}$. Integrals where $\mathbf t_1\to t$.} Equality 
(\ref{sn0}) implies that the lower limit $\beta_0$ tends to $0$ if 
$k_1^0=t-t_1$ vanishes. The upper limit is equal to zero too. Then the 
integral over parameter $\beta$ vanishes whenever an expression under 
integral sign is smooth. So, we must limit our computations to the 
singular terms only. They are performed in  \ref{pink}; these integrals 
do not contribute in the energy-momentum at all.

{\bf $\mathbf 2^{\rm o}$. Integrals where $\mathbf t_1=t_2$.} According 
to eq.(\ref{sn0}), the equality $t_1=t_2$ yields $\sin\vartheta_0=1$ and 
lower limit $\beta_0=-\tan\vartheta_0$ tends to $-\infty$. The small 
parameter is the positively valued difference $\triangle t=t_1-t_2$. The 
integration is performed in  \ref{blue}; the resulting terms belong 
to the bound part of energy-momentum (to that which is permanently 
``attached'' to the charge and is carried along with it.)

{\bf $\mathbf 3^{\rm o}$. Integrals where $\mathbf t_2\to-\infty$.} 
Equality (\ref{sn0}) implies that the lower limit $\beta_0$ tends to $0$ 
if $k_2^0=t-t_2$ increases extremely. Then the integral over parameter 
$\beta$ vanishes whenever an expression under integral sign is smooth. 
So, we must limit our computations to the singular terms only. They are 
performed in  \ref{blue}; the resulting terms belong to the bound 
electromagnetic ``cloud'' which can not be separated from the charged 
particle.

{\bf $\mathbf 4^{\rm o}$. Integrals at point where $\mathbf 
\beta=\beta_0$.} In this case the radius of the smallest circle pictured 
in figure \ref{bt} vanishes and it reduces to the point $A$. The 
contribution in $p_{\rm em}^\alpha$ is given by the last line of 
eq.(\ref{pG}). In \ref{yellow} we present the integrand as the combination 
of partial derivatives in time variables and nonderivative tail. After 
integration over $t_1$ or $t_2$, the derivatives are coupled with bound 
terms obtained in  \ref{blue}; the sum is absorbed by three-momentum of 
bare particle within renormalization procedure. The tail contains 
radiative terms which detach themselves from the charge and lead 
independent existence. 

Summing up all the contributions $2^{\rm o}-4^{\rm o}$ we finally obtain
\begin{eqnarray}\label{p0em}
p^0_{\rm em}(t)&=&e^2\left.
\frac{1+1/2\sqrt{1-{\bf v}_1^2}}{1+\sqrt{1-{\bf v}_1^2}}\frac{1}{\sqrt{1-{\bf v}_1^2}}
\right|_{t_1\to -\infty}^{t_1=t}+
\frac{e^2}{2}\int_{-\infty}^tdt_2\frac{1}{\sqrt{2\sigma(t,t_2)}}
\nonumber\\
&+&e^2\int_{-\infty}^t dt_1\int_{-\infty}^{t_1}dt_2
\left[
-\frac{(v_1\cdot v_2)q^0}{(2\sigma)^{3/2}}+
\frac12\frac{(v_1\cdot q)}{(2\sigma)^{3/2}}+
\frac12\frac{(v_2\cdot q)}{(2\sigma)^{3/2}}
\right]\\\label{piem}
p^i_{\rm em}(t)&=&e^2\left.
\frac{1+1/2\sqrt{1-{\bf v}_1^2}}{1+\sqrt{1-{\bf 
v}_1^2}}\frac{v_1^i}{\sqrt{1-{\bf v}_1^2}}
\right|_{t_1\to -\infty}^{t_1=t}+
\frac{e^2}{2}\int_{-\infty}^tdt_2\frac{v_2^i}{\sqrt{2\sigma(t,t_2)}}
\nonumber\\
&+&e^2\int_{-\infty}^t dt_1\int_{-\infty}^{t_1}dt_2
\left[
-\frac{(v_1\cdot v_2)q^i}{(2\sigma)^{3/2}}+
\frac12\frac{(v_1\cdot q)v_2^i}{(2\sigma)^{3/2}}+
\frac12\frac{(v_2\cdot q)v_1^i}{(2\sigma)^{3/2}}
\right].
\end{eqnarray}
The finite terms which depend on the end points only are noncovariant. 
They express the ``deformation'' of electromagnetic cloud due to the 
choice of coordinate-dependent hole around the particle in the 
integration surface $\Sigma_t$. We neglect these structureless terms. The 
single integrals describe the covariant singular part of energy-momentum 
carried by electromagnetic field. The first term in between the square 
brackets of eqs.(\ref{p0em}) and (\ref{piem}) cannot be rewritten 
as the partial derivative in $t_1$ or $t_2$. It determines the radiation 
reaction in $2+1$ electrodynamics.

\section{Conclusions}\label{concl}

In the present paper, we calculate the total flows of (retarded) 
electromagnetic field energy, momentum and angular momentum which flow 
across the plane $\Sigma_t=\{y\in{\mathbb M}_{\,3}:y^0=t\}$. The 
field is generated by a point-like electric charge arbitrarily moving in 
flat space-time of three dimensions. The computation is not a trivial 
matter, since the Maxwell energy-momentum tensor density evaluated at field 
point $y\in\Sigma_t$ is nonlocal. In odd dimensions the retarded field is 
generated by the portion of the world line $\zeta$ that corresponds to the 
particle's history {\it before} $t^{ret}(y)$. Since the stress-energy 
tensor is quadratic in field strengths, we should {\it twice} integrate it 
over $\zeta$. We integrate it also over two variables which parametrize 
$\Sigma_t$ in order to calculate energy-momentum and angular momentum 
which flow across this plane. Thanks to the integration we reduce the 
support of the retarded and advanced Green's functions to particle's 
trajectory.

The Dirac scheme which manipulates fields {\it on the world line only} is 
the key point of investigation. By fields we mean the convolution 
of three-velocity and nonlocal part of the retarded strength 
tensor evaluated at point $z(\tau_1)\in\zeta$; the torque of this ``Lorentz 
$\theta$-force'' arises in electromagnetic field's total angular 
momentum. (The singular $\delta$-term (\ref{Fd}) is defined on the light 
cone; it is meaningless since both the field point, $z(t_1)$, and the 
emission point, $z(t_2)$, lie on the time-like world line). The retarded 
and the advanced quantities arise naturally. The retarded Lorentz 
self-force, as well as its torque, contains integration over the portion of 
the world line that corresponds to the interval $-\infty<t_2\le t_1$.  
Domain of integration of their advanced counterparts corresponds to the 
interval $t_1\le t_2\le t$.

Noether quantity $G^\alpha_{\rm em}$ carried by the electromagnetic field 
consists of terms of two quite different types: (i) singular, 
$G^\alpha_{\rm S}$, which is permanently attached to the source and 
carried along with it, and (ii) radiative, $G^\alpha_{\rm R}$, which 
detaches itself from the charge and leads independent existence.
The former is the half-sum of retarded and advanced expressions,
integrated over $\zeta$, while the latter is the integral of one-half of 
the retarded quantity minus one-half of the advanced one. Within 
regularization procedure the bound terms $G_{\rm S}^\alpha$ are coupled 
with energy-momentum and angular momentum of the bare source, so that 
already renormalized characteristics $G^\alpha_{\rm part}$ of charged 
particle are proclaimed to be finite. Noether quantities which are 
properly conserved become
$$
G^\alpha=G^\alpha_{\rm part}+G^\alpha_{\rm R}.
$$

The regularization procedure which relies on energy-momentum and angular 
momentum balance equations is proposed. Energy-momentum balance equations 
define the change of particle's three-momentum under the influence of an 
external electromagnetic field where loss of energy due to radiation is 
taken into account. The angular momentum balance equations explain how this 
already renormalized three-momentum depends on particle's individual 
characteristics. They constitute the system of three linear equations in 
three components of particle's momentum. Its rank is equal to 2, so that 
arbitrary scalar function arises naturally. It can be interpreted as a {\it 
dynamical} mass of dressed charge which is proclaimed to be finite. A 
surprising feature is that this mass depends on the particle's history 
before the instant of observation when the charge is accelerated. Already 
renormalized particle's momentum contains, apart from the usual velocity 
term, also nonlocal contribution from point-like particle's 
electromagnetic field.

Having substituted this expression in the energy-momentum balance equations 
we derive a three-dimensional analogue of the Lorentz-Dirac equation 
$$
ma^\mu_\tau=eu_{\tau,\alpha} F^{\mu\alpha}_{\rm ret}(\tau)+ 
\frac{e^2}{2}a^\mu_\tau\int_{-\infty}^\tau 
\frac{ds}{\sqrt{2\sigma(\tau,s)}}
+eu_{\tau,\alpha} F^{\mu\alpha}_{\rm ext}.
$$
The loss of energy due to radiation is determined by work done by Lorentz 
force of point-like charge acting upon itself. The nonlocal term which is 
proportional to the particle's acceleration provides finiteness of the 
self-action. The third term describes the influence of an external field.

In this paper we develop a convenient technique which allows us to 
integrate nonlocal stress-energy tensor over the spacelike plane. The next 
step will be to implement this technique to a point particle coupled to 
massive scalar field following an arbitrary trajectory on a flat 
space-time. The Klein-Gordon field generated by the scalar charge holds 
energy near the particle. This circumstance makes the procedure of 
decomposition of the energy-momentum into bound and radiative parts unclear.

In Ref. \cite{AHNS} the remarkable correspondence is established between 
dynamical equations which govern behaviour of superfluid 
$\!\!\!\phantom{I}^4$He films and Maxwell equations for electrodynamics 
in 2+1 dimensions (see also refs.\cite{FL,Z})\footnote{I wish to 
thank O.Derzhko for drawing these papers to my attention.}. Perhaps the 
effective equation of motion (\ref{mext}) will be useful in study of 
phenomena in superfluid dynamics which correspond to the radiation 
friction in $2+1$ electrodynamics.

\section*{Acknowledgments}
I am grateful to Professor V.Tre\-tyak for continuous encouragement and for 
a helpful reading of this manuscript. I would like to thank A.Du\-vi\-ryak 
and R.Matsyuk for many useful discussions.

\subsubsection{Uniformly moving charge (conserved quantities)}\label{unif}
\renewcommand{\theequation}{\Alph{subsubsection}.\arabic{equation}}
\setcounter{equation}{0}

The simplest field is generated by an unmoved charge placed at the 
coordinate origin. Setting $z=(t,0,0)$ and $u=(1,0,0)$ in eq.(\ref{Fret}), 
one can derive that the only nontrivial components of static field are:
\begin{eqnarray}\label{A1}
F_{i0}&=&e\int\limits_{-\infty}^{y^0-r}\frac{dt}{\sqrt{(y^0-t)^2-r^2}}
\frac{y^i}{(y^0-t)^2}\\
&=&\left.-e\frac{y^i}{r^2}
\frac{\sqrt{(y^0-t)^2-r^2}}{y^0-t}\right|_{t\to -\infty}^{t=y^0-r}
=e\frac{y^i}{r^2}.\nonumber
\end{eqnarray}
$r:=\sqrt{(y^1)^2+(y^2)^2}$ is the distance to the charge. Having performed 
Poincar\'e transformation, the combination of translation and Lorentz 
transformation, we find the field generated by a uniformly moving charge:
\begin{equation}\label{AF}
F_{\alpha\beta}=e\frac{u_\alpha k_\beta - u_\beta k_\alpha}{r}.
\end{equation}
Here $r=-\eta_{\alpha\beta}(y^\alpha -z^\alpha(s))u^\beta$ is the retarded 
distance where the particle's position $z^\alpha(\tau)=z_0^\alpha 
+u^\alpha\tau$ is referred to the retarded instant of time $s$. We denote 
$k^\alpha$ the null vector $y^\alpha-z^\alpha(s)$ rescaled by the retarded 
distance, i.e.
\begin{equation}
k^\alpha=\frac{y^\alpha-z^\alpha(s)}{r}.
\end{equation}

It is straightforward to substitute eq.(\ref{AF}) in eq.(\ref{T}) to 
calculate the electromagnetic field's stress-energy tensor:
\begin{equation}\label{AT}
2\pi T^{\alpha\beta}=\frac{e^2}{r^2}\left(
u^\alpha k^\beta+u^\beta k^\alpha -k^\alpha k^\beta 
+\frac12\eta^{\alpha\beta}
\right).
\end{equation}
Now we calculate the electromagnetic field momentum 
\begin{equation}\label{ppem}
p^\nu_{\rm em}(t)=\int_{\Sigma_t} d\sigma_\mu T^{\mu\nu}
\end{equation}
where an integration surface $\Sigma_t$ is a surface of constant $y^0$. 
We start with coordinate transformation $(y^0,y^1,y^2)\mapsto 
(r,s,\varphi)$ locally given by
\begin{equation}
y^\alpha=z^\alpha(s)+rk^\alpha,\qquad 
k^\alpha=\Lambda^\alpha{}_{\alpha'}n^{\alpha'}
\end{equation}
where $n^{\alpha'}=(1,\cos\varphi,\sin\varphi)$. The Lorentz matrix 
$\Lambda$ determines the transformation to the particle's comoving 
Lorentz frame where the particle is at rest. To adopt these curvilinear 
coordinates to the integration surface $\Sigma_t=\{y\in{\mathbb 
M}_{\,3}:y^0=t\}$ we replace the retarded distance $r$ by the expression
\begin{equation}\label{Ar}
r=\frac{t-s}{k^0}
\end{equation}
where $t$ is the observation time. On rearrangement, the final coordinate 
transformation \linebreak$(y^0,y^1,y^2)\mapsto (t,s,\varphi)$ looks as 
follows:
\begin{equation}\label{Atr}
y^0=t,\qquad y^i=z^i(s)+(t-s)\frac{k^i}{k^0}.
\end{equation}
Differentiation of this coordinate transformation yields the differential 
chart
\begin{eqnarray}
\vec{e}_t&=&\frac{1}{k^0}\left(k^0\vec{e}_0+k^i\vec{e}_i\right)\\
\vec{e}_s&=&\left(v^i-\frac{k^i}{k^0}\right)\vec{e}_i\nonumber\\
\vec{e}_\varphi&=&(t-s)\left(\frac{k^i_\varphi}{k^0}-
\frac{k^i}{(k^0)^2}k^0_\varphi\right)\vec{e}_i\nonumber
\end{eqnarray}
where $k_\varphi^\alpha=\partial k^\alpha/\partial\varphi$ and 
$v^i=\gamma^{-1}u^i$. Their scalar products are the components of 
metric tensor $g$ of Minkowski space ${\mathbb M}_{\,3}$ as it is viewed 
in curvilinear coordinates (\ref{Atr}). To calculate the determinant of 
$g$ it is sufficient to know that
\begin{equation}
g_{tt}=0,\quad g_{t\varphi}=0,\quad g_{ts}=-\frac{\gamma^{-1}}{k^0},\quad
g_{\varphi\varphi}=\frac{(t-s)^2}{(k^0)^2}.
\end{equation}

The surface element is given by $d\sigma_0=\sqrt{-g}dsd\varphi$ where
\begin{equation}
\sqrt{-g}=\gamma^{-1}\frac{t-s}{(k^0)^2}
\end{equation}
is the Jacobian of coordinate transformation (\ref{Atr}). 
Electromagnetic field momentum (\ref{ppem}) takes the form:
\begin{equation}
p_{\rm em}^\beta=\frac{e^2}{2\pi}\int\limits_{-\infty}^tds
\int\limits_0^{2\pi}d\varphi \frac{\gamma^{-1}}{t-s}\left(
u^0k^\beta +u^\beta k^0-k^0k^\beta+(1/2)\eta^{0\beta}
\right).
\end{equation}
The angular integration can be handled via the relations
\begin{equation}
\int\limits_0^{2\pi}d\varphi k^\alpha=2\pi u^\alpha,\qquad
\int\limits_0^{2\pi}d\varphi k^\alpha k^\beta=3\pi u^\alpha u^\beta +
\pi\eta^{\alpha\beta}.
\end{equation}
After trivial calculations we arrive at the logarithmic divergence 
\begin{equation}\label{Apem}
p_{\rm em}^\beta=\frac{e^2}{2}\int\limits_{-\infty}^t 
ds\frac{u^\beta}{t-s}
\end{equation}
as could be expected for the two-dimensional Coulomb potential.

\subsubsection{Coordinate system}\label{coord}
\renewcommand{\theequation}{\Alph{subsubsection}.\arabic{equation}}
\setcounter{equation}{0}

The coordinate transformation (\ref{ct}) is associated with two points, 
$z(t_1)$
and $z(t_2)$, on an accelerated world line $\zeta$ (see figure \ref{K1K2}).
Its differentiation yields differential chart
\begin{eqnarray}
{\vec e}_t&=&{\vec 
e}_0-q\frac{\partial\beta}{\partial t}{\vec e}_{1'}\nonumber\\
{\vec e}_R&=&n^{j'}{\vec 
e}_{j'}-q\frac{\partial\beta}{\partial R}{\vec e}_{1'}\nonumber\\
{\vec e}_{\varphi}&=&Rn^{j'}_\varphi {\vec e}_{j'}
\end{eqnarray}
where ${\mathbf n}=(\cos\varphi,\sin\varphi)$, 
${\mathbf n}_\varphi =(-\sin\varphi,\cos\varphi)$ and ${\vec e}_{j'}={\vec 
e}_i\omega^i{}_{j'}$. The orthogonal matrix
\begin{equation}\label{om}
\omega=\left(
\begin{array}{cc}
n_q^1& -n_q^2\\[0.5ex]
n_q^2& n_q^1
\end{array}
\right)
\end{equation}
where $n_q^i=q^i/q$ rotates space axes until a new $x-$axis be directed 
along two-vector ${\mathbf q}:={\mathbf z}(t_1)-{\mathbf z}(t_2)$ (we 
denote $q:=\sqrt{{\mathbf q}^2}$). In new coordinates, three-vectors $K_a=y 
-z(t_a), a=1,2,$ pointing from points of emanation $z(t_a):=z_a$ to an 
observation point $y\in\Sigma_t$ (see figure \ref{K1K2}) have the 
following forms:
\begin{equation}
K_a^0=t-t_a:=k_a^0,\qquad K_a^i=\omega^i{}_jk_a^j
\end{equation}
where
\begin{eqnarray}\label{kk}
k^1_1&=&-\beta q+R\cos\varphi, \qquad k^2_1=R\sin\varphi\\
k^1_2&=&\alpha q+R\cos\varphi, \qquad k^2_2=R\sin\varphi .\nonumber
\end{eqnarray} 

\begin{figure}[t]
\begin{center}
\epsfclipon
\epsfclipon
\epsfig{file=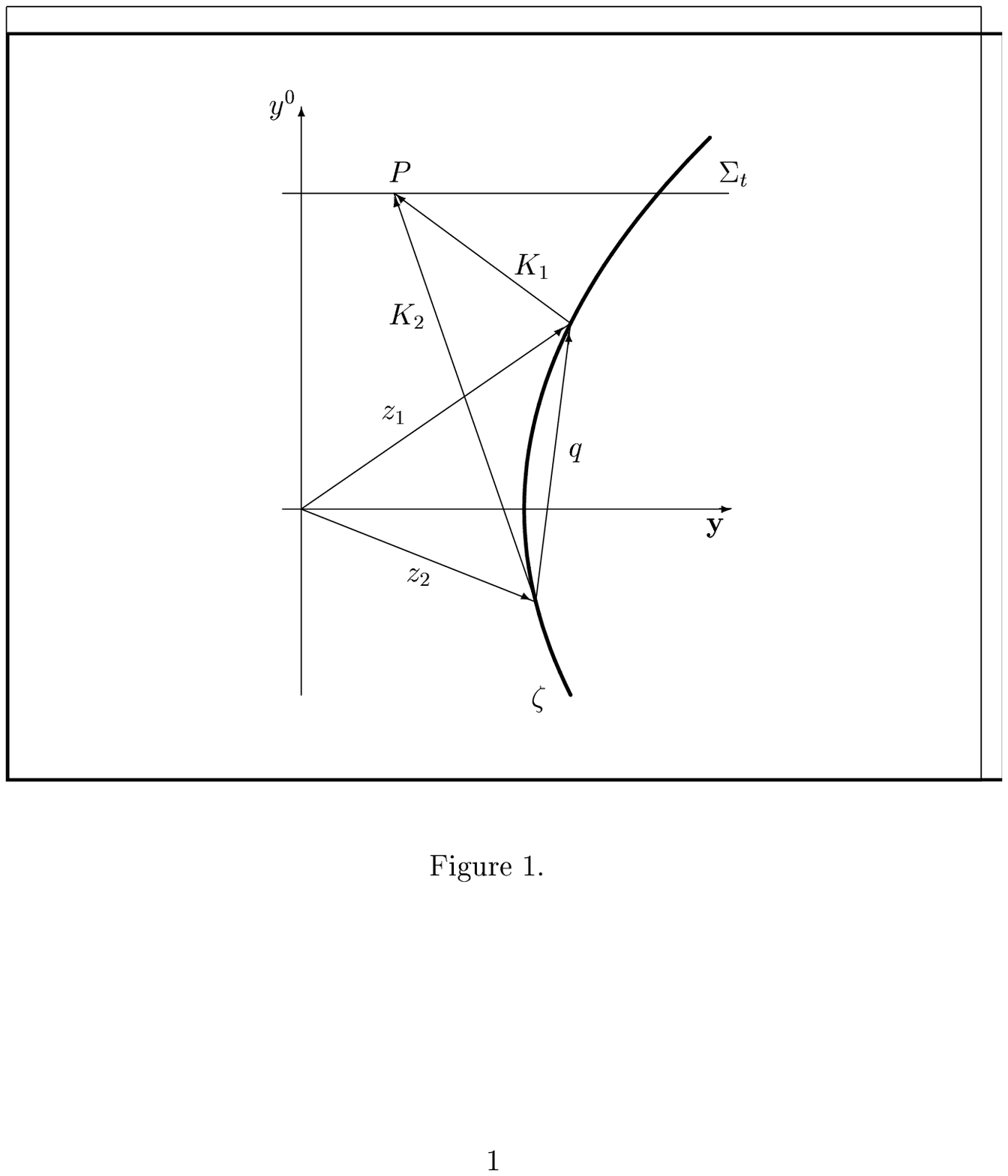,width=6cm}
\end{center}
\caption{\label{K1K2}
\small The separation vector $K_a$ is a vector pointing from point of 
emission $z(t_a):=z_a$ to point of observation $P\in\Sigma_t$ with 
coordinates $(y^0,y^1,y^2)$. The integrand $t^{0\alpha}$ involves also 
one-half of square of three-vector $q:=z_1-z_2=K_2-K_1$ (double Synge's 
function) and its partial derivatives with respect to time variables. 
Space components of $q$ determine the orthogonal matrix (\ref{om}).
}
\end{figure} 

The norms $\|K_a\|=\sqrt{-(K_a\cdot K_a)}$ of the separation vectors 
$K_1$ and $K_2$ pictured in figure \ref{K1K2} contain the angular 
variable:
\begin{eqnarray}\label{KKn}
-(K_1\cdot K_1)&=&(k_1^0)^2-\beta^2{\mathbf q}^2+2\beta qR\cos\varphi 
-R^2,\nonumber\\
-(K_2\cdot K_2)&=&(k_2^0)^2-\alpha^2{\mathbf q}^2-2\alpha qR\cos\varphi 
-R^2.
\end{eqnarray}
Substituting the right-hand side of eq.(\ref{r2}) for $R^2$ in these
expressions and comparing them with Jacobian (\ref{Jc}) leads to 
the important relations:
\begin{equation}\label{KKJ}
\|K_1\|^2=-2\beta J,\qquad \|K_2\|^2=2\alpha J.
\end{equation}
They immediately follow:
\begin{equation}
\frac{J}{\sqrt{-(K_1\cdot K_1)}\sqrt{-(K_2\cdot 
K_2)}}=\frac{1}{2\sqrt{-\beta\alpha}}.
\end{equation}

To concretely compute integrals over $\beta$ we clarify the mathematical 
sense of this parameter. Since eq.(\ref{KKJ}), we parametrize the ratio 
$\|K_1\|/\|K_2\|$ by angle variable $\vartheta$:
\begin{equation}
\sin\vartheta=\sqrt{\frac{-\beta}{\alpha}}.
\end{equation}
Having inserted it in expression (\ref{r2}) we obtain algebraic equation on 
$\sin^2\vartheta$. Its solution is as follows:
\begin{equation}
\cos2\vartheta =\frac{-(k_1^0)^2+{\mathbf q}^2+R^2+\sqrt{D}}{(k_2^0)^2-R^2}
\end{equation}
where $D$ is defined by eq.(\ref{beta}). If $R=k_1^0$ then $\vartheta=0$ 
and parameter $\beta$ vanishes. If $R=0$ we obtain the lower limit of the
integral over $\beta$. We denote it as $\beta_0=-\tan^2\vartheta_0$ where
\begin{equation}\label{sn0}
\sin\vartheta_0=\frac{1}{2k_2^0}\left(
\sqrt{2\Sigma}-\sqrt{2\sigma}
\right)
\end{equation}
and
\begin{equation}\label{Ss}
\Sigma=\frac12\left(k_2^0+k_1^0\right)^2 - \frac12{\mathbf q}^2,\qquad
\sigma=\frac12\left(k_2^0-k_1^0\right)^2 - \frac12{\mathbf q}^2.
\end{equation}

\subsubsection{Integration over angular variable}\label{varphi}
\renewcommand{\theequation}{\Alph{subsubsection}.\arabic{equation}}
\setcounter{equation}{0}
To calculate the energy-momentum (\ref{pnt}) carried by the electromagnetic 
field we should perform the integration over angle first. When facing 
this problem it is convenient to mark out $\varphi$-dependent terms 
in expressions under the integral sign. In the Maxwell 
energy-momentum tensor density we distinguish the second-order 
differential operator ${\hat T}_a$ with $\varphi$-dependent 
coefficients (see eq.(\ref{Ta})). Having integrated this operator 
over $\varphi$ we obtain the operator ${\hat{\cal T}}_a$ which can be 
decomposed into a combination of partial derivatives in time variables 
${\hat \Pi}_a$ given by eq.(\ref{Pa}) and tail $\pi_a$ of the type 
in eq.(\ref{pa}). 

This Appendix is concerned with the computation of the tails. Equipped 
with them we express the integrand as a combination of partial derivatives 
in $t_1$ and $t_2$. It allows us to integrate the electromagnetic field's 
stress-energy tensor over the time variables, as well as over $\beta$.

To implement this strategy we must first integrate the coefficients 
(\ref{ABC}) over the angle variable. We start with the simplest one
\begin{equation}\label{BD}
{\cal D}^a=\frac{1}{2\pi}\int_0^{2\pi}d\varphi\frac{a}{r_1r_2}
\end{equation}
where $\varphi$-dependent numerator $a$ is equal to the Jacobian (\ref{Jc}) 
or to $1$. Our task is to rewrite the integrand as a sum of term with 
denominator $r_1$ and term with denominator $r_2$. To do it we introduce a 
new layer of mathematical formalism and develop convenient technique.

We introduce null vector $n=(1,\cos\varphi,\sin\varphi)$ which belongs to 
the vector space, say $V$, such that ${\mathbf i}_0, {\mathbf i}_1$, 
and ${\mathbf i}_2$ is its linear basis. We shall use 
$\eta_{\alpha\beta}={\rm diag}(-1,1,1)$ and its inverse 
$\eta^{\alpha\beta}={\rm diag}(-1,1,1)$ to lower and raise indices, 
respectively. We introduce the pairing
\begin{eqnarray}
(\cdot)&:&V\times V\to \mathbb R\\
&& ({\sf a}\cdot{\sf b})\mapsto\eta_{\alpha\beta}{\sf a}^\alpha{\sf b}^\beta
\end{eqnarray}
which will be called the ``scalar product''.

We express the $\varphi$-dependent constructions 
\begin{equation}\label{rc}
r_a=-(K_a\cdot v_a), \qquad c_a=\gamma_a^{-2}+(K_a\cdot {\dot v}_a)
\end{equation}
as the scalar products $-({\sf r_a}\cdot n)$ and $({\sf c_a}\cdot 
n)$, respectively. We shall use {\sf sans-serif} letters for the 
components of timelike three-vectors ${\sf r}_a\in V$ and ${\sf 
c}_a\in V$ \begin{eqnarray}\label{rc0}
{\sf r}_1^0&=&k_1^0+\beta ({\mathbf v}_1{\mathbf q}),\qquad
{\sf r}_2^0=k_2^0-\alpha ({\mathbf v}_2{\mathbf q}),\qquad
{\sf r}_a^j=Rv_a{}^i\omega_i{}^j;\\
{\sf c}_1^0&=&-\gamma_1^{-2}+\beta ({\mathbf\dot v}_1{\mathbf q}),\qquad
{\sf c}_2^0=-\gamma_2^{-2}-\alpha ({\mathbf\dot v}_2{\mathbf q}),\qquad
{\sf c}_a^j=R{\dot v}_a{}^i\omega_i{}^j.\nonumber
\end{eqnarray}
The Jacobian (\ref{Jc}) becomes the scalar product $({\sf J}\cdot n)={\sf 
J}_0+{\sf J}_1\cos\varphi +{\sf J}_2\sin\varphi$, where 
\begin{equation}\label{b}
{\sf J}_0=\beta {\mathbf q}^2 
+(1/2)\left[(k_2^0)^2-(k_1^0)^2-{\mathbf q}^2\right],\qquad {\sf J}_1=-qR,
\qquad {\sf J}_2=0.
\end{equation}

We introduce the dual space of one-forms, say $W$, with basis 
${\hat\omega}^0,{\hat\omega}^1$ and ${\hat\omega}^2$ such that 
${\hat\omega}^\mu({\mathbf i}_\nu)=\delta_\nu^\mu$ where ${\mathbf i}_0, 
{\mathbf i}_1, {\mathbf i}_2$ constitute the basis of $V$. The wedge 
product ${\hat L}={\hat a}\wedge{\hat b}$ of two one forms ${\hat a}$ and 
${\hat b}$ constitutes two-form
\begin{equation}
{\hat 
L}=\left(a_0b_1-a_1b_0\right){\hat\omega}^0\wedge{\hat\omega}^1+
\left(a_0b_2-a_2b_0\right){\hat\omega}^0\wedge{\hat\omega}^2+
\left(a_1b_2-a_2b_1\right){\hat\omega}^1\wedge{\hat\omega}^2.
\end{equation}
We introduce dual three-vector ${\sf L}=^*\!\!{\hat L}$ with components
\begin{eqnarray}\label{hatL}
{\sf L}^\alpha&=&\frac{1}{2!}\varepsilon^{\alpha\beta\gamma}L_{\beta\gamma}
\\
&=&\varepsilon^{\alpha\beta\gamma}a_\beta b_\gamma.\nonumber
\end{eqnarray}
$\varepsilon^{\alpha\beta\gamma}$ denotes the Ricci symbol in three 
dimensions:
\begin{equation}\label{eps}
\varepsilon^{\alpha\beta\gamma}=\left\{
\begin{array}{ccccccccc}
1 &{\rm when}& \alpha\beta\gamma& {\rm is}& {\rm an}& {\rm even}& 
{\rm permutation}& {\rm of}& 0,1,2\\
-1& {\rm when}& \alpha\beta\gamma& {\rm is}& {\rm an}& {\rm odd}& 
{\rm permutation}& {\rm of}& 0,1,2\\
0& {\rm otherwise}.&&&&&&&
\end{array}
\right.
\end{equation}

We raise indices in eq.(\ref{hatL}) and define the vector product of two 
vectors, ${\sf a}$ and ${\sf b}$:
\begin{equation}\label{sfL}
{\sf L}^\alpha=\varepsilon^\alpha{}_{\mu\nu}{\sf a}^\mu{\sf b}^\nu.
\end{equation}
Tensor
\begin{equation}\label{ep}
\varepsilon^\alpha{}_{\mu\nu}=\varepsilon^{\alpha\beta\gamma}\eta_{\beta\mu}
\eta_{\gamma\nu}
\end{equation}
has the components
\begin{equation}\label{epp}
\varepsilon^0{}_{\mu\nu}=\varepsilon^{0\mu\nu},\qquad
\varepsilon^1{}_{\mu\nu}=-\varepsilon^{1\mu\nu},\qquad
\varepsilon^2{}_{\mu\nu}=-\varepsilon^{2\mu\nu}.
\end{equation}
It is interesting that tensor
\begin{equation}
\varepsilon_{\lambda\mu\nu}=\varepsilon^{\alpha\beta\gamma}
\eta_{\alpha\lambda}\eta_{\beta\mu}\eta_{\gamma\nu}
\end{equation}
is equal to $\varepsilon^{\lambda\mu\nu}$ taken with opposite sign.

Now we calculate the double vector product
\begin{equation}
[{\sf A}[{\sf B}{\sf C}]]^\alpha=
\varepsilon^\alpha{}_{\beta\gamma}{\sf A}^\beta\varepsilon^\gamma{}_{\mu\nu}
{\sf B}^\mu{\sf C}^\nu.
\end{equation}
Since
\begin{equation}
\varepsilon^\alpha{}_{\beta\gamma}\varepsilon^\gamma{}_{\mu\nu}=
-\delta^\alpha{}_\mu\eta_{\beta\nu}+\delta^\alpha{}_\nu\eta_{\beta\mu}
\end{equation}
we arrive to the unusual rule
\begin{equation}\label{vpr}
[{\sf A}[{\sf B}{\sf C}]]=-{\sf B}({\sf A}\cdot{\sf C})+
{\sf C}({\sf A}\cdot{\sf B})
\end{equation}
instead of the well-known law acting in space with Euclidean metric.

To simplify the denominator $r_1r_2$ in the integrand of eq.(\ref{BD}) as 
much as possible we rewrite $2\pi$-periodic functions (\ref{rc}) as 
follows:
\begin{equation}\label{Bra}
r_a=-{\sf r}_{a,0}-\rho_a\sin(\varphi+\phi_a),\qquad
\rho_a=\sqrt{{\sf r}_{a,1}^2+{\sf r}_{a,2}^2}.
\end{equation}
(We recall that $r_a$ is the scalar products $({\sf r}_a\cdot n)$ taken 
with opposite sign, components ${\sf r}_a^\mu$ are given 
by eqs.(\ref{rc0}).) Shift in argument of $a$-th function is determined 
by the relations
\begin{equation}
{\sf r}_{a,1}=\rho_a\sin\phi_a,\qquad 
{\sf r}_{a,2}=\rho_a\cos\phi_a.
\end{equation}
After some algebra we rewrite the integrand of eq.(\ref{BD}) as the 
following sum
\begin{equation}\label{BI}
\frac{a}{r_1r_2}=\frac{A_{12}^a-C_{12}^a\rho_1\cos(\varphi+\phi_1)}{r_1}+
\frac{A_{21}^a+C_{12}^a\rho_2\cos(\varphi+\phi_2)}{r_2}.
\end{equation}
Coefficients $A_{12}^a, A_{21}^a$ and $C_{12}^a$ satisfy the vector equation
\begin{equation}\label{be}
- A_{12}^a{\sf r}_2 - A_{21}^a{\sf r}_1 + C_{12}^a{\sf L}_{12}={\sf a}
\end{equation}
where by ${\sf r}_1$ and ${\sf r}_2$ we mean three-vectors with components 
in eq.(\ref{rc0}) and ${\sf L}_{12}=[{\sf r}_1{\sf r}_2]$.

To solve equation (\ref{be}) we postmultiply it on the vector product 
$[{\sf r}_1{\sf L}_{12}]$, then on the vector product
$[{\sf r}_2{\sf L}_{21}]$ and, finally, on ${\sf L}_{12}$. After some 
algebra we obtain
\begin{equation}\label{AAC}
A_{12}^a=\frac{\left([{\sf a}{\sf r}_1]\cdot{\sf L}_{12}\right)}{D_{12}},
\qquad
A_{21}^a=\frac{\left([{\sf a}{\sf r}_2]\cdot{\sf L}_{21}\right)}{D_{21}},
\qquad
C_{12}^a=\frac{\left({\sf a}\cdot{\sf L}_{12}\right)}{D_{12}}
\end{equation}
where the denominator $D_{12}=({\sf L}_{12}\cdot{\sf L}_{12})$ is symmetric 
in its indices.

Substituting eq.(\ref{BI}) into eq.(\ref{BD}) and using the identities
\begin{eqnarray}\label{B1}
\frac{1}{2\pi}\int_0^{2\pi}
\frac{d\varphi}{{\sf r}_a^0-\rho_a\sin(\varphi+\phi_a)}&=&\frac{1}{\|{\sf 
r}_a\|}\\
\frac{1}{2\pi}\int_0^{2\pi}d\varphi
\frac{\cos(\varphi+\phi_a)}{{\sf r}_a^0-\rho_a\sin(\varphi+\phi_a)}&=&0
\nonumber
\end{eqnarray}
yields
\begin{equation}\label{Dfin}
{\cal D}^a=\frac{A_{12}^a}{\|{\sf r}_1\|}+\frac{A_{21}^a}{\|{\sf r}_2\|}
\end{equation}
after integration over $\varphi$.

Now we turn to the calculation of the coefficient
\begin{equation}\label{B}
{\cal B}^a=\frac{1}{2\pi}\int_0^{2\pi}d\varphi\frac{ac_2}{r_1(r_2)^2}.
\end{equation}
Equipped with the relations in eq.(\ref{BI}) we rewrite the integrand as a 
sum of terms which are proportional to the $1/r_1$, $1/r_2$, and 
$1/(r_2)^2$, respectively. Using the identities
\begin{eqnarray}\label{B2}
\frac{1}{2\pi}\int_0^{2\pi}
\frac{d\varphi}{\left[{\sf 
r}_a^0-\rho_a\sin(\varphi+\phi_a)\right]^2}&=&\frac{{\sf r}_a^0}{\|{\sf 
r}_a\|^3}\\
\frac{1}{2\pi}\int_0^{2\pi}d\varphi
\frac{\cos(\varphi+\phi_a)}{\left[{\sf r}_a^0-\rho_a\sin(\varphi+
\phi_a)\right]^2}&=&0
\nonumber
\end{eqnarray}
and taking into account the relations in eq.(\ref{B1}) gives
\begin{eqnarray}\label{Bfin}
{\cal B}^a&=&-\frac{1}{\|{\sf r}_2\|^3}
\frac{({\sf a}\cdot{\sf r}_2)({\sf c}_2\cdot{\sf r}_2)}{D_{21}}
({\sf r}_2\cdot{\sf r}_1)
+\frac{1}{\|{\sf r}_2\|}\left[A_{12}^{c_2}A_{21}^a+
A_{12}^aA_{21}^{c_2}-\frac{({\sf a}\cdot{\sf c}_2)({\sf r}_1\cdot{\sf r}_2)}
{D_{21}}\right]\nonumber\\
&+&\frac{1}{\|{\sf r}_1\|}\left[2A_{12}^aA_{12}^{c_2}-\frac{([{\sf 
a}{\sf r}_1]\cdot[{\sf c}_2{\sf r}_1])}
{D_{12}}\right].
\end{eqnarray}
The resulting expression for the term
\begin{equation}\label{BC}
{\cal C}^a=\frac{1}{2\pi}\int_0^{2\pi}d\varphi\frac{ac_1}{(r_1)^2r_2}
\end{equation}
can be obtained by interchanging indices 1 and 2 in the right-hand side of 
eq.(\ref{Bfin}).

After a routine computation based on the repeated usage of relation 
(\ref{BI}) we find the last term
\begin{eqnarray}\label{BA}
{\cal 
A}^a&=&\frac{1}{2\pi}\int_0^{2\pi}d\varphi\frac{ac_1c_2}{(r_1)^2(r_2)^2}\\
&=&\frac{B_{12}}{\|{\sf r}_1\|}+\frac{B_{21}}{\|{\sf r}_2\|}+
J_1\frac{({\sf a}\cdot{\sf r}_1)}{\|{\sf r}_1\|^3}+
J_2\frac{({\sf a}\cdot{\sf r}_2)}{\|{\sf r}_2\|^3}\nonumber
\end{eqnarray}
where
\begin{eqnarray}
J_1&=&2A_{12}^{c_1}A_{12}^{c_2}-
\frac{([{\sf c}_1{\sf r}_1]\cdot[{\sf c}_2{\sf r}_1])}{D_{12}}\\
B_{12}&=&3A_{12}^aA_{12}^{c_1}A_{21}^{c_2}+
3A_{12}^aA_{12}^{c_2}A_{21}^{c_1}+
2A_{12}^{c_1}A_{12}^{c_2}A_{21}^a+
A_{12}^a\left\{
\frac{([{\sf c}_1{\sf r}_1]\cdot[{\sf c}_2{\sf r}_2])}{D_{12}}+
\frac{([{\sf c}_1{\sf r}_2]\cdot[{\sf c}_2{\sf r}_1])}{D_{12}}
\right\}\nonumber\\
&-&A_{21}^{c_1}\frac{([{\sf c}_2{\sf r}_1]\cdot[{\sf a}{\sf r}_1])}{D_{12}}
-A_{21}^{c_2}\frac{([{\sf c}_1{\sf r}_1]\cdot[{\sf a}{\sf r}_1])}{D_{12}}
+A_{12}^{c_1}\frac{([{\sf c}_2{\sf r}_1]\cdot[{\sf a}{\sf r}_2])}{D_{12}}
+A_{12}^{c_2}\frac{([{\sf c}_1{\sf r}_1]\cdot[{\sf a}{\sf r}_2])}{D_{12}}
\nonumber
\end{eqnarray}
and the others, $B_{21}$ and $J_2$, can be obtained by interchanging 
indices $1$ and $2$.

We now turn to the differentiation of coefficient (\ref{Dfin}) 
with respect to time variables $t_1$ and $t_2$. We will use Latin indices 
$a$ and $b$ which run from $1$ to $2$ ($a\ne b$). We introduce new 
denotations $\kappa_1=\alpha$ and $\kappa_2=\beta$ for 
time-independent variables $\beta$ and $\alpha=1-\beta$. 
Differentiation of ${\cal D}^a$ is based on the relations obtained 
via differentiation of zeroth components (\ref{rc0}), (\ref{b}) and 
square of radius (\ref{r2}):
\begin{eqnarray}
\frac{\partial {\sf r}_a^0}{\partial t_a}&=&{\sf c}_a^0-\kappa_a{\mathbf 
v}_a^2,\qquad
\frac{\partial {\sf r}_a^0}{\partial t_b}=-\kappa_b({\mathbf 
v}_a{\mathbf v}_b)\nonumber\\
\frac{\partial {\sf J}^0}{\partial t_a}&=&-(-1)^a{\sf r}_a^0-\kappa_a
({\mathbf v}_a{\mathbf q}),\qquad
\frac{\partial R^2}{\partial t_a}=-2\kappa_a{\sf r}_a^0.\nonumber
\end{eqnarray}
They immediately give
\begin{eqnarray}\label{dra}
\frac{\partial ({\sf r}_a\cdot{\sf r}_a)}{\partial t_a}&=&
2({\sf r}_a\cdot{\sf c}_a),\qquad
\frac{\partial ({\sf r}_a\cdot{\sf r}_a)}{\partial t_b}=
\frac{2\kappa_b}{R^2}
\left[{\sf r}_a^0({\sf r}_a\cdot{\sf r}_b)-
{\sf r}_b^0({\sf r}_a\cdot{\sf r}_a)\right]\\
\frac{\partial ({\sf r}_b\cdot{\sf r}_a)}{\partial t_b}&=&
({\sf r}_a\cdot{\sf c}_b)+
\frac{\kappa_b}{R^2}
\left[{\sf r}_a^0({\sf r}_b\cdot{\sf r}_b)-
{\sf r}_b^0({\sf r}_a\cdot{\sf r}_b)\right]
\nonumber
\end{eqnarray}
and eventually gives 
\begin{equation}\label{d1ra}
\frac{\partial}{\partial t_a}\left(\frac{1}{\|{\sf r}_a\|}
\right)=\frac{({\sf c}_a\cdot{\sf r}_a)}{\|{\sf r}_a\|^3},\qquad
\frac{\partial}{\partial t_b}\left(
\frac{1}{\|{\sf r}_a\|}
\right)=\frac{\kappa_b}{\|{\sf r}_a\|^3}
\frac{{\sf r}_a^0({\sf r}_a\cdot{\sf r}_b)
-{\sf r}_b^0({\sf r}_a\cdot{\sf r}_a)}{R^2}.
\end{equation}
After some algebra we also obtain
\begin{eqnarray}
\frac{\partial ({\sf J}\cdot{\sf r}_a)}{\partial t_a}&=&(-1)^a
({\sf r}_a\cdot{\sf r}_a)+({\sf J}\cdot{\sf c}_a)+
\frac{\kappa_a}{R^2}\left[{\sf J}^0({\sf r}_a\cdot{\sf r}_a)-
{\sf r}_a^0({\sf J}\cdot{\sf r}_a)\right]\\
\frac{\partial ({\sf J}\cdot{\sf r}_a)}{\partial t_b}&=&(-1)^b
({\sf r}_b\cdot{\sf r}_a)+
\frac{\kappa_b}{R^2}\left[{\sf J}^0({\sf r}_a\cdot{\sf r}_b)+
{\sf r}_a^0({\sf J}\cdot{\sf r}_b)-2{\sf r}_b^0({\sf J}\cdot{\sf 
r}_a)\right].\nonumber
\end{eqnarray}
Usage of these relations allows us to calculate the following derivatives
\begin{eqnarray}
\frac{\partial A_{ab}^J}{\partial t_a}&=&A_{ab}^JA_{ba}^{c_a}+
A_{ab}^{c_a}A_{ba}^J+\frac{([{\sf J}{\sf r}_a]\cdot[{\sf c}_a{\sf 
r}_b])}{D_{ab}}\\
\frac{\partial A_{ab}^J}{\partial t_b}&=&a-b+ 
2A_{ab}^{c_b}A_{ab}^J-\frac{([{\sf J}{\sf r}_a]\cdot[{\sf c}_b{\sf 
r}_a])}{D_{ab}}-\frac{\kappa_b}{R^2}
\left({\sf J}^0+{\sf r}_2^0A_{12}^J+{\sf r}_1^0A_{21}^J
\right)\nonumber
\end{eqnarray}
where latin indices $a$ and $b$ run from $1$ to $2, a\ne b$.

Having differentiated eq.(\ref{Dfin}), after a straightforward  
calculations we derive the following relations
\begin{equation}\label{DJ}
\frac{\partial{\cal D}^J}{\partial t_1}={\cal C}^J- 
\frac{\partial}{\partial\beta}\left(\frac{\alpha}{\|{\sf r}_2\|}\right),
\qquad
\frac{\partial{\cal D}^J}{\partial t_2}={\cal B}^J 
-\frac{\partial}{\partial\beta}\left(\frac{\beta}{\|{\sf r}_1\|}\right).
\end{equation}

Further we find out the expression $\partial{\cal C}^J/\partial t_2$ and 
prove the identity
\begin{equation}\label{piJ}
{\cal A}^J-\frac{\partial{\cal C}^J}{\partial t_2}=
\frac{\partial}{\partial t_1}\left(
{\cal B}^J-\frac{\partial{\cal D}^J}{\partial t_2}
\right)\quad {\rm i.e.} \quad
\pi^J=0.
\end{equation}
(One can derive $\partial{\cal B}^J/\partial t_1$, subtract it from ${\cal 
A}^J$ and compare the result with $\partial/\partial t_2({\cal 
C}^J-\partial{\cal D}^J/\partial t_1)$.) 

Similarly we derive an analogous equality where Jacobian ${\sf J}$ 
with components (\ref{b}) is replaced by unit three-vector ${\sf 
o}=(-1,0,0)$. Having substituted ${\sf o}$ for ${\sf a}$ in the 
expressions (\ref{Dfin}), (\ref{Bfin}) and (\ref{BA}) we obtain the terms 
${\cal D}^0$, ${\cal B}^0$ and ${\cal A}^0$, respectively. The remaining 
term, ${\cal C}^0$, can be obtained from ${\cal B}^0$ via reciprocity. 
The derivatives of coefficients $A_{ab}^0$ are as follows:
\begin{eqnarray}
\frac{\partial A_{ab}^0}{\partial t_a}&=&A_{ab}^0A_{ba}^{c_a}+
A_{ab}^{c_a}A_{ba}^0+\frac{([{\sf o}{\sf r}_a]\cdot[{\sf c}_a{\sf 
r}_b])}{D_{ab}}\\
\frac{\partial A_{ab}^0}{\partial t_b}&=& 
2A_{ab}^{c_b}A_{ab}^0-\frac{([{\sf o}{\sf r}_a]\cdot[{\sf c}_b{\sf 
r}_a])}{D_{ab}}-\frac{\kappa_b}{R^2}
\left({\sf o}^0+{\sf r}_2^0A_{12}^0+{\sf r}_1^0A_{21}^0
\right).\nonumber
\end{eqnarray}
Substituting these into equality
\begin{equation}
\frac{\partial{\cal D}^0}{\partial t_a}=\frac{\partial}{\partial t_a}
\left(
\frac{A_{12}^0}{\|{\sf r}_1\|}+\frac{A_{21}^0}{\|{\sf r}_2\|}
\right)
\end{equation}
and using the identities (\ref{d1ra}) yields
\begin{equation}\label{D0}
\frac{\partial{\cal D}^0}{\partial t_1}={\cal C}^0- 
\frac{\alpha{\mathbf v}_2^2}{\|{\sf r}_2\|^3},
\qquad
\frac{\partial{\cal D}^0}{\partial t_2}={\cal B}^0 
-\frac{\beta{\mathbf v}_1^2}{\|{\sf r}_1\|^3}.
\end{equation}
And, finally, we calculate the partial derivative $\partial{\cal 
C}^0/\partial t_2$, subtract it from ${\cal A}^0$ and compare the result 
with  $\partial/\partial t_1({\cal B}^0-\partial{\cal D}^0/\partial t_2)$.
We obtain
\begin{equation}\label{pi0}
{\cal A}^0-\frac{\partial{\cal B}^0}{\partial t_1}-
\frac{\partial{\cal C}^0}{\partial t_2}+
\frac{\partial^2 {\cal D}^0}{\partial t_1\partial t_2}=0
\quad {\rm i.e.} \quad
\pi^0=0.
\end{equation}

Now, we calculate the tail
\begin{equation}\label{pi1al}
\pi_a^\alpha={\cal A}_a^\alpha-\frac{\partial{\cal B}_a^\alpha}{\partial 
t_1}-
\frac{\partial{\cal C}_a^\alpha}{\partial t_2}+
\frac{\partial^2 {\cal D}_a^\alpha}{\partial t_1\partial t_2}
\end{equation}
where
\begin{eqnarray}\label{ABCK}
{\cal D}_a^\alpha&=&\frac{1}{2\pi}\int_0^{2\pi}d\varphi 
\frac{K_a^\alpha}{r_1r_2},\qquad
{\cal B}_a^\alpha=\frac{1}{2\pi}\int_0^{2\pi}d\varphi 
\frac{K_a^\alpha c_2}{r_1(r_2)^2}\\
{\cal C}_a^\alpha&=&\frac{1}{2\pi}\int_0^{2\pi}d\varphi\frac{K_a^\alpha 
c_1}{(r_1)^2r_2},
\qquad
{\cal A}_a^\alpha=\frac{1}{2\pi}\int_0^{2\pi}d\varphi 
\frac{K_a^\alpha c_1c_2}{(r_1)^2(r_2)^2}.
\nonumber
\end{eqnarray}
The zeroth component, $K_a^0=k_a^0$, does not depend on $\varphi$. 
Inserting relations ${\cal D}_a^0=k_a^0{\cal D}^0$, ${\cal 
B}_a^0=k_a^0{\cal B}^0$, ${\cal C}_a^0=k_a^0{\cal C}^0$, and ${\cal 
A}_a^0=k_a^0{\cal A}^0$ into eq.(\ref{pi1al}) and taking into account 
identity (\ref{pi0}) yields
\begin{equation}\label{piK0}
\pi_1^0={\cal B}^0-\frac{\partial {\cal D}^0}{\partial t_2},\quad
\pi_2^0={\cal C}^0-\frac{\partial {\cal D}^0}{\partial t_1}.
\end{equation}

Space components, $K_a^i$, depend on $\varphi$. They can be expressed as 
the scalar product $({\sf K}_a^i\cdot n)$ where components of three-vectors
${\sf K}_a^i\in V$ are as follows:
\begin{equation}
{\sf K}_{1,0}^i=-\beta q^i,\quad {\sf K}_{2,0}^i=\alpha q^i,\quad
{\sf K}_{a,1}^i=R\omega^i{}_1,\quad {\sf K}_{a,2}^i=R\omega^i{}_2,
\end{equation}
where $\omega^i{}_j$ are components of the orthogonal matrix (\ref{om}).
Having substituted ${\sf K}_a^i$ for ${\sf a}$ in expressions (\ref{Dfin}), 
(\ref{Bfin}), and (\ref{BA}) we obtain the terms ${\cal D}_a^i$, ${\cal 
B}_a^i$ and ${\cal A}_a^i$, respectively. The last term, ${\cal C}_a^i$, 
can be obtained from ${\cal B}_a^i$ via reciprocity.
To differentiate them we need the equalities
\begin{eqnarray}
\frac{\partial ({\sf K}_a^i\cdot {\sf r}_a)}{\partial t_a}&=&
({\sf K}_a^i\cdot {\sf c}_a) -v_a^i{\sf r}_a^0 - 
\frac{\kappa_a}{R^2}\left[
{\sf K}_{a,0}^i({\sf r}_a\cdot {\sf r}_a)+
{\sf r}_a^0({\sf K}_a^i\cdot {\sf r}_a)\right]\nonumber\\
\frac{\partial ({\sf K}_a^i\cdot {\sf r}_b)}{\partial t_a}&=&
-v_a^i{\sf r}_b^0+
\frac{\kappa_a}{R^2}\left[{\sf r}_b^0({\sf K}_a^i\cdot {\sf r}_a)-
{\sf K}_{a,0}^i({\sf r}_b\cdot {\sf r}_a)-
2{\sf r}_a^0({\sf K}_a^i\cdot {\sf r}_b)\right]\nonumber\\
\frac{\partial ({\sf K}_a^i\cdot {\sf r}_b)}{\partial t_b}&=&
({\sf K}_a^i\cdot {\sf c}_b) - \frac{\kappa_b}{R^2}\left[
{\sf K}_{a,0}^i({\sf r}_b\cdot {\sf r}_b)+
{\sf r}_b^0({\sf K}_a^i\cdot {\sf r}_b)\right]\nonumber\\
\frac{\partial ({\sf K}_a^i\cdot {\sf r}_a)}{\partial t_b}&=&
\frac{\kappa_b}{R^2}\left[{\sf r}_a^0({\sf K}_a^i\cdot {\sf r}_b)-
{\sf K}_{a,0}^i({\sf r}_b\cdot {\sf r}_a)-
2{\sf r}_b^0({\sf K}_a^i\cdot {\sf r}_a)\right]\nonumber
\end{eqnarray}
in addition to eqs.(\ref{dra}) and (\ref{d1ra}). 

The derivation of equalities
\begin{eqnarray}\label{DBC1i}
{\cal C}_1^i-\frac{\partial {\cal D}_1^i}{\partial t_1}&=&v_1^i{\cal D}^0+
\frac{\alpha}{\|{\sf r}_2\|^3}\left(-\beta q^i{\mathbf v}_2^2+{\sf 
r}_2^0v_2^i\right)\\
{\cal B}_1^i-\frac{\partial {\cal D}_1^i}{\partial t_2}&=&
\frac{\beta}{\|{\sf r}_1\|^3}\left(-\beta q^i{\mathbf v}_1^2+{\sf 
r}_1^0v_1^i\right)\nonumber\\
{\cal C}_2^i-\frac{\partial {\cal D}_2^i}{\partial t_1}&=&
\frac{\alpha}{\|{\sf r}_2\|^3}\left(\alpha q^i{\mathbf v}_2^2+{\sf 
r}_2^0v_2^i\right)\nonumber\\
{\cal B}_2^i-\frac{\partial {\cal D}_2^i}{\partial t_2}&=&v_2^i{\cal D}^0+
\frac{\beta}{\|{\sf r}_1\|^3}\left(\alpha q^i{\mathbf v}_1^2+{\sf 
r}_1^0v_1^i\right)\nonumber
\end{eqnarray}
is virtually identical to that presented above, and we shall not bother 
with the details. Finally, after a straightforward (but fairly lengthy) 
calculations we derive the following relations
\begin{equation}\label{pii}
\pi_1^i=v_1^i\left({\cal B}^0-\frac{\partial{\cal D}^0}{\partial 
t_2}\right),\qquad
\pi_2^i=v_2^i\left({\cal C}^0-\frac{\partial{\cal D}^0}{\partial 
t_1}\right)
\end{equation}
which generalize eqs.(\ref{piK0}).

In analogous way one can derive the equalities
\begin{equation}\label{piJi}
\pi_1^{iJ}=v_1^i\left({\cal B}^J-\frac{\partial{\cal D}^J}{\partial 
t_2}\right),\qquad
\pi_2^{iJ}=v_2^i\left({\cal C}^J-\frac{\partial{\cal D}^J}{\partial 
t_1}\right)
\end{equation}
which arise in $\varphi-$integration of angular momentum carried by the 
electromagnetic field. 

We will need also the tail
\begin{equation}\label{pialb}
\pi_{12}^{\alpha\beta}=
\frac{\partial^2 {\cal D}_{12}^{\alpha\beta}}{\partial t_1\partial t_2}-
\frac{\partial{\cal B}_{12}^{\alpha\beta}}{\partial t_1}-
\frac{\partial{\cal C}_{12}^{\alpha\beta}}{\partial t_2}+{\cal 
A}_{12}^{\alpha\beta}
\end{equation}
where
\begin{eqnarray}\label{ABCKK}
{\cal 
D}_{12}^{\alpha\beta}&=&\frac{\displaystyle 1}{\displaystyle 
2\pi}\int_0^{2\pi}d\varphi \frac{K_1^\alpha K_2^\beta}{r_1r_2},\qquad
{\cal B}_{12}^{\alpha\beta}=\frac{1}{2\pi}\int_0^{2\pi}d\varphi 
\frac{K_1^\alpha K_2^\beta c_2}{r_1(r_2)^2}\\
{\cal C}_{12}^{\alpha\beta}&=&\frac{1}{2\pi}\int_0^{2\pi}d\varphi
\frac{K_1^\alpha K_2^\beta c_1}{(r_1)^2r_2},
\qquad
{\cal A}_{12}^{\alpha\beta}=\frac{1}{2\pi}\int_0^{2\pi}d\varphi 
\frac{K_1^\alpha K_2^\beta c_1c_2}{(r_1)^2(r_2)^2}.
\nonumber
\end{eqnarray}
It can be obtained by means of covariant generalization of previous 
relations. Setting $\alpha=0$ and $\beta=0$ and taking into account 
eq.(\ref{pi0})
we obtain
\begin{eqnarray}\label{pi00}
\pi_{12}^{00}&=&k_1^0\left({\cal C}^0-\frac{\partial {\cal D}^0}{\partial 
t_1}\right)+
k_2^0\left({\cal B}^0-\frac{\partial {\cal D}^0}{\partial t_2}\right)
+{\cal D}^0\nonumber\\
&=&{\cal C}_1^0-\frac{\partial {\cal D}_1^0}{\partial t_1}+
{\cal B}_2^0-\frac{\partial {\cal D}_2^0}{\partial t_2}-{\cal D}^0
\end{eqnarray}
where relations ${\cal D}_a^0=k_a^0{\cal D}^0$, 
${\cal B}_a^0=k_a^0{\cal B}^0$, and ${\cal C}_a^0=k_a^0{\cal 
C}^0$ are taken into account. If $\alpha=i$ and $\beta=0$ we have
\begin{eqnarray}\label{pii0}
\pi_{12}^{i0}&=&{\cal C}_1^i-\frac{\partial {\cal D}_1^i}{\partial 
t_1}+
k_2^0v_1^i\left({\cal B}^0-\frac{\partial {\cal D}^0}{\partial t_2}\right)
\nonumber\\
&=&{\cal C}_1^i-\frac{\partial {\cal D}_1^i}{\partial t_1}+v_1^i\left(
{\cal B}_2^0-\frac{\partial {\cal D}_2^0}{\partial t_2}\right)
-v_1^i{\cal D}^0.
\end{eqnarray}
If $\alpha=0$ and $\beta=j$ we arrive at
\begin{eqnarray}\label{pi0j}
\pi_{12}^{0j}&=&k_1^0v_2^j\left({\cal C}^0-\frac{\partial {\cal 
D}^0}{\partial t_1}\right)+
{\cal B}_2^j-\frac{\partial {\cal D}_2^j}{\partial t_2}
\nonumber\\
&=&v_2^j\left({\cal C}_1^0-\frac{\partial {\cal D}_1^0}{\partial 
t_1}\right)+
{\cal B}_2^j-\frac{\partial {\cal D}_2^j}{\partial t_2}
-v_2^j{\cal D}^0.
\end{eqnarray}
An obvious generalization of expressions (\ref{pi00})-(\ref{pi0j}) is
\begin{equation}\label{pi_alb}
\pi_{12}^{\alpha\beta}=
v_1^\alpha\left({\cal B}_2^\beta-\frac{\partial {\cal D}_2^\beta}{\partial 
t_2}\right)
+v_2^\beta\left({\cal C}_1^\alpha-\frac{\partial {\cal 
D}_1^\alpha}{\partial t_1}\right)-v_1^\alpha v_2^\beta {\cal D}^0.
\end{equation}

\subsubsection{Calculation of integrals where $t_1\to t$}\label{pink}
\renewcommand{\theequation}{\Alph{subsubsection}.\arabic{equation}}
\setcounter{equation}{0}

In this Appendix we compute the integrals in eq.(\ref{pG}) where time 
parameter $t_1$ tends to the observation time $t$:
\begin{eqnarray}\label{Gat}
p_t^\alpha&=&e^2\left[\lim_{k_1^0\to 0}
\int_{\beta_0}^0d\beta G_{12}^\alpha\right]_{t_2\to -\infty}^{t_2=t}
+e^2\int_{-\infty}^tdt_2\lim_{k_1^0\to 0}\left[
\left.G_{12}^\alpha\right|_{\beta=\beta_0}
\frac{\partial \beta_0}{\partial t_2}\right]
\\
&+&e^2\int_{-\infty}^tdt_2\lim_{k_1^0\to 0}\left[
\int_{\beta_0}^0d\beta G_1^\alpha\right]
\nonumber
\end{eqnarray}
Equality (\ref{sn0}) implies that the lower limit $\beta_0$ tends to $0$ 
if $k_1^0=t-t_1$ vanishes. With a degree of accuracy sufficient for our 
purpose,
\begin{equation}
\beta_0=-\frac{(k_1^0)^2}{(k_2^0)^2-{\mathbf q}^2(t,t_2)}.
\end{equation}
Integrals over parameter $\beta$ vanish whenever an expression under 
integral sign is smooth. So, we must limit our computations to the singular 
terms only. We expand expressions in powers of the small parameter $\beta$ 
and then integrate them.

Let us consider contribution $p_t^0$ to the energy $p_{\rm em}^0$ carried 
by electromagnetic field. Setting $k_1^0=0$ in two-point functions 
(\ref{km}) we obtain
\begin{eqnarray}
\kappa(t,t_2)&=&\nu(t,t_2)\frac{\partial\sigma}{\partial 
t_2}+\sigma\mu(t,t_2)\\
\mu(t,t_2)&=&-\frac{\partial\nu(t,t_2)}{\partial t_2}
\nonumber
\end{eqnarray}
where
\begin{equation}
\nu(t,t_2)=\frac12\left[k_2^0+({\mathbf q}{\mathbf v}_t)\right].
\end{equation}
Integration of the singular part of $G_{12}^0$ given by eq.(\ref{G00}) gives
the regular expression:
\begin{equation}\label{G00t}
\lim_{t_1\to t}\int_{\beta_0}^0d\beta G_{12}^0=
-\frac{1}{|{\mathbf v}_t|}\ln\frac{1+|{\mathbf v}_t|}{1-|{\mathbf 
v}_t|}
\frac{\nu(t,t_2)}{\sqrt{2\sigma(t,t_2)}}
\end{equation}
where ${\mathbf v}_t$ denotes the particle's velocity referred to the time 
of observation. According to eq.(\ref{Gat}), we should take the function 
in eq.(\ref{G00t}) at the end points, i.e. its value at the remote past 
$t_2\to -\infty$ should be subtracted from its value near the observation 
instant $t$.

Similarly we calculate the integral
\begin{eqnarray}\label{G01t}
\lim_{t_1\to t}\int_{\beta_0}^0d\beta G_1^0&=&\left(
2-\frac{1}{|{\mathbf v}_t|}\ln\frac{1+|{\mathbf v}_t|}{1-|{\mathbf 
v}_t|}\right)\frac{\kappa(t,t_2)}{[2\sigma(t,t_2)]^{3/2}}\\
&-&\left(
1+\frac{1}{2|{\mathbf v}_t|}\ln\frac{1+|{\mathbf v}_t|}{1-|{\mathbf 
v}_t|}\right)\frac{\mu(t,t_2)}{\sqrt{2\sigma(t,t_2)}}.\nonumber
\end{eqnarray}
According to eq.(\ref{Gat}), the result should be added to the limit
\begin{equation}
\lim_{t_1\to t}\left[
\left.G_{12}^0\right|_{\beta=\beta_0}\frac{\partial\beta_0}{\partial 
t_2}\right]=
\frac{-2\kappa(t,t_2)}{[2\sigma(t,t_2)]^{3/2}}+
\frac{\mu(t,t_2)}{\sqrt{2\sigma(t,t_2)}}
\end{equation}
and the sum should be integrated over $t_2$. After a straightforward 
calculations we obtain the expression (\ref{G00t}), taken with opposite 
sign. Their sum vanishes. Therefore $p_t^0$ does not contribute 
in the energy carried by electromagnetic field.

In an analogous way we calculate contribution $p_t^i$ to the momentum of 
the electromagnetic field. Integration of the singular part of function 
(\ref{Gi}) over $\beta$ gives
\begin{equation}\label{Gi12}
\left[\lim_{t_1\to t}\int_{\beta_0}^0d\beta 
G_{12}^i\right]_{t_2\to -\infty}^{t_2=t}=
-\frac{1}{2|{\mathbf v}_t|}\ln\frac{1+|{\mathbf v}_t|}{1-|{\mathbf 
v}_t|}\left[\frac{q^i+v_t^ik_2^0}{\sqrt{2\sigma(t,t_2)}}
\right]_{t_2\to -\infty}^{t_2=t}.
\end{equation}
It is the first term in eq.(\ref{Gat}). The limit under the integral 
sign (second term) is as follows:
\begin{equation}
\lim_{t_1\to t}\left[
\left.G_{12}^i\right|_{\beta=\beta_0}\frac{\partial\beta_0}{\partial 
t_2}\right]=-\frac{q^i+v_t^ik_2^0}{\left[2\sigma(t,t_2)\right]^{3/2}}
\frac{\partial\sigma}{\partial t_2}.
\end{equation}
We add it to the integral
\begin{eqnarray}\label{Gi1t}
\lim_{t_1\to t}\int_{\beta_0}^0d\beta G_1^i&=&\left(
1-\frac{1}{2|{\mathbf v}_t|}\ln\frac{1+|{\mathbf v}_t|}{1-|{\mathbf 
v}_t|}\right)\frac{q^i+v_t^ik_2^0}{[2\sigma(t,t_2)]^{3/2}}
\frac{\partial\sigma}{\partial t_2}\\
&-&\frac{1}{2|{\mathbf v}_t|}\ln\frac{1+|{\mathbf v}_t|}{1-|{\mathbf 
v}_t|}\frac{v_2^i+v_t^i}{\sqrt{2\sigma(t,t_2)}}\nonumber
\end{eqnarray}
and integrate over $t_2$. We arrive at the function of the end points 
only which annuls eq.(\ref{Gi12}).

\subsubsection{Calculation of integrals where $t_2\to t_1$ and $t_2\to 
-\infty$}\label{blue}
\renewcommand{\theequation}{\Alph{subsubsection}.\arabic{equation}}
\setcounter{equation}{0}

In this Appendix we compute the integrals in eq.(\ref{pG}) where time 
parameters $t_1$ and $t_2$ are equal to each other. We add also 
the integral evaluated at the remote past:
\begin{eqnarray}\label{pdt}
p_\triangle^\alpha(t)&=&-e^2\int_{-\infty}^tdt_2\lim_{\triangle t\to 
0}\left[
\int_{\beta_0}^0d\beta\left(\frac{\partial G_{12}^\alpha}{\partial 
t_2}+G_1^\alpha\right)\right]_{t_1=t_2+\triangle t}\\
&+&e^2\int_{-\infty}^tdt_1\lim_{\triangle t\to 0}\left[
\int_{\beta_0}^0d\beta G_2^\alpha\right]_{t_2=t_1-\triangle t}
-e^2\int_{-\infty}^tdt_1\lim_{t_2\to 
-\infty}\int_{\beta_0}^0d\beta G_2^\alpha .\nonumber
\end{eqnarray}
For fixed instant $t_1$ we assume that the limit
\begin{equation}\label{As}
A^i=\lim_{t_2\to -\infty}\frac{q^i(t_1,t_2)}{t_1-t_2}
\end{equation}
is finite if the motion is infinite. (In the specific case of finite 
motion, $A^i$=0.) If $k_2^0\to +\infty$ the lower limit of $\beta$-integrals
\begin{equation}
\beta_0=-\left(\frac{k_1^0}{k_2^0}\right)^2\frac{1}{1-{\mathbf A}^2},
\end{equation}
tends to the upper limit (zero) and we must limit our computations to the 
singular terms only.

According to eq.(\ref{sn0}), the equality $t_1=t_2$ yields
$\sin\vartheta_0=1$ and the lower limit, $\beta_0=-\tan\vartheta_0$, tends 
to $-\infty$. In this case the small parameter is the difference 
$\triangle t=t_1-t_2$. If the instants $t_1$ and $t_2$ are close to each 
other, function $\beta_0(t,t_1,t_2)$ raises as $(\triangle t)^{-1}$:
the product $\beta_0\triangle t$ possesses finite limit. We expand the
expressions under integral sign in powers of $\triangle t$ and 
thereafter we integrate the series. The integration over $\beta$ can be 
handled via the relations
\begin{eqnarray}\label{intb}
\int_{\beta_0}^0\frac{d\beta}{\sqrt{-\beta\alpha}}&=&2\ln\left(
\sqrt{-\beta_0}+\sqrt{\alpha_0}\right)\\
\int_{\beta_0}^0\frac{d\beta}{\sqrt{-\beta\alpha}}\beta&=&
-\alpha_0\sqrt{\frac{-\beta_0}{\alpha_0}}+
\ln\left(\sqrt{-\beta_0}+\sqrt{\alpha_0}\right)\nonumber\\
\int_{\beta_0}^0\frac{d\beta}{\sqrt{-\beta\alpha}}\beta^2&=&
-\frac12\beta_0\alpha_0\sqrt{\frac{-\beta_0}{\alpha_0}}
-\frac34\alpha_0\sqrt{\frac{-\beta_0}{\alpha_0}}+
\frac34\ln\left(\sqrt{-\beta_0}+\sqrt{\alpha_0}\right).\nonumber
\end{eqnarray}
Next we take the limit $\triangle t\to 0$. Suffice it to know that 
\begin{eqnarray}\label{intdt}
\lim_{\triangle t\to 0}\triangle t
\int_{\beta_0}^0\frac{d\beta}{\sqrt{-\beta\alpha}}\beta&=&
-\frac{k_2^0}{1+\sqrt{1-{\mathbf v}_2^2}}\\
\lim_{\triangle t\to 0}(\triangle t)^2
\int_{\beta_0}^0\frac{d\beta}{\sqrt{-\beta\alpha}}\beta^2&=&\frac12
\frac{\left(k_2^0\right)^2}{\left[1+\sqrt{1-{\mathbf 
v}_2^2}\right]^2}.\nonumber
\end{eqnarray} 

We begin with the zeroth component. Setting eqs.(\ref{G00}) in 
eq.(\ref{pdt}), after some algebra we arrive at
\begin{eqnarray}\label{p0blue}
p_\triangle^0(t)&=&-e^2\int_{-\infty}^tdt_2\lim_{\triangle 
t\to 0}\left[
\int_{\beta_0}^0\frac{d\beta}{\sqrt{-\beta\alpha}}
\left(\frac{\partial\kappa}{\partial 
t_2}{\cal D}^0+\kappa{\cal B}^0\right)
\right.\\
&-&\left.\int_{\beta_0}^0d\beta\frac{\alpha-\beta}{\sqrt{-\beta\alpha}}
\left(\frac{\partial\mu}{\partial 
t_2}{\cal D}^J+\mu{\cal B}^J\right)
\right]_{t_1=t_2+\triangle t}\nonumber\\
&+&e^2\int_{-\infty}^tdt_1
\left[\kappa\int_{\beta_0}^0d\beta 
\sqrt{\frac{\alpha}{-\beta}}\frac{{\bf v}_2^2}{\|r_2\|^3}-
\mu\int_{\beta_0}^0d\beta 
I'\frac{\partial}{\partial\beta}
\left(\frac{\alpha}{\|r_2\|}
\right)\right]_{t_2\to -\infty}^{t_2=t_1-\triangle t}.\nonumber
\end{eqnarray}
Recall that 
\begin{equation}\label{DB}
{\cal D}^a=\frac{1}{2\pi}\int_0^{2\pi}d\varphi\frac{a}{r_1r_2},\qquad
{\cal B}^a=\frac{1}{2\pi}\int_0^{2\pi}d\varphi\frac{ac_2}{r_1(r_2)^2}
\end{equation}
where $a=1$ for ${\cal D}^0,{\cal B}^0$ and $a$ is equal to Jacobian 
(\ref{Jc}) for coefficients labeled by $J$.

The quantity $r_2=-({\sf r}_2\cdot n)$ can be related to the quantity
$r_1=-({\sf r}_1\cdot n)$ by Taylor expansion in powers of $\triangle t$. 
With a degree of accuracy sufficient for our purposes we obtain 
\begin{equation}
r_2=r_1+\triangle tc_2
\end{equation}
where $c_2=({\sf c}_2\cdot n)$. Integration of (\ref{DB}) over the 
angular variable gives
\begin{eqnarray}
{\cal D}^a&=&\frac{({\sf a}\cdot{\sf r}_2)}{\|{\sf r}_2\|^3}+\triangle t
\left[\frac32\frac{({\sf a}\cdot{\sf r}_2)({\sf c}_2\cdot{\sf r}_2)}{\|{\sf 
r}_2\|^5}+\frac12\frac{({\sf a}\cdot{\sf c}_2)}{\|{\sf r}_2\|^3}
\right]\\
{\cal B}^a&=& \frac32\frac{({\sf a}\cdot{\sf r}_2)({\sf c}_2\cdot{\sf r}_2)}{\|{\sf 
r}_2\|^5}+\frac12\frac{({\sf a}\cdot{\sf c}_2)}{\|{\sf r}_2\|^3}
\nonumber\\
&+&
\triangle t
\left[\frac52\frac{({\sf a}\cdot{\sf r}_2)({\sf c}_2\cdot{\sf 
r}_2)^2}{\|{\sf r}_2\|^7}+
\frac12\frac{({\sf a}\cdot{\sf r}_2)({\sf c}_2\cdot{\sf c}_2)+
2({\sf a}\cdot{\sf c}_2)({\sf c}_2\cdot{\sf 
r}_2)}{\|{\sf r}_2\|^5}
\right].\nonumber
\end{eqnarray}

Expanding the integrands in eq.(\ref{p0blue}) in powers of $\triangle t$ 
and using the relations in eq.(\ref{intdt}) we finally obtain
\begin{eqnarray}\label{p0blfn}
p_\triangle^0(t)&=&e^2\int_{-\infty}^tdt_2
\frac{({\mathbf v}_2\mathbf{\dot v}_2)}{\left[1-{\mathbf 
v}_2^2\right]^{3/2}}
\lim_{\triangle t\to 0}
\left.\ln\left(\sqrt{\alpha_0}+\sqrt{-\beta_0}\right)\right|_{t_1=t_2+\triangle 
t}\\
&-&e^2\int_{-\infty}^tdt_1
\frac{{\mathbf v}_1^2}{k_1^0\sqrt{1-{\mathbf 
v}_1^2}}\lim_{\triangle t\to 0}\left.\sqrt{-\beta_0\alpha_0}
\right|_{t_2=t_1-\triangle t}\nonumber\\
&+&e^2\int_{-\infty}^tdt_2
\frac{({\mathbf v}_2\mathbf{\dot v}_2)}{\left[1-{\mathbf 
v}_2^2\right]^{3/2}}\frac{1}{1+\sqrt{1-{\mathbf v}_2^2}}
\left[2-\frac12\frac{1+{\mathbf v}_2^2}{1+\sqrt{1-{\mathbf 
v}_2^2}}\right]\nonumber\\
&+&e^2\int_{-\infty}^tdt_1\left[
\mu(t_1,t_2)I'_0\frac{\alpha_0}{{\sf r}_2^0}\right]_{t_2\to 
-\infty}^{t_2=t_1-\triangle t}+
e^2\int_{-\infty}^tdt_1\lim_{k_2^0\to\infty}
\frac{\mu(t_1,t_2)\sqrt{1-{\mathbf A}^2}}{k_1^0\left[1-({\mathbf A}{\mathbf 
v}_2)\right]}\nonumber
\end{eqnarray}
after integration by parts of the last term in eq.(\ref{p0blue}). 

The first and the second terms are singular. To deal with 
divergences it is efficient to introduce the hyperbolic angles $\Psi$ 
and $\psi$: 
\begin{eqnarray}
\cosh\Psi&=&\frac{k_2^0+k_1^0}{\mathbf q},\qquad 
\sinh\Psi=\frac{\sqrt{2\Sigma}}{\mathbf q}\\
\cosh\psi&=&\frac{k_2^0-k_1^0}{\mathbf q},\qquad 
\sinh\psi=\frac{\sqrt{2\sigma}}{\mathbf q}.\nonumber
\end{eqnarray}
(Function $\Sigma(t,t_1,t_2)$ is introduced in \ref{coord}.) In 
these notations
\begin{eqnarray}\label{Psp}
\beta_0&=&-\frac12\left[\cosh(\Psi-\psi)-1\right]=
-\sinh^2\frac{\Psi-\psi}{2}\\
\alpha_0&=&\frac12\left[\cosh(\Psi-\psi)+1\right]=
\cosh^2\frac{\Psi-\psi}{2}.\nonumber
\end{eqnarray}
so that
\begin{eqnarray}
\ln\left(\sqrt{\alpha_0}+\sqrt{-\beta_0}\right)&=&\frac{\Psi-\psi}{2}\\
\sqrt{-\beta_0\alpha_0}&=&\frac12\sinh(\Psi-\psi).\nonumber
\end{eqnarray}

Since the factor before the sign of limit is the total time 
derivative, the logarithmic divergence in eq.(\ref{p0blfn}) can be 
integrated by parts:
\begin{eqnarray}
&&\int_{-\infty}^tdt_2\frac{d}{dt_2}\left(
\frac{1}{\sqrt{1-{\mathbf v}_2^2}}\right)\lim_{\triangle t\to 0}
\left.\frac{\Psi-\psi}{2}\right|_{t_1=t_2+\triangle t}=
\frac{1}{\sqrt{1-{\mathbf v}_2^2}}
\left.\lim_{\triangle t\to 0}\frac{\Psi-\psi}{2}\right|_{t_2\to 
-\infty}^{t_2=t}\\
&-&\frac12\int_{-\infty}^t\frac{dt_2}{\sqrt{1-{\mathbf v}_2^2}}
\lim_{\triangle t\to 0}\left[
\frac{\partial(\Psi-\psi)}{\partial t_1}
+\frac{\partial(\Psi-\psi)}{\partial t_2}
\right]_{t_1=t_2+\triangle t}.\nonumber
\end{eqnarray} 
Taking into account that at the end points hyperbolic angles vanish, we 
finally obtain
\begin{eqnarray}
&&e^2\int_{-\infty}^tdt_2 \frac{({\mathbf v}_2\mathbf{\dot 
v}_2)}{\left[1-{\mathbf v}_2^2\right]^{3/2}}
\lim_{\triangle t\to 0}
\left.\ln\left(\sqrt{\alpha_0}+
\sqrt{-\beta_0}\right)\right|_{t_1=t_2+\triangle t}\\
&=&\frac{e^2}{2}\int_{-\infty}^tdt_2\left[
\frac{1}{k_2^0\sqrt{1-{\mathbf v}_2^2}}-
\frac{({\mathbf v}_2\mathbf{\dot v}_2)}{(1-{\mathbf v}_2^2)(1+
\sqrt{1-{\mathbf v}_2^2})}\right].\nonumber
\end{eqnarray} 

Now we rewrite the second divergent term involved in eq.(\ref{p0blfn}). 
We expand the integrand in powers of $\triangle t$. Passing to the limit
$\triangle t\to 0$, we arrive at:
\begin{eqnarray}
&&\int_{-\infty}^tdt_1 \frac{{\mathbf v}_1^2}{k_1^0\sqrt{1-{\mathbf 
v}_1^2}}\lim_{\triangle t\to 0}\left.\sqrt{-\beta_0\alpha_0}
\right|_{t_2=t_1-\triangle t}=
\int_{-\infty}^tdt_1\lim_{\triangle t\to 0}\frac{1-\sqrt{1-{\mathbf 
v}_1^2}}{\triangle t\sqrt{1-{\mathbf v}_1^2}}\\
&+&\int_{-\infty}^tdt_1\left[
\frac{1-\sqrt{1-{\mathbf v}_1^2}}{2k_1^0\sqrt{1-{\mathbf v}_1^2}}
+\frac{({\mathbf v}_1\mathbf{\dot v}_1)}{\sqrt{1-{\mathbf v}_1^2}(1+
\sqrt{1-{\mathbf v}_1^2})}-
\frac{({\mathbf v}_1\mathbf{\dot v}_1)}{2(1-{\mathbf v}_1^2)}
\right].\nonumber
\end{eqnarray} 
Inserting these expressions in eq.(\ref{p0blfn}) we finally obtain
\begin{eqnarray}\label{p0fn}
p_\triangle^0(t)&=&\frac{e^2}{2}\int_{-\infty}^tdt_2\left[
\frac{1}{k_2^0\sqrt{1-{\mathbf v}_2^2}}-
\frac{({\mathbf v}_2\mathbf{\dot v}_2)}{(1-{\mathbf v}_2^2)(1+
\sqrt{1-{\mathbf v}_2^2})}\right]\\
&-&e^2\int_{-\infty}^tdt_1\lim_{\triangle t\to 0}\frac{1-\sqrt{1-{\mathbf 
v}_1^2}}{\triangle t\sqrt{1-{\mathbf v}_1^2}}\nonumber\\
&-&e^2\int_{-\infty}^tdt_1\left[
\frac{1-\sqrt{1-{\mathbf v}_1^2}}{2k_1^0\sqrt{1-{\mathbf v}_1^2}}
+\frac{({\mathbf v}_1\mathbf{\dot v}_1)}{\sqrt{1-{\mathbf v}_1^2}(1+
\sqrt{1-{\mathbf v}_1^2})}-
\frac{({\mathbf v}_1\mathbf{\dot v}_1)}{2(1-{\mathbf v}_1^2)}
\right]\nonumber\\
&+&e^2\int_{-\infty}^tdt_2
\frac{({\mathbf v}_2\mathbf{\dot v}_2)}{\left[1-{\mathbf 
v}_2^2\right]^{3/2}}\frac{1}{1+\sqrt{1-{\mathbf v}_2^2}}
\left[2-\frac12\frac{1+{\mathbf v}_2^2}{1+\sqrt{1-{\mathbf 
v}_2^2}}\right]\nonumber\\
&+&e^2\int_{-\infty}^tdt_1\left[
\mu(t_1,t_2)I'_0\frac{\alpha_0}{{\sf r}_2^0}\right]_{t_2\to 
-\infty}^{t_2=t_1-\triangle t}+
e^2\int_{-\infty}^tdt_1\lim_{k_2^0\to\infty}
\frac{\mu(t_1,t_2)\sqrt{1-{\mathbf A}^2}}{k_1^0\left[1-({\mathbf A}{\mathbf 
v}_2)\right]}.\nonumber
\end{eqnarray}

The calculation of momentum corrections is virtually identical to what 
is presented here, and we shall not worry with the details. It suffices 
to present the resulting expression
\begin{eqnarray}\label{pifn}
p_\triangle^i(t)&=&\frac{e^2}{2}\int_{-\infty}^tdt_2\left[
\frac{v_2^i}{k_2^0\sqrt{1-{\mathbf v}_2^2}}-
\frac{v_2^i({\mathbf v}_2\mathbf{\dot v}_2)}{(1-{\mathbf v}_2^2)(1+
\sqrt{1-{\mathbf v}_2^2})}\right]
\\
&-&e^2\int_{-\infty}^tdt_1\lim_{\triangle t\to 0}\frac{v_1^i}{\triangle 
t\sqrt{1-{\mathbf v}_1^2}(1+\sqrt{1+{\mathbf v}_1^2})}\nonumber\\
&-&\frac{e^2}{2}\int_{-\infty}^tdt_1\left[
\frac{v_1^i}{k_1^0\sqrt{1-{\mathbf v}_1^2}(1+\sqrt{1-{\mathbf v}_1^2})}
-\frac{v_1^i({\mathbf v}_1\mathbf{\dot v}_1)}{(1-{\mathbf v}_1^2)(1+
\sqrt{1-{\mathbf v}_1^2})^2}
\right]\nonumber\\
&+&e^2\int_{-\infty}^tdt_2\frac{1}{1+\sqrt{1-{\mathbf v}_2^2}}\left\{
\frac{({\mathbf v}_2\mathbf{\dot v}_2)v_2^i}{\left[1-{\mathbf 
v}_2^2\right]^{3/2}}
\left[2-\frac12\frac{1+{\mathbf 
v}_2^2}{1+\sqrt{1-{\mathbf v}_2^2}}\right]
+\frac{{\dot v}_2^i}{\sqrt{1-{\mathbf v}_2^2}}
\right\}\nonumber\\
&+&\frac{e^2}{2}\int_{-\infty}^tdt_1\left[
\left(v_1^i+v_2^i\right)I'_0\frac{\alpha_0}{{\sf r}_2^0}\right]_{t_2\to 
-\infty}^{t_2=t_1-\triangle t}+
\frac{e^2}{2}\int_{-\infty}^tdt_1\lim_{k_2^0\to\infty}
\frac{\left(v_1^i+v_2^i\right)\sqrt{1-{\mathbf A}^2}}{k_1^0\left[1-({\mathbf A}{\mathbf 
v}_2)\right]}.\nonumber
\end{eqnarray}
In the next Appendix the contribution from integrals at point 
$\beta=\beta_0$ is found. Its bound part contains the terms which 
annihilate the divergent terms in expressions (\ref{p0fn}) and 
(\ref{pifn}).

\subsubsection{Calculation of integrals at point where 
$\beta=\beta_0$}\label{yellow}
\renewcommand{\theequation}{\Alph{subsubsection}.\arabic{equation}}
\setcounter{equation}{0}

We now would like to extract the partial derivatives with respect to 
time variables from the integrand of the following double integral
\begin{equation}\label{Galph}
p^\alpha_0(t)=e^2
\begin{array}{c}
\displaystyle 
\int_{-\infty}^tdt_1\int_{-\infty}^{t_1}dt_2\\
\\[-1em]
\displaystyle 
\int_{-\infty}^tdt_2\int_{t_2}^tdt_1
\end{array}
\left(
\left[\frac{\partial G_{12}^\alpha}{\partial t_2}+G_1^\alpha\right]_{\beta_0}
\frac{\partial\beta_0}{\partial t_1} + 
\left.G_2^\alpha\right|_{\beta_0}\frac{\partial\beta_0}{\partial t_2}
\right)
\end{equation}
Note that
\begin{equation}
\frac{\partial\beta_0}{\partial t_1}=\alpha_0\frac{{\sf r}_1^0}{{\sf J}_0},
\qquad 
\frac{\partial\beta_0}{\partial t_2}=\beta_0\frac{{\sf r}_2^0}{{\sf J}_0}.
\end{equation}
and functions $G_{12}^\alpha,G_1^\alpha$, and $G_2^\alpha$ are given by 
eqs.(\ref{G00}) and (\ref{Gi}). 

If $\beta=\beta_0$ the radius $R$ of the smallest circle pictured at 
figure \ref{bt} vanishes and it reduces to point $A$. Norms $\|r_a\|$
and $\|c_a\|$ become zeroth component ${\sf r}_a^0$ and ${\sf c}_a^0$, 
respectively. Hence the coefficients (\ref{ABC}) get simplified, e.g.
\begin{equation}\label{BCD}
{\cal D}^0=\frac{1}{{\sf r}_1^0{\sf r}_2^0},\qquad
{\cal D}^J=\frac{{\sf J}_0}{{\sf r}_1^0{\sf r}_2^0}\qquad
{\cal D}_1^i=\frac{-\beta_0 q^i}{{\sf r}_1^0{\sf r}_2^0},\qquad
{\cal D}_2^i=\frac{\alpha_0 q^i}{{\sf r}_1^0{\sf r}_2^0}.
\end{equation}
In terms of two-point functions $\sigma=-1/2(q\cdot q)$ and 
$\Sigma=\sigma+k_1^0k_2^0$ the angle-free functions $\kappa$, $\mu$ and 
$\lambda_a$ involved in $p_{\rm em}^\alpha$ are as follows:
\begin{eqnarray}
\kappa&=&\frac12\left(\Sigma\frac{\partial^2\Sigma}{\partial 
t_1\partial t_2}-\frac{\partial\Sigma}{\partial 
t_1}\frac{\partial\Sigma}{\partial t_2}\right),\qquad
\mu=\frac12\frac{\partial^2\Sigma}{\partial t_1\partial t_2}\nonumber\\
\lambda_a&=&k_a^0\frac{\partial\Sigma}{\partial t_a}+\Sigma\nonumber\\
&=&k_a^0\frac{\partial\sigma}{\partial t_a}+\sigma.
\end{eqnarray}

First we set $\alpha=0$. Routine scrupulous calculations allow us to 
rewrite the expression under the integral signs in eq.(\ref{Galph}) as 
follows:
\begin{eqnarray}\label{p0_y}
&&\left[\frac{\partial G_{12}^0}{\partial 
t_2}+G_1^0\right]_{\beta_0}
\frac{\partial\beta_0}{\partial t_1} + 
\left.G_2^0\right|_{\beta_0}\frac{\partial\beta_0}{\partial t_2}=
\frac{\partial}{\partial t_2}\left(
\sqrt{\frac{\alpha_0}{-\beta_0}}\frac{\kappa}{{\sf r}_2^0{\sf J}_0}
-I_0'\mu\frac{\alpha_0}{{\sf r}_2^0}\right)
\\
&+&\frac12\frac{\partial}{\partial t_2}\left[
\frac{k_2^0+({\mathbf q}{\mathbf v}_1)}{k_1^0}
\left(\frac{1}{\sqrt{2\Sigma}}+\frac{1}{\sqrt{2\sigma}}
\right)+\frac{1}{\sqrt{2\Sigma}}\right]\nonumber\\
&+&\frac12\frac{\partial}{\partial t_1}\left[
\frac{k_1^0-({\mathbf q}{\mathbf v}_2)}{k_2^0}
\left(\frac{1}{\sqrt{2\Sigma}}-\frac{1}{\sqrt{2\sigma}}
\right)+\frac{1}{\sqrt{2\Sigma}}\right]\nonumber\\
&-&\frac{\partial^2\sigma}{\partial t_1\partial 
t_2}\frac{q^0}{(2\sigma)^{3/2}}+\frac12
\left(\frac{\partial}{\partial t_1}-\frac{\partial}{\partial t_2}\right)
\frac{1}{\sqrt{2\sigma}}.\nonumber
\end{eqnarray}
This expression contains the term which is proportional to the mixed 
second-order partial derivative of $\sigma$ which can not be rewritten as 
a derivative with respect to $t_1$ or $t_2$.

In analogous way we rewrite the $\beta_0$ part of space components $p_{\rm 
em}^i$ of the momentum carried by the electromagnetic field:
\begin{eqnarray}\label{pi_y}
&&\left[\frac{\partial G_{12}^i}{\partial 
t_2}+G_1^i\right]_{\beta_0}
\frac{\partial\beta_0}{\partial t_1} + 
\left.G_2^i\right|_{\beta_0}\frac{\partial\beta_0}{\partial t_2}=
\\
&=&\frac12\frac{\partial}{\partial t_2}\left[
\sqrt{\frac{\alpha_0}{-\beta_0}}
\frac{1}{{\sf r}_2^0{\sf J}_0}\left(
\frac{\partial\lambda_1}{\partial t_2}\alpha_0q^i
-\frac{\partial\lambda_2}{\partial t_1}\beta_0q^i
+v_1^i\lambda_2+v_2^i\lambda_1
\right)-I_0'\frac{\alpha_0}{{\sf r}_2^0}\left(v_1^i+v_2^i\right)\right]
\nonumber\\
&+&\frac12\frac{\partial}{\partial t_2}\left[
\frac{v_1^ik_2^0+q^i}{k_1^0}
\left(\frac{1}{\sqrt{2\Sigma}}+\frac{1}{\sqrt{2\sigma}}
\right)+\frac{v_1^i}{\sqrt{2\Sigma}}\right]\nonumber\\
&+&\frac12\frac{\partial}{\partial t_1}\left[
\frac{v_2^ik_1^0-q^i}{k_2^0}
\left(\frac{1}{\sqrt{2\Sigma}}-\frac{1}{\sqrt{2\sigma}}
\right)+\frac{v_2^i}{\sqrt{2\Sigma}}\right]
\nonumber\\
&-&\frac{\partial^2\sigma}{\partial t_1\partial 
t_2}\frac{q^i}{(2\sigma)^{3/2}}+
\frac12\frac{\partial}{\partial 
t_1}\left(\frac{v_2^i}{\sqrt{2\sigma}}\right)-\frac12
\frac{\partial}{\partial t_2}\left(\frac{v_1^i}{\sqrt{2\sigma}}\right).
\nonumber
\end{eqnarray}

Having integrated expressions (\ref{p0_y}) and (\ref{pi_y}) over time 
variables $t_1$ and $t_2$ according to the rule (\ref{Galph}) we obtain
\begin{eqnarray}\label{p0bfin}
p_0^0(t)&=&e^2\int_{-\infty}^tdt_1\left[
\sqrt{\frac{\alpha_0}{-\beta_0}}\frac{\kappa}{{\sf r}_2^0{\sf J}_0}
-I_0'\mu\frac{\alpha_0}{{\sf r}_2^0}
\right]_{t_2\to -\infty}^{t_2=t_1}\\
&+&\frac{e^2}{2}\int_{-\infty}^tdt_1\left[
\frac{k_2^0+({\mathbf q}{\mathbf v}_1)}{k_1^0}
\left(\frac{1}{\sqrt{2\Sigma}}+\frac{1}{\sqrt{2\sigma}}\right)
+\frac{1}{\sqrt{2\Sigma}}
\right]_{t_2\to -\infty}^{t_2=t_1}\nonumber\\
&+&\frac{e^2}{2}\int_{-\infty}^tdt_2\left[
\frac{k_1^0-({\mathbf q}{\mathbf v}_2)}{k_2^0}
\left(\frac{1}{\sqrt{2\Sigma}}-\frac{1}{\sqrt{2\sigma}}\right)
+\frac{1}{\sqrt{2\Sigma}}
\right]^{t_1=t}_{t_1=t_2}\nonumber\\
&+&e^2\int_{-\infty}^tdt_1\int_{-\infty}^{t_1}dt_2\left[
-\frac{q^0(v_1\cdot v_2)}{(2\sigma)^{3/2}}+
\frac12\frac{(q\cdot v_1)}{(2\sigma)^{3/2}}+\frac12\frac{(q\cdot 
v_2)}{(2\sigma)^{3/2}}
\right]\nonumber\\\label{pibfin}
p_0^i(t)&=&\frac{e^2}{2}\int_{-\infty}^tdt_1\left[
\sqrt{\frac{\alpha_0}{-\beta_0}}
\frac{1}{{\sf r}_2^0{\sf J}_0}\left(
\frac{\partial\lambda_1}{\partial t_2}\alpha_0q^i
-\frac{\partial\lambda_2}{\partial t_1}\beta_0q^i
+v_1^i\lambda_2+v_2^i\lambda_1
\right)\right.\\
&-&\left.I_0'\frac{\alpha_0}{{\sf r}_2^0}\left(v_1^i+v_2^i\right)
\right]_{t_2\to -\infty}^{t_2=t_1}\nonumber\\
&+&\frac{e^2}{2}\int_{-\infty}^tdt_1\left[
\frac{v_1^ik_2^0+q^i}{k_1^0}
\left(\frac{1}{\sqrt{2\Sigma}}+\frac{1}{\sqrt{2\sigma}}\right)
+\frac{v_1^i}{\sqrt{2\Sigma}}
\right]_{t_2\to -\infty}^{t_2=t_1}\nonumber\\
&+&\frac{e^2}{2}\int_{-\infty}^tdt_2\left[
\frac{v_2^ik_1^0-q^i}{k_2^0}
\left(\frac{1}{\sqrt{2\Sigma}}-\frac{1}{\sqrt{2\sigma}}\right)
+\frac{v_2^i}{\sqrt{2\Sigma}}
\right]^{t_1=t}_{t_1=t_2}\nonumber\\
&+&e^2\int_{-\infty}^tdt_1\int_{-\infty}^{t_1}dt_2\left[
-\frac{q^i(v_1\cdot v_2)}{(2\sigma)^{3/2}}+\frac12
\frac{v_2^i(q\cdot v_1)}{(2\sigma)^{3/2}}+\frac12\frac{v_1^i(q\cdot 
v_2)}{(2\sigma)^{3/2}}
\right]\nonumber
\end{eqnarray}
So, besides the double integral which describes the self-action which 
depends not only on the current state of motion of the particle but also 
on its past history, we have the integrals of functions of the end points 
only. 

According to eqs.(\ref{Ss}), function 
$\left.\Sigma(t,t_1,t_2)\right|_{t_1=t}$ is equal to the function 
$\sigma(t,t_2)$. This circumstance simplifies evaluation of the terms 
referred to this end point. We expand the terms near $t_2=t_1$ in powers of 
$\triangle t=t_1-t_2$ and take the limit $\triangle t\to 0$. We use the 
assumption in eq.(\ref{As}) when $t_2\to -\infty$. After some algebra we 
finally obtain:
\begin{eqnarray}\label{p0bfn}
p_0^0(t)&=&-e^2\int_{-\infty}^tdt_1\left[
\lim_{\triangle t\to 0}\frac{1}{\triangle t}+\frac{1}{2k_1^0}-
\frac{({\mathbf v}_1\mathbf{\dot v}_1)}{2\sqrt{1-{\mathbf 
v}_1^2}(1+\sqrt{1-{\mathbf v}_1^2})}
\right]\\
&-&e^2\int_{-\infty}^tdt_1\lim_{t_2\to 
-\infty}\frac{\mu(t_1,t_2)\sqrt{1-{\mathbf A}^2}}{k_1^0\left[1- 
({\mathbf A}{\mathbf v}_2)\right]}
-e^2\int_{-\infty}^tdt_1\left[
\mu(t_1,t_2)I'_0\frac{\alpha_0}{{\sf r}_2^0}
\right]_{t_2\to -\infty}^{t_2=t_1}\nonumber\\
&+&\frac{e^2}{2}\int_{-\infty}^tdt_1
\lim_{\triangle t\to 0}\frac{1}{\triangle t\sqrt{1-{\mathbf v}_1^2}}
+\frac{e^2}{2}\int_{-\infty}^tdt_2
\lim_{\triangle t\to 0}\frac{1}{\triangle t\sqrt{1-{\mathbf v}_2^2}}
\nonumber\\
&+&\frac{e^2}{2}\int_{-\infty}^tdt_2\frac{1}{\sqrt{2\sigma(t,t_2)}}
\nonumber\\
&+&e^2\int_{-\infty}^tdt_1\int_{-\infty}^{t_1}dt_2\left[
-\frac{q^0(v_1\cdot v_2)}{(2\sigma)^{3/2}}
+\frac12\frac{(q\cdot v_1)}{(2\sigma)^{3/2}}
+\frac12\frac{(q\cdot v_2)}{(2\sigma)^{3/2}}
\right]\nonumber
\nonumber\\\label{pibfn}
p_0^i(t)&=&e^2\int_{-\infty}^t
\frac{dt_1}{1+\sqrt{1-{\mathbf v}_1^2}}
\left[
-\lim_{\triangle t\to 0}\frac{v_1^i}{\triangle t}-\frac{v_1^i}{2k_1^0}+
\frac{v_1^i({\mathbf v}_1\mathbf{\dot v}_1)}{\sqrt{1-{\mathbf 
v}_1^2}(1+\sqrt{1-{\mathbf v}_1^2})}+\frac{{\dot v}_1^i}{2}
\right]\nonumber\\
&-&\frac{e^2}{2}\int_{-\infty}^tdt_1\lim_{t_2\to 
-\infty}\frac{\left(v_1^i+v_2^i\right)\sqrt{1-{\mathbf 
A}^2}}{k_1^0\left[1- ({\mathbf A}{\mathbf v}_2)\right]}
-\frac{e^2}{2}\int_{-\infty}^tdt_1\left[
\left(v_1^i+v_2^i\right)I'_0\frac{\alpha_0}{{\sf r}_2^0}
\right]_{t_2\to -\infty}^{t_2=t_1}\nonumber\\
&+&\frac{e^2}{2}\int_{-\infty}^tdt_1
\lim_{\triangle t\to 0}\frac{v_1^i}{\triangle t\sqrt{1-{\mathbf v}_1^2}}
+\frac{e^2}{2}\int_{-\infty}^tdt_2
\lim_{\triangle t\to 0}\frac{v_2^i}{\triangle t\sqrt{1-{\mathbf v}_2^2}}\\
&+&\frac{e^2}{2}\int_{-\infty}^tdt_2\frac{v_2^i}{\sqrt{2\sigma(t,t_2)}}
\nonumber\\
&+&e^2\int_{-\infty}^tdt_1\int_{-\infty}^{t_1}dt_2\left[
-\frac{q^i(v_1\cdot v_2)}{(2\sigma)^{3/2}}
+\frac12\frac{v_2^i(q\cdot v_1)}{(2\sigma)^{3/2}}
+\frac12\frac{v_1^i(q\cdot v_2)}{(2\sigma)^{3/2}}
\right]\nonumber.
\end{eqnarray}
Divergent terms annul their counterparts from eqs.(\ref{p0fn}) and 
(\ref{pifn}). After that only one term remains:
\begin{equation}
\frac{e^2}{2}\int_{-\infty}^tdt_2\frac{v_2^\mu}{\sqrt{2\sigma(t,t_2)}}=
\frac{e^2}{2}\int_{-\infty}^t dt_2 
\left.
\frac{v_2^\mu}{\sqrt{2\Sigma}}
\right|_{t_1=t_2}^{t_1=t}
+\frac{e^2}{2}\int_{-\infty}^t dt_1
\left.
\frac{v_1^\mu}{\sqrt{2\Sigma}}
\right|_{t_2\to -\infty}^{t_2=t_1}.
\end{equation}
It is the singular part of energy-momentum carried by the
electromagnetic field. (It is worth noting that inverse square root
$[\left.2\Sigma(t,t_1,t_2)\right|_{t_2\to -\infty}]^{-1/2}$ vanishes 
even if $t_1\to -\infty$, see eqs.(\ref{Ss}).)

Final expressions are presented in Section \ref{trace} (see 
eqs.(\ref{p0em}) and (\ref{piem})).

\subsubsection{Angular momentum in 2+1 electrodynamics}\label{angul}
\renewcommand{\theequation}{\Alph{subsubsection}.\arabic{equation}}
\setcounter{equation}{0}

We now turn to the calculation of the angular momentum tensor
\begin{equation}\label{Mem}
M^{\mu\nu}_{\rm em}(t)=\int_{\Sigma_t} 
d\sigma_0\left(y^\mu T^{0\nu} - y^\nu T^{0\mu}\right)
\end{equation}
carried by the electromagnetic field due to a pointlike charge. We apply 
the convenient coordinate system introduced in Section \ref{trace} and 
detailed in \ref{coord}. 

We present the integrand of eq.(\ref{Mem}) in the following form:
\begin{equation}\label{mm}
m_{\rm em}^{\mu\nu}= 
m_{12}^{\mu\nu}+m_{21}^{\mu\nu}-m_{12}^{\nu\mu}-m_{21}^{\nu\mu}
\end{equation}
where
\begin{equation}\label{m12}
m_{12}^{\mu\nu}=\left(z_1^\mu+K_1^\mu\right)\frac{1}{2\pi}
\left[
f_{(1)}^{0\lambda}f_{(2)\lambda}^\nu -\frac14\eta^{0\nu}
f_{(1)}^{\alpha\beta}f^{(2)}_{\alpha\beta}
\right].
\end{equation}
It is straightforward to substitute the fields (\ref{F1F2}) into this 
expression to calculate the first term of the integrand (\ref{mm}). The 
others can be obtained by interchanging of the pair of indices $(1,2)$ 
and $(\mu,\nu)$.
 
Having integrated expression $Jm_{12}^{\mu\nu}$ over $\varphi$ we 
obtain 
\begin{eqnarray}\label{m-12}
{\cal M}_{12}^{\mu\nu}&=&\frac{1}{2}I\left\{
{\hat{\cal T}}_{12}^{\mu\nu}\left(\frac{\partial\lambda_1}{\partial 
t_2}\right)+{\hat{\cal T}}_1^\mu\left(v_2^\nu\lambda_1\right)-
v_2^\nu\frac{\partial\lambda_1}{\partial t_2}{\cal C}_1^\mu-
v_2^\nu\frac{\partial^2\lambda_1}{\partial t_1\partial t_2}{\cal D}_1^\mu
\right\}-\frac12I'{\hat{\cal T}}_1^{\mu J}\left(v_2^\nu\right)\nonumber\\
&+&\frac{z_1^\mu}{2}I\left\{
{\hat{\cal T}}_2^\nu\left(\frac{\partial\lambda_1}{\partial 
t_2}\right)+{\hat{\cal T}}^0\left(v_2^\nu\lambda_1\right)-
v_2^\nu\frac{\partial\lambda_1}{\partial t_2}{\cal C}^0-
v_2^\nu\frac{\partial^2\lambda_1}{\partial t_1\partial t_2}{\cal D}^0
\right\}-\frac{z_1^\mu}{2}I'
{\hat{\cal T}}^J\left(v_2^\nu\right)\nonumber\\
&-&\frac{\eta^{0\nu}}{4}\left\{
I\left[
{\hat{\cal T}}_1^\mu\left(\lambda\right)+
z_1^\mu{\hat{\cal T}}^0\left(\lambda\right)
\right]-
I'\left[
{\hat{\cal T}}_1^{\mu J}\left(\lambda_0\right)+
z_1^\mu{\hat{\cal T}}^J\left(\lambda_0\right)
\right]
\right\}
\end{eqnarray}
where functions $\lambda$ and $\lambda_0$ are given by eqs.(\ref{lbd}).

Usage of the equalities in eq.(\ref{cli}) derived in \ref{varphi} allows us 
to rewrite the integrand (\ref{m-12}) as the following sum:
\begin{eqnarray}\label{M-12}
{\cal M}_{12}^{\mu\nu}&=&\frac{1}{2}I\left\{
{\hat\Pi}_{12}^{\mu\nu}\left(\frac{\partial\lambda_1}{\partial 
t_2}\right)+{\hat\Pi}_1^\mu\left(v_2^\nu\lambda_1\right)-
\frac{\partial}{\partial t_1}\left(v_2^\nu\frac{\partial\lambda_1}{\partial 
t_2}{\cal D}_1^\mu\right)
\right.\\
&+&\left.{\hat\Pi}_2^\nu\left(z_1^\mu\frac{\partial\lambda_1}{\partial 
t_2}\right)-
\frac{\partial}{\partial t_2}\left(v_1^\mu\frac{\partial\lambda_1}{\partial 
t_2}{\cal D}_2^\nu\right)
\right.\nonumber\\
&+&\left.{\hat\Pi}^0\left(z_1^\mu v_2^\nu\lambda_1\right)-
\frac{\partial}{\partial 
t_1}\left(z_1^\mu v_2^\nu\frac{\partial\lambda_1}{\partial t_2}{\cal 
D}^0\right)-
\frac{\partial}{\partial t_2}\left(v_1^\mu 
v_2^\nu\lambda_1{\cal D}^0\right)
\right\}\nonumber\\
&-&\frac12I'\left\{
{\hat\Pi}_1^{\mu J}\left(v_2^\nu\right)
+{\hat\Pi}^J\left(z_1^\mu v_2^\nu\right)-
\frac{\partial}{\partial t_2}\left(v_1^\mu 
v_2^\nu{\cal D}^J\right)
\right\}\nonumber\\
&-&\frac{\eta^{0\nu}}{4}\left\{
I\left[
{\hat\Pi}_1^\mu\left(\lambda\right)+
{\hat\Pi}^0\left(z_1^\mu\lambda\right)-
\frac{\partial}{\partial t_2}\left(v_1^\mu\lambda{\cal D}^0\right)
\right]\right.\nonumber\\
&-&\left.I'\left[
{\hat\Pi}_1^{\mu J}\left(\lambda_0\right)+
{\hat\Pi}^J\left(z_1^\mu\lambda_0\right)-
\frac{\partial}{\partial t_2}\left(v_1^\mu\lambda_0{\cal D}^J\right)
\right]
\right\}.\nonumber
\end{eqnarray}
Operators $\Pi^a$ are combinations of partial derivatives (\ref{Pa}).

Further we perform the integration over the time variables and $\beta$ 
according to the rule in eq.(\ref{pG}). It results in functions of the end 
points only. We deal with  four types of integrals described in 
subsections \ref{trace}.$1^o$-\ref{trace}.$4^o$. All of them possess a
specific small parameter. Near the observation time $t$ the small 
parameter is $\beta$ as well as when $t_2\to -\infty$. If $t_2$ tends 
to $t_1$ (or vice versa), their difference $t_1-t_2$ tends to zero. 
These circumstances simplify the computation of integrals of types 
\ref{trace}.$1^o$-\ref{trace}.$3^o$ which is virtually identical to 
that presented in \ref{pink} and \ref{blue}, and we shall not bother with 
the details. We obtain the bound terms only which should be absorbed 
within the renormalization procedure. Radiative terms arise from the 
integration near the point $\beta=\beta_0$ where radial variable $R=0$ 
(see integral type \ref{trace}.$4^o$).

So, having computed the radiative angular momentum we are not going 
beyond the limit $R\to 0$. The terms involved in the final expression
(\ref{M-12}) get simplified sufficiently:
\begin{eqnarray}
{\cal D}_{12}^{ij}&=&-\frac{\alpha_0\beta_0q^iq^j}{{\sf r}_1^0{\sf r}_2^0}\\
{\cal C}_{12}^{ij}-\frac{\partial {\cal D}_{12}^{ij}}{\partial t_1}&=&
-\frac{\alpha_0^2\beta_0q^iq^j}{({\sf r}_2^0)^3}{\mathbf v}_2^2 +
\alpha_0\frac{v_1^iq^j}{{\sf r}_1^0{\sf r}_2^0}-
\alpha_0\beta_0\frac{q^iv_2^j}{({\sf r}_2^0)^2}+
\alpha_0^2\frac{v_2^iq^j}{({\sf r}_2^0)^2}+
\alpha_0\frac{\delta^{ij}}{{\sf r}_2^0}\nonumber\\
{\cal B}_{12}^{ij}-\frac{\partial {\cal D}_{12}^{ij}}{\partial t_2}&=&
-\frac{\alpha_0\beta_0^2q^iq^j}{({\sf r}_1^0)^3}{\mathbf v}_1^2 -
\beta_0\frac{v_2^jq^i}{{\sf r}_1^0{\sf r}_2^0}+
\alpha_0\beta_0\frac{q^jv_1^i}{({\sf r}_1^0)^2}-
\beta_0^2\frac{v_1^jq^i}{({\sf r}_1^0)^2}+
\beta_0\frac{\delta^{ij}}{{\sf r}_1^0};\nonumber\\
{\cal D}_{1}^{iJ}&=&-\frac{\beta_0q^i{\sf J}_0}{{\sf r}_1^0{\sf r}_2^0}
\qquad
{\cal D}_{2}^{iJ}=\frac{\alpha_0q^i{\sf J}_0}{{\sf r}_1^0{\sf r}_2^0}\\
{\cal C}_{1}^{iJ}-\frac{\partial{\cal D}_{1}^{iJ}}{\partial t_1}&=&
v_1^i\frac{{\sf J}_0}{{\sf r}_1^0{\sf r}_2^0}+
\alpha_0v_2^i\frac{{\sf J}_0}{({\sf r}_2^0)^2}-
\left[\frac{\partial}{\partial\beta}\left(\frac{\alpha\beta 
q^i}{\|{\sf r}_2\|}\right)\right]_{\beta_0}\nonumber\\
{\cal B}_{1}^{iJ}-\frac{\partial{\cal D}_{1}^{iJ}}{\partial t_2}&=&
\beta_0v_1^i\frac{{\sf J}_0}{({\sf r}_1^0)^2}-
\left[\frac{\partial}{\partial\beta}\left(\frac{\beta^2
q^i}{\|{\sf r}_1\|}\right)\right]_{\beta_0}\nonumber\\
{\cal C}_{2}^{iJ}-\frac{\partial{\cal D}_{2}^{iJ}}{\partial t_1}&=&
\alpha_0v_2^i\frac{{\sf J}_0}{({\sf r}_2^0)^2}+
\left[\frac{\partial}{\partial\beta}\left(\frac{\alpha^2
q^i}{\|{\sf r}_2\|}\right)\right]_{\beta_0}\nonumber\\
{\cal B}_{2}^{iJ}-\frac{\partial{\cal D}_{2}^{iJ}}{\partial t_2}&=&
v_2^i\frac{{\sf J}_0}{{\sf r}_1^0{\sf r}_2^0}+
\beta_0v_1^i\frac{{\sf J}_0}{({\sf r}_1^0)^2}+
\left[\frac{\partial}{\partial\beta}\left(\frac{\alpha\beta 
q^i}{\|{\sf r}_1\|}\right)\right]_{\beta_0}.\nonumber
\end{eqnarray}
They are supplemented with expressions (\ref{Dfin}), (\ref{D0}), 
and (\ref{DJ}) taken in the point where radial variable $R=0$.

First we put $\mu=0$ and $\nu=i$ into eq.(\ref{M-12}). The other terms 
which constitute the mixed spacetime components 
${\cal M}_{\rm em}^{0i}$ 
are obtained by interchanging of indices. A direct consequence of the 
reciprocity is the following combination of partial derivatives in $t_1$ 
and $t_2$:
\begin{eqnarray} \label{M0if}
{\cal M}_{\rm em}^{0i}&=&\frac{e^2}{2}I\left[
{\hat\Pi }_1^i\left(t\frac{\partial\lambda_2}{\partial t_1}\right) +
{\hat\Pi }_2^i\left(t\frac{\partial\lambda_1}{\partial t_2}\right)+
{\hat\Pi 
}^0\left[t\left(v_2^i\lambda_1+v_1^i\lambda_2\right)\right]\right.\\
&-&\left.\frac{\partial}{\partial t_1}\left(tv_2^i
\frac{\partial\lambda_1}{\partial t_2}{\cal D}^0
\right)-
\frac{\partial}{\partial t_2}\left(tv_1^i
\frac{\partial\lambda_2}{\partial t_1}{\cal D}^0
\right)
\right]\nonumber\\
&-&\frac{e^2}{2}I'
{\hat\Pi}^J\left[t\left(v_1^i+v_2^i\right)\right]\nonumber\\
&-&\frac{e^2}{2}I\left[
{\hat\Pi }_1^i\left(\Lambda\right) +
{\hat\Pi }_2^i\left(\Lambda\right)+
{\hat\Pi 
}^0\left[(z_1^i+z_2^i)\Lambda\right]-\frac{\partial}{\partial 
t_1}\left(v_2^i
\Lambda{\cal D}^0
\right)-
\frac{\partial}{\partial t_2}\left(v_1^i
\Lambda{\cal D}^0
\right)
\right]\nonumber\\
&+&\frac{e^2}{2}I'\left[
{\hat\Pi}_1^{iJ}(1)+{\hat\Pi}_2^{iJ}(1)+{\hat\Pi}^J\left(z_1^i+z_2^i\right)
-\frac{\partial}{\partial t_1}\left(v_2^i{\cal D}^J\right)
-\frac{\partial}{\partial t_2}\left(v_1^i{\cal D}^J\right)
\right]
\nonumber\\
&-&\frac{e^2}{4}\left\{
I\left[
{\hat\Pi}_1^i(\lambda)+{\hat\Pi}_2^i(\lambda)+
{\hat\Pi}^0\left[\lambda\left(z_1^i+z_2^i\right)\right]
-\frac{\partial}{\partial t_1}\left(v_2^i\lambda{\cal D}^0\right)
-\frac{\partial}{\partial t_2}\left(v_1^i\lambda{\cal D}^0\right)
\right]
\right.\nonumber\\
&-&\left.
I'\left[
{\hat\Pi}_1^{iJ}(\lambda_0)+{\hat\Pi}_2^{iJ}(\lambda_0)+
{\hat\Pi}^J\left[\lambda_0\left(z_1^i+z_2^i\right)\right]
-\frac{\partial}{\partial t_1}\left(v_2^i\lambda_0{\cal D}^J\right)
-\frac{\partial}{\partial t_2}\left(v_1^i\lambda_0{\cal D}^J\right)
\right]
\right\}\nonumber
\end{eqnarray} 
where
\begin{equation}
\Lambda=k_1^0k_2^0\frac{\partial^2\sigma}{\partial t_1\partial t_2}
+k_1^0\frac{\partial\sigma}{\partial t_1}
+k_2^0\frac{\partial\sigma}{\partial t_2}
+\sigma .
\end{equation} 

Now we turn to the integration over times $t_1$ and $t_2$. It is 
sufficient to examine the integrals near the point $R=0$. The computation
is virtually identical to that presented in \ref{yellow}. After a 
tedious calculation we obtain the following cumbersome expression:
\begin{eqnarray}\label{M0ifin}
M_{\beta_0}^{0i}&=&\frac{e^2}{2}\int_{-\infty}^tdt_1
t\frac{\alpha_0}{r_2^0}\left[
\frac{I_0}{{\sf J}_0}
\left(\frac{\partial\lambda_1}{\partial t_2}\alpha_0q^i
-\frac{\partial\lambda_2}{\partial t_1}\beta_0q^i
+v_1^i\lambda_2+v_2^i\lambda_1\right)
-I_0'\left(v_1^i+v_2^i\right)
\right]_{t_2\to -\infty}^{t_2=t_1}\nonumber\\
&-&e^2\int_{-\infty}^tdt_1\left[
\frac{\alpha_0}{r_2^0}\left(
\frac{I_0}{{\sf J}_0}\kappa-I_0'\mu\right)
\left(\alpha_0z_1^i+\beta_0z_2^i\right)
\right]_{t_2\to -\infty}^{t_2=t_1}\\
&+&\frac{e^2}{2}\int_{-\infty}^tdt_2\left[
\left(
t_2\frac{v_2^ik_1^0-q^i}{k_2^0}-z_2^i\frac{k_1^0-({\mathbf 
q}{\mathbf v}_2)}{k_2^0}
\right)
\left(\frac{1}{\sqrt{2\Sigma}}-\frac{1}{\sqrt{2\sigma}}\right)
+\frac{2k_1^0v_2^i}{\sqrt{2\sigma}}
\right]^{t_1=t}_{t_1=t_2}\nonumber\\
&+&\frac{e^2}{2}\int_{-\infty}^tdt_1\left[
\left(
t_1\frac{v_1^ik_2^0+q^i}{k_1^0}-z_1^i\frac{k_2^0+({\mathbf 
q}{\mathbf v}_1)}{k_1^0}
\right)
\left(\frac{1}{\sqrt{2\Sigma}}+\frac{1}{\sqrt{2\sigma}}\right)
+\frac{2k_2^0v_1^i}{\sqrt{2\sigma}}
\right]_{t_2\to -\infty}^{t_2=t_1}\nonumber\\
&+&
\frac{e^2}{2}\int_{-\infty}^tdt_2\left[
\frac{t_1v_2^i-z_1^i}{\sqrt{2\Sigma}}
\right]^{t_1=t}_{t_1=t_2}
+\frac{e^2}{2}\int_{-\infty}^tdt_1\left[
\frac{t_2v_1^i-z_2^i}{\sqrt{2\Sigma}}
\right]_{t_2\to -\infty}^{t_2=t_1}\nonumber\\
&+&\frac{e^2}{2}\int_{-\infty}^tdt_1\int_{-\infty}^{t_1}dt_2\left[
2\frac{\partial^2\sigma}{\partial t_1\partial t_2}
\frac{t_1z_2^i-t_2z_1^i}{(2\sigma)^{3/2}}
+t_1\frac{\partial}{\partial t_1}\left(\frac{v_2^i}{\sqrt{2\sigma}}\right)
-z_1^i\frac{\partial}{\partial t_1}\left(\frac{1}{\sqrt{2\sigma}}\right)
\right.\nonumber\\
&-&\left.t_2\frac{\partial}{\partial 
t_2}\left(\frac{v_1^i}{\sqrt{2\sigma}}\right)
+z_2^i\frac{\partial}{\partial t_2}\left(\frac{1}{\sqrt{2\sigma}}\right)
\right].\nonumber
\end{eqnarray} 
The single integrals belong to the boundary part of angular momentum 
carried by the electromagnetic field. We couple them with integrals 
over $\beta$ taken at the end points $t_1=t_2$ and $t_2\to -\infty$. 
(Such integrals are described in subsections 
\ref{trace}.$1^o$-\ref{trace}.$4^o$.) The result is as follows:
\begin{equation}
M_S^{0i}=
\frac{e^2}{2}t\int_{-\infty}^tds\frac{v^i(s)}{\sqrt{2\sigma(t,s)}}-
\frac{e^2}{2}z^i(t)\int_{-\infty}^t\frac{ds}{\sqrt{2\sigma(t,s)}}.
\end{equation}
The double integral in eq.(\ref{M0ifin}) describes the radiative part; it 
can be rewritten as follows:
\begin{eqnarray}\label{MR0i}
M_R^{0i}&=&\frac{e^2}{2}\int_{-\infty}^tdt_1\int_{-\infty}^{t_1}dt_2\left[
t_1v_{1,\alpha}\frac{-v_2^\alpha q^i+v_2^iq^\alpha}{(2\sigma)^{3/2}}
-z_1^iv_{1,\alpha}\frac{-v_2^\alpha q^0+q^\alpha}{(2\sigma)^{3/2}}
\right.\nonumber\\
&+&\left.
t_2v_{2,\alpha}\frac{-v_1^\alpha q^i+v_1^iq^\alpha}{(2\sigma)^{3/2}}
-z_2^iv_{2,\alpha}\frac{-v_1^\alpha q^0+q^\alpha}{(2\sigma)^{3/2}}
\right].
\end{eqnarray} 
Taking $t_2\to t_1$ limit reveals the proper short-distance 
behaviour.

Now we calculate the space component $M_{\rm em}^{ij}$. Setting 
$\mu=i$ and $\nu=j$ into eq.(\ref{M-12}) we obtain ${\cal M}_{12}^{ij}$.
Having interchanged upper and lower indices we find all the terms which 
constitute the expression ${\cal M}_{\rm em}^{ij}$ obtained from 
eq.(\ref{mm}) via integration over $\varphi$. Further we integrate 
them over times $t_1$ and $t_2$. After tedious calculations we finally 
obtain:
\begin{eqnarray}\label{Mijfin}
M_{\beta_0}^{ij}&=&\frac{e^2}{2}\int_{-\infty}^tdt_1\left\{
\frac{\alpha_0}{r_2^0}\left[
\frac{I_0}{{\sf J}_0}
\left(\frac{\partial\lambda_1}{\partial t_2}\alpha_0(z_1^iq^j-z_1^jq^i)
-\frac{\partial\lambda_2}{\partial t_1}\beta_0(z_2^iq^j-z_2^jq^i)
\right.\right.\right.\\[-0.5ex]
&+&\left.\left.\left.(\alpha_0z_1^i+\beta_0z_2^i)(v_1^j\lambda_2+v_2^j\lambda_1)
-(\alpha_0z_1^j+\beta_0z_2^j)(v_1^i\lambda_2+v_2^i\lambda_1)
\right)\right.\right.\nonumber\\
&-&\left.\left.I_0'\left((\alpha_0z_1^i+\beta_0z_2^i)(v_1^j+v_2^j)
-(\alpha_0z_1^j+\beta_0z_2^j)(v_1^i+v_2^i)
\right)\phantom{\frac{I_0}{{\sf J}_0}}\!\!\!\!\!\!\!\!
\right]
\right\}_{t_2\to -\infty}^{t_2=t_1}\nonumber\\[-1ex]
&+&\frac{e^2}{2}\int_{-\infty}^tdt_2\left[
\frac{1}{k_2^0}\left(
z_1^iz_2^j-z_1^jz_2^i+k_1^0(z_2^iv_2^j-z_2^jv_2^i)
\right)
\left(\frac{1}{\sqrt{2\Sigma}}-\frac{1}{\sqrt{2\sigma}}\right)
\right]^{t_1=t}_{t_1=t_2}\nonumber\\
&+&\frac{e^2}{2}\int_{-\infty}^tdt_1\left[
\frac{1}{k_1^0}\left(
-z_1^iz_2^j+z_1^jz_2^i+k_2^0(z_1^iv_1^j-z_1^jv_1^i)
\right)
\left(\frac{1}{\sqrt{2\Sigma}}+\frac{1}{\sqrt{2\sigma}}\right)
\right]_{t_2\to -\infty}^{t_2=t_1}\nonumber\\
&+&
\frac{e^2}{2}\int_{-\infty}^tdt_2\left[
\frac{z_1^iv_2^j-z_1^jv_2^i}{\sqrt{2\Sigma}}
\right]^{t_1=t}_{t_1=t_2}
+\frac{e^2}{2}\int_{-\infty}^tdt_1\left[
\frac{z_2^iv_1^j-z_2^jv_1^i}{\sqrt{2\Sigma}}
\right]_{t_2\to -\infty}^{t_2=t_1}\nonumber\\
&+&\frac{e^2}{2}\int_{-\infty}^tdt_1\int_{-\infty}^{t_1}dt_2\left[
2\frac{\partial^2\sigma}{\partial t_1\partial t_2}
\frac{z_1^iz_2^j-z_1^jz_2^i}{(2\sigma)^{3/2}}
+z_1^i\frac{\partial}{\partial t_1}\left(\frac{v_2^j}{\sqrt{2\sigma}}\right)
-z_1^j\frac{\partial}{\partial t_1}\left(\frac{v_2^i}{\sqrt{2\sigma}}\right)
\right.\nonumber\\
&-&\left.z_2^i\frac{\partial}{\partial 
t_2}\left(\frac{v_1^j}{\sqrt{2\sigma}}\right)
+z_2^j\frac{\partial}{\partial t_2}\left(\frac{v_1^i}{\sqrt{2\sigma}}\right)
\right].\nonumber
\end{eqnarray} 
All the single integrals should be added to the integrals over $\beta$ 
evaluated at limit points $t_1=t_2$ and $t_2\to -\infty$; the sum is
the singular part of angular momentum of the electromagnetic field:
\begin{equation}
M_S^{ij}=
\frac{e^2}{2}z^i(t)\int_{-\infty}^tds\frac{v^j(s)}{\sqrt{2\sigma(t,s)}}-
\frac{e^2}{2}z^j(t)\int_{-\infty}^tds\frac{v^i(s)}{\sqrt{2\sigma(t,s)}}.
\end{equation}
The integrand of the radiative part is symmetric in indices $(12)$ 
and antisymmetric in indices $(ij)$:
\begin{eqnarray}\label{MRij}
M_R^{ij}&=&\frac{e^2}{2}\int_{-\infty}^tdt_1\int_{-\infty}^{t_1}dt_2\left[
z_1^iv_{1,\alpha}\frac{-v_2^\alpha q^j+v_2^jq^\alpha}{(2\sigma)^{3/2}}
-z_1^jv_{1,\alpha}\frac{-v_2^\alpha q^i+v_2^iq^\alpha}{(2\sigma)^{3/2}}
\right.\nonumber\\
&+&\left.
z_2^iv_{2,\alpha}\frac{-v_1^\alpha q^j+v_1^jq^\alpha}{(2\sigma)^{3/2}}
-z_2^jv_{2,\alpha}\frac{-v_1^\alpha q^i+v_1^iq^\alpha}{(2\sigma)^{3/2}}
\right].
\end{eqnarray} 
The resulting expressions (\ref{MR0i}) and (\ref{MRij}) can be 
rewritten in a manifestly covariant fashion.


\begin{thebibliography}{99}
\bibitem{Gl}
D. V. Gal'tsov, Phys. Rev. D {\bf 66}, 025016 (2002).

\bibitem{KLS}
P. O. Kazinski, S. L. Lyakhovich, and A. A. Sharapov, Phys. Rev. D 
{\bf 66}, 025017 (2002).

\bibitem{Dir}
P. A. M. Dirac, Proc. R. Soc. London, Ser. A {\bf 167}, 148 (1938).

\bibitem{Rohr}
F. Rohrlich, {\it Classical Charged Particles} (Addison-Wesley, Redwood, 
CA, 1990).

\bibitem{PsPr}
E. Poisson, e-print arXiv:gr-qc/9912045.

\bibitem{TVW}
C. Teitelboim, D. Villarroel, and C. C. van Weert, Riv.Nuovo Cimento {\bf 
3}, 9 (1980).

\bibitem{WB}
B. S. DeWitt and R. W. Brehme, Ann.Phys. (N.Y.) {\bf 9}, 220 (1960).

\bibitem{Hb}
J. M. Hobbs, Ann.Phys. (N.Y.) {\bf 47}, 141 (1968).

\bibitem{DW}
S. Detweiler and B. F, Whiting, Phys. Rev. D {\bf 67}, 024025 (2003).

\bibitem{Teit}
C. Teitelboim, Phys. Rev. D {\bf 1}, 1572 (1970).

\bibitem{LV}
C. A. L\'opez and D. Villarroel, Phys. Rev. D {\bf 11}, 2724 (1975).

\bibitem{Yar03}
Yu. Yaremko, J.Phys.A {\bf 36}, 5149 (2003).

\bibitem{Yar04}
Yu. Yaremko, J.Phys.A {\bf 37}, 1079 (2004).

\bibitem{Kos}
B. P. Kosyakov, Teor. Mat. Fiz. {\bf 119}, 119 (1999); [Theor. Math. Phys. 
{\bf 119}, 493 (1999)].

\bibitem{QW}
T. C. Quinn and R. M. Wald, Phys. Rev D {\bf 60}, 064009 (1999).

\bibitem{MST}
Y. Mino, M. Sasaki, and T. Tanaka, Phys. Rev. D {\bf 55}, 3457 (1997).

\bibitem{Q}
T. C. Quinn, Phys. Rev. D {\bf 62}, 064029 (2000).

\bibitem{Pois}
E. Poisson, Living Rev. Relativ. {\bf 7} (2004).

\bibitem{Br}
L. M. Burko, Class. Quant. Grav. {\bf 19}, 3745 (2002).

\bibitem{AHNS}
V. Ambegaokar, B. I. Halperin, D. R. Nelson, and E. D. Siggia, Phys. Rev. B 
{\bf 21}, 1806 (1980).

\bibitem{FL}
M. P. A. Fisher and D. H. Lee, Phys. Rev. B {\bf 39}, 2756 (1989).

\bibitem{Z}
S.-C. Zhang, e-print arXiv:hep-th/0210162.
\end{thebibliography}
\end{document}